\documentclass[a4paper,12pt]{article}
\usepackage[utf8]{inputenc}

\usepackage{amsmath,amsfonts,amsthm,amssymb,dsfont}
\usepackage{graphicx,wrapfig,lipsum}
\usepackage[section]{placeins}
\usepackage{indentfirst}
\usepackage{cite}

\usepackage{tikz}
\usetikzlibrary{shapes.geometric,decorations.markings}
\usepackage{subfigure}

\usepackage[hidelinks]{hyperref}
\hypersetup{colorlinks = true, linkcolor = red, linktocpage = true, citecolor = blue}

\usepackage{geometry}
\numberwithin{equation}{section}
\newcommand{\Vol}{\text{Vol}}
\newcommand{\Trh}[1]{h^{#1}_{\phantom{#1}#1}}
\newcommand{\GUdd}[3]{\Gamma^{#1}_{\phantom{#1}#2 #3}}
\newcommand{\hGUdd}[3]{\hat{\Gamma}^{#1}_{\phantom{#1}#2 #3}}
\newcommand{\RUddd}[4]{R^{#1}_{\phantom{#1}#2 #3 #4}}

\newcommand{\hRUddd}[4]{\hat{R}^{#1}_{\phantom{#1}#2 #3 #4}}
\newcommand{\hRdddd}[4]{\hat{R}_{#1 #2 #3 #4}}
\newcommand{\dg}{\delta g}
\newcommand{\Id}{\mathbb{I}}

\newcommand{\hypgeom}[4]{\left._{2}F_{1}\right.\left( #1,#2;#3;#4\right)}
\newcommand{\IntegerPart}[1]{\left\lfloor #1\right\rfloor}
\newcommand{\bvarphi}{\bar{\varphi}}

\begin{document}

\begin{titlepage}

\begin{center}

$\phantom{.}$\\ \vspace{2cm}
\noindent{\Large{\textbf{Confinement and D5-Branes}}}

\vspace{1cm}

Carlos Nunez$^a$\footnote{c.nunez@swansea.ac.uk}, 
Marcelo Oyarzo$^b$\footnote{moyarzoca1@gmail.com} and Ricardo Stuardo$^a$\footnote{ricardostuardotroncoso@gmail.com}

\vspace{0.5cm}

\textit{$^a$Department of Physics, Swansea University, Swansea SA2 8PP, United Kingdom\\
$^b$Departamento de Física, Universidad de Concepción, Casilla 160-C, Concepción, Chile }

\end{center}

\vspace{0.5cm}
\centerline{\textbf{Abstract}} 

\vspace{0.5cm}

\noindent{In this work we present new solutions of type IIB supergravity based on wrapped D5 branes. We propose that two of these backgrounds are holographically dual to Quantum Field Theories that confine. The high energy regime of the field theories is that of a Little String Theory. We study various observables (Wilson and 't Hooft loops, Entanglement entropy, density of degrees of freedom and the spectrum of spin-two glueballs, among others). We also present two new black membrane backgrounds and analyse some thermodynamic aspects of these solutions.}

\vspace{0.5cm}

\flushright{\textit{Dedicated to the memory of Fidel A. Schaposnik}}

\vspace*{\fill}

\end{titlepage}

\newpage

\tableofcontents
\thispagestyle{empty}

\newpage
\setcounter{page}{1}
\setcounter{footnote}{0}

\section{Introduction}
Maldacena presented the AdS/CFT conjecture  in \cite{Maldacena:1997re}, and very relevant refinements appeared in \cite{Gubser:1998bc}, \cite{Witten:1998qj}. Various papers elaborated on different aspects of the correspondence. In particular, the beautiful works \cite{Itzhaki:1998dd},  \cite{Witten:1998zw}, \cite{Boonstra:1998mp} explored the idea of applying the duality in non-conformal settings.

This lead to the study of different set-ups, exploring the duality in phenomenologically appealing situations (by this we mean, string backgrounds dual to minimally SUSY or SUSY-breaking QFTs). 
There are two  well-developed  lines of work:

\begin{itemize}
    \item{ A very fertile line of work, dealing with a two-node quiver field and deformation/resolution of the conifold was developed in various works. See, for example \cite{Klebanov:1998hh}-\cite{Butti:2004pk}.}
    \item{Considering higher dimensional p-branes D$_{p\geq 4}$, with some compact directions, such that the compactified lower dimensional (low energy) QFT is either non-SUSY or with small amount of SUSY preserved. See for example  \cite{Maldacena:2000yy}-\cite{Aharony:2002up}.}
\end{itemize}
The papers \cite{Maldacena:2009mw}-\cite{Gaillard:2010gy} make contact between the geometries describing the first and second lineages of models. The geometric connection implies a relation between the associated QFTs in each case. The geometrisation of a number of field theoretical aspects was achieved: confinement, symmetries breaking, glueball and meson like excitations, other non-perturbative excitations in the field theory and  finite temperature aspects, among other things.

In this paper we present new type IIB backgrounds that are examples of the second type of models. We work with D5 branes, compactify them on circles or two-spheres and generate new solutions of the Type IIB equations of motion. The solutions are particularly simple, the dilaton is exactly linear, the radial functions of the circle and two-sphere are analytic. The backgrounds presented in this paper can be used as starting point for calculations that in the past gave information about non-perturbative aspects of interesting QFTs. Various computations are explicit and analytic thanks to the simplicity of the background. 

The material in this paper is organised as follows.
\\
\\
{\bf --}
In Section \ref{sectionBIxx} we study a new smooth type IIB solutions dual to a $(4+1)$ dimensional QFT with minimal amount of SUSY (or zero SUSY, depending on parameters). This background is inspired on a kind of solutions recently found by Anabal\'on and Ross \cite{Anabalon:2021tua}, see also \cite{Canfora:2021nca}. The UV completion in terms of a Little String Theory (LST) is discussed. Wilson loops, 't Hooft loops and Entanglement Entropy are calculated, together with a quantity that gives and estimate of the number of degrees of freedom as a function of the energy (density of states). Masses for spin-two glueballs are calculated. Depending on the parameters of the model we find a spectrum that starts discrete and then becomes continuous, or is purely continuous (as the energy is increased). A notion of gauge coupling is defined that is in agreement with the confining behaviour displayed by Wilson loops and screening behaviour shown by 't Hooft loops.
\\
\\
{\bf --}
In Section \ref{section2+1qft} we write a new non-SUSY and smooth type IIB solution representing D5 branes wrapping $S^1\times S^2$. Whilst the $S^1$ shrinks smoothly, the $S^2$ remains of finite size. The finite size of the $S^2$ (which  is held stable by a meron-type gauge field) is responsible for certain non-field theoretic behaviour of observables of the QFT. We compare with the results of Section \ref{sectionBIxx}.
\\
\\
{\bf --}
Section \ref{sectioncomparison} makes a comparison between models presented in past bibliography and the models in Sections \ref{sectionBIxx} and \ref{section2+1qft}. Depending on the fate (shrinking or stabilised) of the space the D5 branes wrap, some physical observables turn out to behave differently. We attempt a geometric classification of these different models. We also discuss the difference and similarities between SUSY preservation via twisting or via the insertion of a Wilson line.
\\
\\
{\bf --}
In Section \ref{bhsectionsx} we present two new black membrane solutions. They are obtained by double Wick rotation of the backgrounds in Sections \ref{sectionBIxx} and \ref{section2+1qft}. Some aspects of the thermodynamics are studied.
\\
\\
{\bf --}
Section \ref{sectionconclusions} gives a summary of our results, conclusions and lines for future research. Various appendices, written with pedagogical intention, discuss explicit technical details of the calculations. Hopefully, these are useful to colleagues wishing to work on the topic.

\section{A dual to a (4+1)d confining SUSY QFT}\label{sectionBIxx}

In this section we present a Type IIB background describing a stack of D5 branes that, in a SUSY preserving fashion, wraps a circle direction, leading to a smooth manifold over the whole space.

To describe this system we use coordinates $[t,x_1,x_2,x_3,x_4, \varphi, r,\theta,\phi,\psi]$. We define the left invariant forms of $SU(2)$, in terms of the 
the coordinates $[\theta,\varphi,\psi]$ that parameterise a three sphere,
\begin{eqnarray}
& & \omega_1= \cos\psi d\theta +\sin\psi \sin\theta d\phi,\;\;\;\; \omega_2= -\sin\psi d\theta+\cos\psi \sin\theta d\phi,\;\;\;\; \omega_3=d\psi+\cos\theta d\phi,\nonumber\\
& &  \Vol(S^{3})= \frac{1}{8}\omega_1\wedge\omega_2\wedge\omega_3=\frac{1}{8}\sin\theta d\theta\wedge d\phi\wedge d\psi.\label{su2-left}
\end{eqnarray}
The range of the coordinates in the three sphere is $0\leq\theta\leq\pi$, $0\leq\phi\leq 2\pi$ and $0\leq\psi\leq 4\pi$.
We set the constants $\alpha'=1, g_s=1$. In string frame the background reads,
    \begin{equation}
    \begin{aligned}
        ds^{2}_{st} &= r \left[ dx_{1,4}^2 + f_{s}(r)d\varphi^{2} 
            + \frac{Ndr^{2}}{ r^{2}f_{s}(r)} 
            + \frac{N}{4}\left(\omega^{2}_{1} + \omega^{2}_{2} + 
        \left(\omega_{3} - \sqrt{\frac{8}{N}} Q \zeta(r) d\varphi\right)^{2}\right) \right],\\
        F_{3} &= 2N \Vol(S^{3}) + \sqrt{\frac{N}{2}}Q~ d\left( \zeta(r) \omega_{3}\wedge d\varphi \right),\;\;C_2 =\frac{N}{4}\psi \sin\theta d\theta \wedge d\phi+{\sqrt{\frac{N}{2}}Q} \zeta(r) \omega_3\wedge d\varphi,\\
        \Phi &= \log(r),\label{BI}
    \end{aligned}
    \end{equation}

the electric RR 7-form is
    \begin{equation}\label{C6potential}
    \begin{aligned}
        F_{7} &= -\star F_{3}=- dt \wedge dx_{1} \wedge dx_{2} \wedge dx_{3} \wedge dx_{4}\wedge  \left[ 2 r dr \wedge  d\varphi +Q\sqrt{\frac{N}{2}} \sin\theta d\theta\wedge d\phi \right]\\
      C_6 &= dt \wedge dx_{1} \wedge dx_{2} \wedge dx_{3} \wedge dx_{4}\wedge\left[r^2d\varphi- Q\sqrt{\frac{N}{2}} \cos\theta d\phi \right].
    \end{aligned}
    \end{equation}

The background is defined in terms of the functions $f_s(r)$, $\zeta(r)$ and three parameters ($N$, $r_\pm$ or $Q,m$),
 \begin{eqnarray}
& & f_s(r)=1-\frac{m}{r^2}-\frac{2Q^2}{r^4}=\frac{(r^2-r_+^2)(r^2-r_-^2)}{r^4},\;\;\;\;\; \zeta(r)=\frac{1}{r^2}-\frac{1}{r_+^2},\nonumber\\
& &  2 r_{\pm}^2= m\pm \sqrt{m^2+ 8 Q^2}.\label{functions4+1}
\end{eqnarray}
This background is a solution to the equations of motion of Type IIB, summarised in Appendix \ref{TypeIIBapp}. For the parameter $m=0$, it can be shown by operating with the SUSY variations in eqs.(\ref{susyIIB}),  that there are eight preserved spinors. See Appendix \ref{BIsusy} for the details\footnote{Niall Macpherson has informed us that the sub-manifold $\Sigma_5=[r,\varphi,\theta,\phi,\psi]$ preserves $SU(2)$ structure. We are thankful for the feedback.}.

To calculate the number of five branes, we integrate $F_3$ on the manifold $\Sigma_3=[\theta,\phi,\psi]$. 
 Using that 
\begin{equation}
(2\pi)^{7-p} g_s \alpha'^{\frac{7-p}{2}}N_{Dp}= \int_{\Sigma_{8-p}}F_{8-p},
\end{equation} 
and   setting $g_s=\alpha'=1$ we find,
\begin{equation}
N_{D5}=\frac{1}{4\pi^2}\int_{\Sigma_3}F_3= N.\label{quantcharge1}
\end{equation}

\underline{\bf Comments on SUSY preservation}

Placing a SUSY QFT on a circle and imposing anti-periodic boundary conditions for the fermions completely breaks SUSY. However, by inserting a Wilson Loop of a constant gauge field around the $S^{1}$,  spinors get charged under the $U(1)$ symmetry of the cycle, such that now is possible to preserve some amount of SUSY. This can be seen explicitly in Appendix \ref{AppendixSpinor}. The mechanism was recently explained and applied by Anabal\'on and Ross \cite{Anabalon:2021tua} (see also \cite{Anabalon:2022aig} and see \cite{Bobev:2020pjk} for a precursor). We explain this below.

The flat D5-brane configuration preserves 16 supercharges which do not depend on the six flat directions of the D5-branes world-volume. Periodically identifying one of the flat directions (for example $x^{5}$) and imposing anti-periodic boundary conditions on this circle  completely breaks SUSY (even when we are not adding the thermal factor of the cigar). By inserting the Wilson line, the spinor is now charged under translations on the $S^{1}$, see \eqref{SpinorWilson1}. This allows the supercharges to satisfy the anti-periodic boundary conditions, leading to SUSY preservation.

In order to obtain a 6D QFT that flows to a 5D confining one, we insert a radially dependent Wilson line for the $U(1)$ symmetry along the $S^{1}$. This is realised holographically by the radially dependent fibration of the 3-sphere over the shrinking $S^{1}$.

\underline{\bf Comments on the field theory}

We interpret the background in eq.(\ref{BI}) as a stack of $N$ D5 branes. On these five branes, the $\varphi$-direction is compactified and fibered over the external three sphere --this is the effect of the function $\zeta(r)$. This fibration is such that some amount of SUSY is preserved and the space time ends at $r=r_+$ in a smooth fashion (we explain this in Appendix \ref{BIsusy}). The smoothness-condition imposes the periodicity $\varphi\sim \varphi + L_{\varphi}$ to be
    \begin{equation}\label{CirclePeriod}
        L_{\varphi} = \frac{2\pi \sqrt{N} r_+^2}{(r_+^2-r_-^2)} 
                  = \sqrt{N} \pi  \left( 1 + \frac{m}{\sqrt{m^{2} + 8Q^{2}}}\right).
    \end{equation}

The Ricci scalar for the background of eq.(\ref{BI}) is,
    \begin{equation}\label{ricci}
        R= -\frac{12}{N r} + \frac{4m}{Nr^{3}} - \frac{8Q^{2}}{N r^{5}},
    \end{equation}
this is bounded as $r$ varies in $[r_+,\infty)$.

The field theory  holographically dual to this background is, when $m=0$, a SUSY QFT in $(4+1)$-dimensions when observed at low energies (the region close to $r\sim r_+$). At higher energies the $\varphi$-direction decompactifies and the QFT is UV completed by the theory on D5 branes $(5+1)$ dimensional Super Yang-Mills and then, at even higher energies, a Little String Theory (after S-duality). In fact, whilst the Ricci scalar and other curvature invariants are bounded in the region of large radial coordinate $r$, the dilaton grows unbounded. An S-duality to the NS-five brane frame is needed. The effect of the fibration is to introduce a Wilson line (in the QFT) along the compactified direction $\varphi$.
%
%
In summary, at low energies, the field theory is $(4+1)$ dimensional, apparently gapped and confining (we substantiate these properties in the forthcoming sections). At higher energies, the theory recovers its $(5+1)$ dimensional character and is UV completed by the theory on NS-five branes, a Little String theory (LST).

We can rescale $e^{\Phi}\to e^{\Phi(r)} e^{\Phi_0}$ and $F_3\to e^{-\Phi_0} F_3$, which rescales the Newton constant but keeps the same equations of motion. We can choose $e^{-\Phi_0}$ to be a large integer number. Hence, the range of the radial coordinate for which $e^{\Phi}<1$ can be made large (at the same time, the number of five branes made large and quantised).

The scales at which these different description take over are: $E\sim r_+$, for the scale that compactifies to $(4+1)$ dimensions. The theory becomes six dimensional when $f_s(r_*)\sim 1$. Then, when $e^{\Phi}\sim 1$ is the scale at which the S-duality is needed and the NS-five brane description takes over.

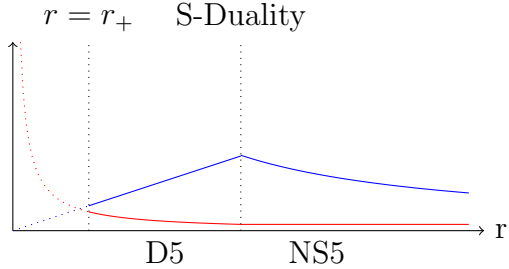
\begin{figure}[h!]
    \centering
    \begin{tikzpicture}
    \draw[->] (0, 0) -- (6.2, 0) node[right] {r};
    \draw[->] (0, 0) -- (0, 2.5) ;
    \draw[scale=1, domain=0:1, dotted, variable=\x, blue] plot ({\x}, {0.33*\x});
    \draw[scale=1, domain=1:3, smooth, variable=\x, blue] plot ({\x}, {0.33*\x});
    \draw[scale=1, domain=3:6, smooth, variable=\x, blue] plot ({\x}, {3*\x^(-1)});
    \draw[scale=1, domain=0.1:1, dotted, variable=\x, red]  plot ({\x}, {0.25*\x^(-1)});
    \draw[scale=1, domain=1:3, smooth, variable=\x, red]  plot ({\x}, {0.25*\x^(-1)});
    \draw[scale=1, domain=3:6, smooth, variable=\x, red]  plot ({\x}, {0.25*3^(-1)});
    \draw[scale=1, domain=0:2.5, dotted, variable=\y, black] plot ({1}, {\y}) node[above] {$r=r_{+}$};
    \draw[scale=1, domain=0:2.5, dotted, variable=\y, black] plot ({3}, {\y}) node[above] {S-Duality};
    \node[below] at (2,0) {D5};
    \node[below] at (4,0) {NS5};
\end{tikzpicture}
    \caption{Behaviour of the string coupling constant (blue) and the Ricci scalar (red) as a function of the holographic coordinate. The two scales of the theory are depicted: at $r=r_{+}$ the compactification circle is effectively of zero size and the theory is (4+1)-dimensional. When the string coupling constant becomes of order one, we S-dualise and the theory is described in terms of the NS5-brane.}
\end{figure}

In what follows, we calculate various observables associated with the QFT. These display behaviours that give account of the IR properties of the field theory (like confinement of quarks, screening of monopoles) and also some of the high energy properties, like the presence of a minimal length, characteristic of the above mentioned  non-local UV-completion. We also propose a definition of an effective gauge coupling, we discuss how to calculate the Entanglement Entropy in a strip region, and describe the calculation and spectrum of spin-two glueballs in the QFT.
\subsection{Observables}
In this section we calculate various observables using the holographic background in eq.(\ref{BI}), by probing it with various available objects in Type IIB.
This  produces values characterising the strongly coupled QFT that we describe above. We start by finding a gauge coupling of the $(4+1)$ QFT.
\subsubsection{Gauge coupling}\label{gaugecoupling1}
To calculate the gauge coupling of the QFT, we probe the background with a D5 brane that extends over $[t,x_1,x_2,x_3,x_4,\varphi]$. We switch on an Abelian gauge field on the brane and calculate the coefficient in front of the gauge field kinetic term (for a small-field expansion). This coefficient in general depends on the radial coordinate, indicating that the gauge coupling is energy dependent. The general procedure was reviewed in \cite{Nunez:2023nnl} and applied in different situations. We give a brief account below.

The induced metric on the D5 brane is 
\[
ds^2_{D5,ind}=rdx_{1,4}^{2}+r\left( f_{s}\left( r\right)
+2Q^{2}\zeta \left( r\right) ^{2}\right) d\varphi ^{2}\ , 
\]%
with determinant%
\begin{equation}
e^{-\Phi }\sqrt{-\det \boldsymbol{g}_{D5,ind}}=r^{2}\left( f_{s}\left( r\right)
+2Q^{2}\zeta \left( r\right) ^{2}\right) ^{1/2}\ .\nonumber
\end{equation}%
The Born-Infeld action is%
\begin{equation}
S_{BI}=T_{D5}\int d^{6}xe^{-\Phi }\sqrt{-\det \left( g_{MN}+2\pi  F_{MN}\right) }\ ,
\end{equation}%
with $\boldsymbol{g}_{D5,ind}=g_{MN}dx^{M}dx^{N}$. When expanded in the weak field
limit  this leads to,
\begin{eqnarray*}
S_{BI} &=&T_{D5}\int d^{6}xe^{-\Phi }\sqrt{-\det \mathbf{g}_{D5,ind}}\left( 1-%
\frac{1}{4}\left( 2\pi \right) ^{2}F^{MN}F_{MN}\right) \ ,
\\
&=&T_{D5}r^{2}\left( f_{s}\left( r\right) +2Q^{2}\zeta \left( r\right)
^{2}\right) ^{1/2}\int d^{5}x d\varphi \left( 1-\frac{1}{4}\left( 2\pi  \right) ^{2}g^{MA}g^{NB}F_{AB}F_{MN}\right) \ .
\end{eqnarray*}%
Turning on $F_{tx}$ on the brane (this can be generalised to all the Minkowski directions), we obtain%
\begin{eqnarray}
S_{BI} &=&T_{D5}r^{2}\left( f_{s}\left( r\right) +2Q^{2}\zeta \left(
r\right) ^{2}\right) ^{1/2}\int d^{5}x d\varphi\left( 1-\frac{1}{4}\left( 2\pi \right) ^{2}2g^{tt}g^{xx}F_{tx}F_{tx}\right) \ ,\nonumber \\
&=&T_{D5} L_\varphi r^{2}\left( f_{s}\left( r\right) +2Q^{2}\zeta \left( r\right)
^{2}\right) ^{1/2}\int d^{5}x\left( 1+\frac{1}{2r^{2}}\left( 2\pi \right) ^{2}F_{tx}F_{tx}\right) \ .\label{borninfeld}
\end{eqnarray}%
We have performed the integration over the $\varphi$-direction (represented by the factor $L_\varphi$ above), to construct the  effective $(4+1)$d gauge coupling. We read the gauge coupling,
\begin{equation}
\frac{1}{4g_{YM}^{2}}=2\pi^2 T_{5}L_{\varphi }\left( f_{s}\left(
r\right) +2Q^{2}\zeta \left( r\right) ^{2}\right) ^{1/2}\ .
\end{equation}%

The asymptotic values of the coupling are%
\begin{equation}
\frac{1}{g_{YM}^{2}}=\left\{ 
\begin{array}{ccc}
8 \pi^2 T_{5}L_{\varphi }\sqrt{1+\frac{2Q^2}{r_{+}^{4} }  } & , & 
r\rightarrow \infty \\ 
0 & , & r\rightarrow r_{+}
\end{array}%
\right. \ .\label{gaugecoupling}
\end{equation}
We interpret this result as follows: at high energies (large values of $r$), the gauge coupling becomes a large constant value. This value is characteristic of the LST. It is the constant coupling of the LST. Note that the constant part of the Wilson line (the term $\frac{2Q^2}{r_+^2}$) enters the coupling above. On the other hand, for low energies, when $r\sim r_+$, we find a gauge coupling that grows unbounded. This suggest that the QFT confines at low energies.

Let us observe that the Wess-Zumino term for the probe D5 branes is,
\begin{eqnarray}
S_{WZ} &=&-T_{D5}\int \left( C_{6}-C_{4}\wedge F_{2}+\frac{1}{2}C_{2}\wedge
F_{2}\wedge F_{2}-\frac{1}{6}C_{0}F_{2}\wedge F_{2}\wedge F_{2}\right) =-T_{D5}\int C_{6}\ ,\nonumber
\end{eqnarray}%
Using the expression for $C_6$ in eq.\eqref{C6potential}, we find
\begin{eqnarray}
S_{WZ} &=&- r^{2}T_{D5}\int dt\wedge d\varphi \wedge dx_{1}\wedge
dx_{2}\wedge dx_{3}\wedge dx_{4}\ ,  \nonumber \\
&=&-r^{2}T_{D5} L_\varphi \int d^{5}x\ .\label{wesszumino}
\end{eqnarray}%
Comparing the expression above with the 'tension' term in eq.(\ref{borninfeld}), we observe that (even in the case $m=0$) this probe is not SUSY preserving. The probe is effectively attracted towards $r_+$ (the tension  is bigger than the charge).

{We close this section with some comments motivated by the above material. First, note that the probe we used is not SUSY. Either there is a different similar D5 probe that is SUSY or the Coulomb branch of the six dimensional QFT is lifted. It would be interesting to further study this.
Second, note that the naive dimensional analysis, indicating that five dimensional QFTs should have weakly coupled IR dynamics is not working here. As mentioned, a VEV for an R-symmetry current might be changing this. Note the same would occur if working with a Witten-like compactification on a circle (breaking SUSY) for Dp branes with $p\geq 5$.}

We go back to the point made below eq.(\ref{gaugecoupling}) that the QFT confines. To substantiate it,  we compute the Wilson loop.
\subsubsection{Wilson Loops}\label{wilsonsectionx}
There is a very developed algorithmic way of computing Wilson loops.  The reader interested in the details should consult the papers \cite{Nunez:2023nnl}-\cite{Sonnenschein:1999if}. 
Here we just summarise the main points briefly.

For observables calculated using branes (usually non-local operators in the dual QFT), it is usual to arrive (after integrating over some coordinates) to an effective action of the form
\begin{equation}
S= T_{eff} \int d\sigma \sqrt{F^2(r) + G^2(r) r'^2}.\label{seffectiva}
\end{equation}
 It is customary to define an ''effective potential'' associated with this effective action. This encodes some of the asymptotic behaviour of the non-local observable (for example, the length). The effective potential $V_{eff}$ is
 \begin{eqnarray}
V_{eff}\left( r\right) &=&\frac{F\left( r\right) }{F\left( r_{0}\right)
G\left( r\right) }\sqrt{F\left( r\right) ^{2}-F\left( r_{0}\right) ^{2}}\ .\label{Veff}
\end{eqnarray}%
Here $r_0$ is the position in the radial coordinate at which the embedding of the extended object has a turning point $\frac{dr}{dx}|_{r=r_0}=0$. We must impose some condition on $V_{eff}$, for example $V_{eff}(r\to\infty)\to\infty$, to achieve good asymptotic behaviour of the holographic calculation of the observables, see \cite{Nunez:2009da}
 for a careful derivation and detailed explanations.

The length of the operator can be calculated by integration of the inverse of $V_{eff}$,
\begin{eqnarray}
L(r_0)= 2\int_{r_0}^\infty \frac{dr}{V_{eff}(r)}.\label{lengthwilson}
\end{eqnarray}
The above integral is usually hard to perform analytically. In such cases, an approximate expression is useful. We denote this approximate expression by $\hat{L}$. It is given by \cite{Kol:2014nqa},
\begin{eqnarray}
L_{app}=\hat{L}\left( r_{0}\right) &=&\left. \pi \frac{G}{F^{\prime }}%
\right\vert _{r=r_{0}}\ .\label{laprox}
\end{eqnarray}%
Similarly, the energy of the observable is calculated as (see \cite{Nunez:2009da}
 for a careful derivation),
 \begin{equation}
 E= F(r_0) L(r_0) + 2 \int_{r_0}^\infty \frac{G(r)}{F(r) }\sqrt{ F^2(r)-F^2(r_0) }  dr -2 \int_{r_+}^\infty G(r) dr.\label{energywilson}
\end{equation}
The stability of the brane embedding used to get eq.(\ref{seffectiva}) is determined by the function
\begin{equation}
Z\left( r_0\right) =\frac{d}{dr_0}\hat{L} \left( r_0\right) \ .\label{zeff}
\end{equation}%
It was suggested in   \cite{Faedo:2014naa}, that $Z\left( r\right) <0$  implies the stability of the embedding. Intuitively, the more the string penetrates into the bulk, the larger $\hat{L}_{app}$ becomes.

Let us now apply the recipe in eqs.(\ref{seffectiva})-(\ref{zeff})  to our background in eq.(\ref{BI}). To calculate the Wilson loop for non-dynamical quarks in the fundamental representation, we follow \cite{Rey:1998ik}, \cite{Maldacena:1998im} and  embed a fundamental string in the configuration,
\begin{equation}
t=\tau \ ,\qquad x=\sigma \ ,\qquad r=r\left( \sigma \right) \ .
\end{equation}%
The Nambu-Goto action is%
\begin{eqnarray}
& & S_{NG}=T_{F1}L_{\tau }\int d\sigma \sqrt{F^{2}+G^{2}r^{\prime 2}}\ ,\;\;\;\;\; F=r\ ,\;\;\;\; G=\frac{\sqrt{N}}{\sqrt{f_{s}\left( r\right) }}\ .
\end{eqnarray}%
The effective potential is
\begin{eqnarray}
V_{eff}\left( r\right) &=&\frac{1}{\sqrt{N} ~r_{0}}\sqrt{r^{2}f_{s}\left( r\right) \left(
r^{2}-r_{0}^{2}\right) }\ .
\end{eqnarray}%
The approximated separation of the quark-anti-quark pair is%
\begin{eqnarray}\label{LapproxWL}
\hat{L}_{QQ}\left( r_{0}\right) &=&\frac{\sqrt{N}~\pi }{\sqrt{f_{s}\left( r_{0}\right) }}\ .
\end{eqnarray}%
This expression for $\hat{L}_{QQ}$ (approximating the separation between the quark pair) becomes divergent in $r_0\sim r_+$. This suggest a confining behaviour, as the pair can be separated to arbitrarily large distances. Below we confirm this when we study the dependence of the energy of the pair, where we find a linear behaviour with the separation. On the other hand, for $r\to\infty$, when studying the high energy regime, we find that there is a minimal separation $\hat{L}_{min}=\pi\sqrt{N}$ between the pair. This indicates a non-local behaviour of the QFT, with a minimal length, associated with the size of little strings.

The stability of this embedding is determined by the function%
\begin{eqnarray}
\frac{d}{dr_0}\hat{L}\left( r_0 \right) &=&-\frac{\sqrt{N}~\pi }{2 \left[ f_{s}\left(
r_0\right) \right] ^{3/2}}f_{s}^{\prime }\left( r_0\right) \ , \\
&=&      -\frac{\sqrt{N}~ \pi r_0\Big(r_0^2(r_+^2+r_-^2) -2 r_+^2r_-^2 \Big)}{\Big[ (r_0^2-r_+^2)(r_0^2-r_-^2)\Big]^{3/2}}
\ ,
\end{eqnarray}%
which is always negative for $r_{0}>r_{+}$. Consequently the proposed configuration for the Wilson loop is
stable.
We can analyse numerically the expressions for the length and the energy of the Wilson loop, see eqs.(\ref{lengthwilson}),(\ref{energywilson}). We have,
\begin{eqnarray}
& &L_{QQ}= 2r_0 \sqrt{N} \int_{r_0}^\infty \frac{r ~dr}{\sqrt{(r^2-r_0^2)(r^2- r_+^2)(r^2-r_-^2)}}\label{energylengthwilson}\\
& & E_{QQ}= r_0 L_{QQ}(r_0)  + 2\sqrt{N} \Big[\int_{r_0}^\infty \sqrt{\frac{r^2(r^2-r_0^2)}{(r^2-r_+^2)(r^2-r_-^2)}} dr     -\int_{r_+}^\infty \frac{r^2~dr}{\sqrt{(r^2-r_+^2)(r^2-r_-^2)}} \Big].\nonumber
\end{eqnarray}
It is interesting to notice that these expressions for the Wilson loop for quarks in the fundamental is exactly the same, up to the replacement $\sqrt{\frac{8}{e_A^2+e_B^2}}\to \sqrt{N}$ as that obtained in equations (4.16) and (4.17) of the paper 
\cite{Nunez:2023nnl}\footnote{This value is the level of the WZW model on the cigar when the sigma model on NS branes is studied. See Section 5 of the paper \cite{Nunez:2023nnl}.}. Hence the analytic result for these integrals and the parametric plot for $E_{QQ}(L_{QQ})$ given in \cite{Nunez:2023nnl} can be translated. In summary, we observe a confining behaviour, together with a minimal separation of the quark-antiquark pair (at large values of $r_0$), that is associated with the scale of the LST. In spite of the field theory analysed here and that in \cite{Nunez:2023nnl} being qualitatively different, the Wilson loop makes no difference between these two effective theories constructed using D5 branes. This feature can only be appreciated using the holographic dual description. 
\begin{figure}
\centering
    \begin{minipage}{.4\textwidth}
    \centering
    \includegraphics[width=.68\linewidth]{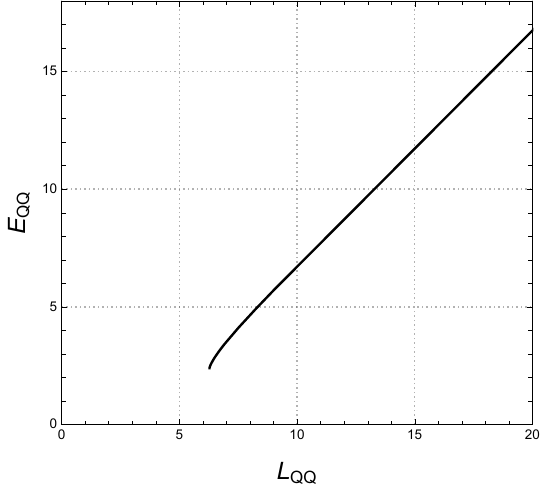}
    \end{minipage}
    \begin{minipage}{.5\textwidth}
    \centering
    \includegraphics[width=1.05\linewidth]{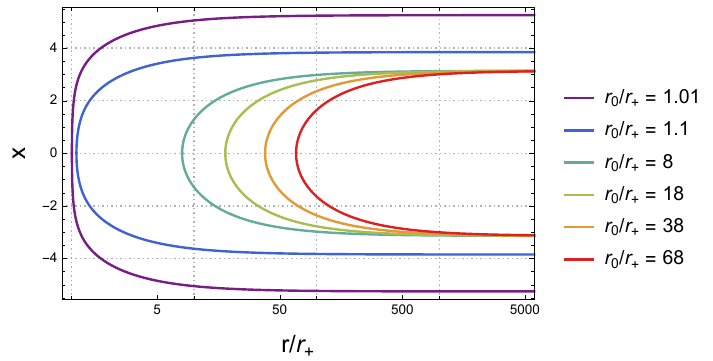}
    \end{minipage}
\caption{Left: Parametric plot of $E_{QQ}(L_{QQ})$ in the BPS bound $m=0$. Bottom Right: the profiles of different strings as they explore the bulk. The longer the separation $L_{QQ}$, the more the string approaches $\frac{r_{0}}{r_{+}}\sim 1$. This is usual of the backgrounds dual to a confining  QFT behaviour.}\label{figura1a}
\end{figure}

 Another interesting observable is the 't Hooft loop, describing the potential between a pair of external magnetic monopoles. We study this next.
\subsubsection{'t Hooft loops}
If the theory confines electric charges, we expect  magnetic monopoles to be screened. We study the monopole-anti-monopole pair by considering an effective magnetically charged string. This object is studied by probing the background with a D5 brane that extends in the directions 
 $\left[ t,x,\varphi ,\theta,\phi,\psi\right] $ and $r=r\left( x\right) $. 
 The induced metric is%
\begin{eqnarray}
ds^2_{D5,ind} = r\left\{ -dt^{2}+\left(\! 1+\frac{Nr^{\prime 2}}{r^{2}f_{s}\left( r\right) }\!\right)\! dx^{2}+\!f_{s}\left( r\right) d\varphi
^{2}\!
 +\!\frac{N}{4}\left[ \!{\omega}_{1}^{2}+\!{\omega}_{2}^{2}+\!\left(\! {\omega}_{3}-\!\sqrt{\frac{8}{N}}Q\zeta \left( r\right) d\varphi\!
\right) ^{2}\right] \right\} \ .  \nonumber
\end{eqnarray}%
When we integrate over the directions $\varphi,\theta,\phi,\psi$ the effective probe is $(1+1)$ dimensional and carries magnetic charge. The effective action takes the form of eq.(\ref{seffectiva}). It reads,
\begin{equation}
S_{eff}=T_{D5}L_{\varphi }16\pi ^{2}\left( \frac{N}{4}\right)^{3/2}\int dt\int dx\sqrt{r^{4}f_{s}\left( r\right) +Nr^{2}r^{\prime 2}}\ .\label{thooftseff}
\end{equation}%
From eq.(\ref{thooftseff}) we identify the functions%
\begin{equation}
F=r^{2}\sqrt{f_{s}\left( r\right) }\ ,\qquad G=\sqrt{N}r\ .\label{thooft1}
\end{equation}%
This object has vanishing tension when it approaches $r=r_+$, as $F(r)=\sqrt{(r^2-r_+^2) (r^2-r_-^2)}$. As a consequence of this, at low energies the monopole anti-monopole pair can be separated at no energy expense,  In contraposition with the Wilson loop, this indicates a screening behaviour.

The effective potential defined in eq.(\ref{Veff}) is for this probe,
\begin{eqnarray}
V_{eff}\left( r\right)  
&=&\frac{r}{\sqrt{N}~ r_{0}^{2}}\sqrt{\frac{f_{s}\left( r\right) }{f_{s}\left(
r_{0}\right) }\left( r^{4}f_{s}\left( r\right) -r_{0}^{4}f_{s}\left(
r_{0}\right) \right) }\ .
\end{eqnarray}%
The approximated  separation for the monopole pair $\hat{L}_{MM}\left(
r_{0}\right) $ is given by%
\begin{eqnarray}
\hat{L}_{MM}\left( r_{0}\right)  &=& 2\pi\sqrt{N} \frac{\sqrt{f_s(r_0) } }{(4 f_s(r_0) + r f_s'(r_0))}= \pi\sqrt{N }\frac{\sqrt{(r_0^2-r_+^2)(r_0^2-r_-^2)}}{2 r_0^2-r_+^2-r_-^2}.\label{lappthoft}
\end{eqnarray}%
This approximate separation for the monopole pair displays  the opposite behaviour to that of the Wilson loop (quark pair). In fact, there is a maximal separation, achieved when $r_0\to\infty$ and the minimal (vanishing) separation is achieved when $r_0\sim r_+$.

The function $Z\left( r_{0}\right) $ associated to the embedding is$_{{}}$%
\begin{eqnarray}
Z\left( r_{0}\right)  &=&\frac{d}{dr_{0}}\left( \hat{L}_{MM}(r_0)     \right) = \pi\sqrt{N}~r_0 \frac{(r_+^2-r_-^2)^2}{(2r_0^2-r_+^2-r_-^2)^2 \sqrt{(r_0^2-r_+^2)(r_0^2-r_-^2)}}.\label{zthooft}
\end{eqnarray}%
This is always positive, indicating that this embedding of the D5 brane is unstable. This instability suggest a transition between a connected embedding, like the one studied above, and a disconnected one. This indicates that the pair of monopoles can be arbitrarily separated at no energy cost. We conclude that the 't Hooft loop displays a screening type of behaviour.

We can write the expressions in eqs. (\ref{lengthwilson}),(\ref{energywilson}). These read, 
    \begin{equation}\label{energylengththooft}
        \begin{aligned}
        &\begin{aligned}
             L_{MM}&= 2\sqrt{N (r_0^2-r_+^2)(r_0^2- r_-^2)} \\
             &\phantom{=} \times
             \int_{r_0}^\infty \frac{r ~ dr}{\sqrt{ (r^2-r_+^2)(r^2-r_-^2) \left[ (r^2-r_+^2)(r^2-r_-^2)-(r_0^2-r_+^2)(r_0^2-r_-^2) \right] }}
        \end{aligned} \\
        &\begin{aligned}
            E_{MM} &= \sqrt{(r_0^2-r_+^2)(r_0^2- r_-^2)} L_{MM}(r_0) \\
            &\phantom{=} 
            + 2\sqrt{N} \left[ \int_{r_0}^\infty \frac{r \sqrt{(r^2-r_+^2)(r^2-r_-^2) - (r_0^2-r_+^2)(r_0^2-r_-^2)}        }{\sqrt{(r^2-r_+^2)(r^2-r_-^2)}        } dr -\int_{r_+}^\infty r ~ dr\right].
        \end{aligned}
        \end{aligned}
    \end{equation}

\begin{figure}[h]
    \centering
    \includegraphics[scale=0.7]{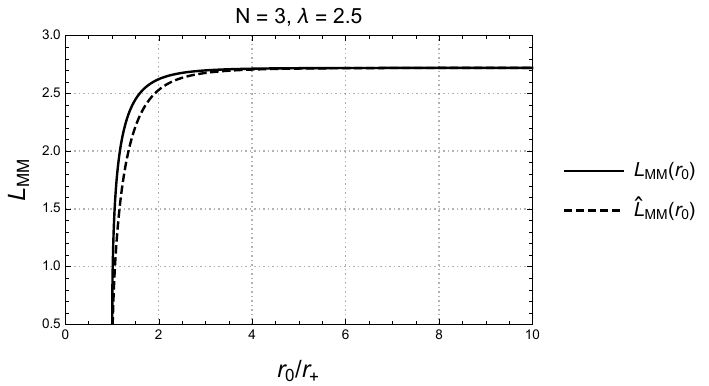}\quad
    \includegraphics[scale=0.7]{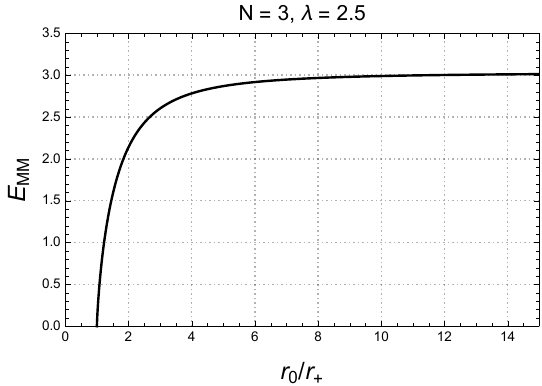}\\
    \includegraphics[scale=0.7]{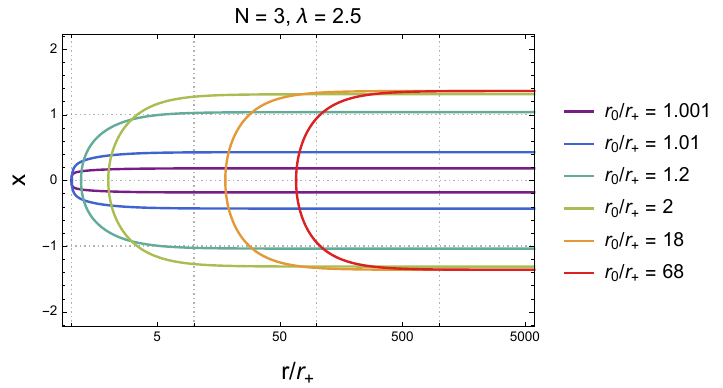}
    \includegraphics[scale=0.85]{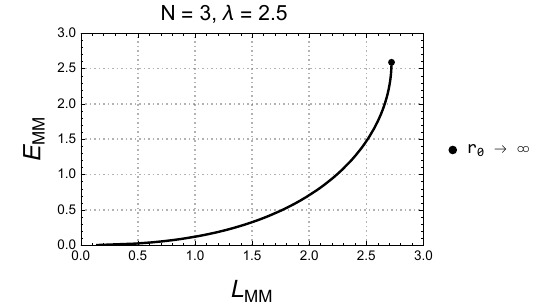}
    \caption{The integrals of the monopole-antimonopole t' Hooft loop \eqref{energylengththooft} can be written in terms of two parameters $\lambda=-r_{-}^2/r_{+}^2$ and $N$. All the plots are considering $\lambda=2.5$ and $N=3$. Upper left: Plot comparing the exact expression for the monopole separation \eqref{energylengththooft} with the approximate expression $\hat{L}_{MM}$ in \eqref{lappthoft}. Upper right: The energy of the t'Hooft loop as a function of the $r_0$ in units of $r_+$. Bottom left: Different  profiles of the macroscopic strings as a function of the turn-around point $r_0$. Bottom Left: Plot of the function $E_{MM}(L_{MM})$. The upwards concavity ascertains the prediction that the embedding is unstable.}
    \label{fig:thooft loop}
\end{figure}

We analyse these integrals by plotting 
$L_{MM}(r_0)$, comparing it with $\hat{L}_{MM}$ in eq.(\ref{lappthoft}), $E_{MM}(r_0)$ and the parametric plot $E_{MM}(L_{MM})$. These are displayed in Figure \ref{fig:thooft loop}.

The Wilson and 't Hooft loops indicate that the background of eq.(\ref{BI}) is dual to a $(4+1)$ dimensional QFT that confines electric charges and screens magnetic ones. An interesting quantity to study is the density of states (the number of degrees of freedom) in terms of the energy (the radial coordinate). Below, we calculate an observable that quantifies this.
\subsubsection{Density of states, free energy or holographic central charge}
In this section we study a quantity that becomes particularly important and well defined in conformal field theories (our QFT is not conformal). In the conformal case, this quantity coincides with the free energy or central charge of the CFT (indicating the number of degrees of freedom in the system). This quantity has been generalised from calculations in backgrounds with an AdS-factor (dual to CFTs) to generic backgrounds  and dilaton, like our case in the background of eq.(\ref{BI}). See \cite{Macpherson:2014eza}, \cite{Bea:2015fja}. For backgrounds and dilaton of the form,
\begin{eqnarray}
ds^{2} &=&\alpha (r,y^{i})\left( dx_{1,d}^{2}+\beta \left( r\right)
dr^{2}\right) +g_{ij}\left( r,y^{i}\right) dy^{i}dy^{j}\ ,
\;\;\;\;\;
\Phi  =\Phi \left( r,y^{i}\right). \label{conventions central charge}
\end{eqnarray}%
Here $y^{i}$ stands for all the internal coordinates. We define the
following quantites%
\begin{equation}
V_{\text{int}}=\int dy^{i}\sqrt{e^{-4\Phi }\det g_{ij}\alpha \left(
r,y^{i}\right) ^{d}}\ ,\qquad H=V_{int}^{2}\ . \label{convv}
\end{equation}%
The holographic central charge is defined as%
\begin{equation}
c_{\text{hol}}=d^{d}\frac{\beta \left( r\right) ^{d/2}H^{\left( 2d+1\right)
/2}}{G_{N}\left( H^{\prime }\right) ^{d}}\ .\label{chol}
\end{equation}
For the background in  eq.(\ref{BI}) we have $ y^{i} \in \left[ \varphi, \theta , \phi ,\psi \right] $ with
\begin{equation}
g_{ij} d y^{i } d y^{j}=
r f_{s} \left( r\right) d\varphi ^{2}+\frac{N r}{4}\left[ {\omega}_{1}^{2}+{\omega}_{2}^{2}+\left( {\omega}_{3}-\sqrt{\frac{8}{N} }  Q\zeta \left( r\right) d\varphi \right) ^{2}\right] \ .
\end{equation}
The functions in eqs.(\ref{conventions central charge})-(\ref{convv}) are
\begin{equation}
\alpha \left( r, y^{i}\right) =r\ ,\quad \beta \left( r\right) =\frac{N}{r^{2}f_{s}\left( r\right) }\ ,\quad d=4\ ,\quad e^{-4\Phi }=\frac{1}{r^{4}}\ .
\end{equation}%
These lead to  $H=\mathcal{N}^{2}r^{4}f_{s}\left( r\right) $ with $\mathcal{N}=2 \pi^{2} \sqrt{N^3}L_{\varphi } $, and the holographic
central charge is
\begin{equation}
c_{\text{hol}}=\frac{256N^2 \mathcal{N} r^{2}f_{s}\left( r\right) ^{5/2}}{G_{N} \left[ f_{s}\left( r\right) +\frac{r}{4}f_{s}^{\prime }\left( r\right) \right] ^{4}}= \frac{16 N^2 {\cal N}}{G_N} \frac{\left[ (r^2-r_+^2)(r^2-r_-^2)\right]^{5/2}}{(2 r^2-r_+^2-r_-^2)^4}\ .\label{chol1}
\end{equation}
This quantity  vanishes for $r=r_+$, suggesting a gapped system, in agreement with the confining behaviour induced from the  results of the Wilson and 't Hooft loops. On the other hand, for large values of $r$ (at high energies), the quantity diverges quadratically. This is in agreement with the non-field theoretical UV-completion provided by the LST.

There is another quantity, called $c_{flow}$ defined  in \cite{Bea:2015fja}. This  considers the system as a  six dimensional non-isotropic QFT. The behaviour of $c_{flow}$ is qualitatively similar to the one described in eq.(\ref{chol1}), but the divergence at high energies is faster. 
\subsubsection{Entanglement Entropy}
A very interesting observable in all QFTs is the Entanglement Entropy (EE). This can be holographically computed following \cite{Ryu:2006bv}. For the case of a strip-region of size $L_{EE}$  and for a confining field theory, it was suggested in \cite{Klebanov:2007ws} that the Entanglement Entropy
 $S_{EE}( L_{EE} )$ 
 presents a phase transition as we vary $L_{EE}$. This proposal was critically analysed in \cite{Kol:2014nqa} and more recently in  \cite{Jokela:2020wgs}. In these papers, it was found that for  gravity backgrounds dual to confining QFTs with non-local UV completion, the phase transition conjectured in \cite{Klebanov:2007ws} is absent. If a UV-cutoff is introduced (avoiding the non-local UV completion), the phase transition is recovered. 

The QFT holographically described by the background in eq.(\ref{BI}), fits the above description. The EE on a strip is calculated by computing the area of an eight-manifold $\Sigma_8$ that is parametrised by,
\begin{eqnarray}
& & \Sigma_8=[x_1,x_2,x_3,x_4,\varphi, \theta,\phi,\psi], \;\;\;\; r(x_1).\label{EE1}\\
& & ds_{\Sigma_8,st}^2=  r \Big[ dx_{2,3,4}^2 + dx_1^2 \left(1+ \frac{N}{ r^{2}f_{s}(r)} r'(x_1)^2\right) +  f_{s}(r)d\varphi^{2} 
            +\nonumber\\ 
& &              \frac{N}{4}\left(\omega^{2}_{1} + \omega^{2}_{2} + 
        \left(\omega_{3} - \sqrt{\frac{8}{N}} Q \zeta(r) d\varphi\right)^{2}\right) \Big],\nonumber
        \end{eqnarray}
and then calculating (for an entangling region of size $L$),
\begin{eqnarray}
& & S_{EE}=\frac{1}{G_N} \int d^8 x\sqrt{e^{-4\Phi} \det g_{\Sigma_8,st}}= \frac{16\pi^2N^{3/2}L_{x_2} L_{x_3}L_{x_4} L_{\varphi} }{8 G_N} \int_{-\frac{L}{2}}^{\frac{L}{2}} dx \sqrt{F^2(r) + G^2(r) r'^2}.\nonumber\\
& & F=r^2\sqrt{f_s(r)},~~~G(r)= \sqrt{N} r.\label{EE2}
\end{eqnarray}
The expression above fits in the generic effective action in eq.(\ref{seffectiva}), hence we can write integral expressions for the separation and EE following the formulas (\ref{lengthwilson}), (\ref{energywilson}). Indeed, we can write an approximate expression for the separation following eq.(\ref{laprox}),
and also evaluate the function $Z(r_0)$ in eq.(\ref{zeff}), indicating stability or instability of the embedding.

 At this point it is useful to observe that the functions $F(r)$ and $G(r)$ that appear in the study of the EE, eq.(\ref{EE2}),  are the same ones that appear in the study of the 't Hooft loop, in eq.(\ref{thooft1}). It follows from this that the approximate expression for the separation is a monotonic function given in eq.(\ref{lappthoft}). This indicates the absence of phase transitions. The function $Z_{EE}(r_0)$ coincides with that in eq.(\ref{zthooft}).

Beyond the use of the approximate expressions, we work with the exact and generic ones derived in \cite{Kol:2014nqa}. We find that our EE does not present a phase transition following the general criterium in \cite{Kol:2014nqa}. The plots are identical to those in Figure \ref{fig:thooft loop}. There is an instability of the embedding--positive $Z(r_0)$, indicating that in the presence of a cutoff, new 'short' configurations appear, leading to the phase transition.

To close this section, we observe that the expressions for the EE obtained in this section, coincide (up to rescaling by constants) with those derived for the same observable in \cite{Nunez:2023nnl}. This indicates deep relations between the very dissimilar QFTs.

\subsubsection{Spin-two glueballs}\label{glueballssection}

We consider glueball-like excitations in the confining $(4+1)$ dimensional QFT holographically described by the background in eq.(\ref{BI}). The glueball excitations are studied by performing a fluctuation of some background field, either Ramond or Neveu-Schwarz  in the Type II action.  The nonlinear character of the equations of motion, implies that this fluctuation sources excitations for other fields in the background. The dynamics is described by linear second order coupled differential equations for the fluctuations. This is characteristically very hard to analyse (diagonalise, resolve, find the normal modes and  spectra, etc)\footnote{ A different approach is to study the equation of motion of a probe scalar in the background. It was recently observed  in \cite{Anabalon:2023lnk}, that a new logarithmic branch of solutions appears in background of the form (\ref{BI}). Whilst we do not refer to these interesting probe scalars as ''glueballs'', their study is interesting.}.

We begin this section by briefly summarising the treatment of a very special kind of fluctuations which are easy to analyse. The detailed derivations are given in Appendix \ref{appendixglue}.

In this section we work in the {\it Einstein frame}, because we are studying particle-like excitations. In Einstein frame, the background metric of eq.(\ref{BI}) reads,
\begin{equation}\label{einstein}
ds^{2}_{E} = \sqrt{r} \left[ dx^{2}_{1,4} + f_{s}(r)d\varphi^{2} 
            + \frac{Ndr^{2}}{ r^{2}f_{s}(r)} 
            + \frac{N}{4}\left(\omega^{2}_{1} + \omega^{2}_{2} + 
        \left(\omega_{3} - \sqrt{\frac{8}{N}} Q \zeta(r) d\varphi\right)^{2}\right) \right].
        \end{equation}
%
Consider metric fluctuations of the form
    \begin{equation}\label{pert}
        \delta g_{\mu\nu} = e^{2A} \bar{h}_{\mu\nu}, \quad 
        \bar{h}_{\mu\nu} = \begin{pmatrix}
                        h_{ab}(x)\tilde{\psi}(y)& 0 \\
                         0 & 0
                        \end{pmatrix},
    \end{equation}
this is, metric fluctuations parallel to the flat space. Here $x^{a}$ denotes the flat space coordinates $[t,x_1,x_2,x_3,x_4]$, while $y^{i}$ correspond to all the transverse ones $[r,\varphi, \theta,\phi,\psi]$. We work in the transverse-traceless gauge
    \begin{equation}
        \Trh{a} = 0, \quad \nabla_{a}h^{a}_{\phantom{a}b}=0.
    \end{equation}
    
Following the appendix of \cite{Caceres:2005yx}, it is consistent to take $\delta F_{\mu\nu\lambda}=0, \delta \Phi  =0 $. In what follows, we rewrite the  Einstein frame  background metric as
    \begin{equation}
        ds^{2} = e^{2A(\vec{y})}\left( ds^{2}(\mathbb{R}^{1,4}) + \hat{g}_{ab}(\vec{y}) dy^a dy^b \right) .\label{einsteinresc}
    \end{equation}
    
Following the treatment in \cite{Bachas:2011xa}, further detailed in Appendix \ref{appendixglue}, we find that changing variables as $\tilde{\psi}(y) = e^{-4A}\Psi (y)$, the function $\Psi(y)$ solves a Schroedinger-like equation 
  \begin{equation}\label{SchroedingerLike}
        - \hat{\square}_{y} \Psi + V(y)\Psi =  M^{2}\Psi,
    \end{equation}
    
with $M^{2}\geq 0$ and where the effective potential is
            \begin{equation}\label{vvv}
                V(y) = e^{-4A}\hat{\square}_{y}  e^{4A} = \frac{e^{-4A} }{ \sqrt{\det \hat{g}_{ab} } }\partial_a \left[ \sqrt{\det \hat{g}_{ab} } \hat{g}^{ab}\partial_b e^{4A}\right].
            \end{equation}
            
Let us apply these formulas to our background in eq.(\ref{BI}), (\ref{einstein}). The warp factors  and internal metric read,
    \begin{eqnarray}\label{metricaIE}
  & &  e^{4A(\vec{y})}= r, \\
        & & \hat{g}_{ab}dy^a dy^b=  f_{s}(r)d\varphi^{2} 
            + \frac{Ndr^{2}}{ r^{2}f_{s}(r)} 
            + \frac{N}{4}\left(\omega^{2}_{1} + \omega^{2}_{2} + 
        \left(\omega_{3} - \sqrt{\frac{8}{N}} Q \zeta(r) d\varphi\right)^{2}\right) . \nonumber
        \end{eqnarray}
        
This gives $\sqrt{\det \hat{g}_{ab}} = \frac{N^{2}}{8r} \sin(\theta)$. In what follows we focus on glueballs that depend on the radial coordinate $r$ and the angular coordinate $\varphi$ as 
    \begin{equation}\label{HarmonicExpansion}
        \Psi(r,\varphi) = e^{i \frac{2\pi}{L_{\varphi}}n \varphi}\Psi(r),~~~n\in \mathbb{Z}
    \end{equation}
but are not excited in the $[\theta,\phi,\psi]$ directions. The  potential in \eqref{vvv} is,
    \begin{equation}\label{potsch}
        V(r)= \frac{1}{N}\partial_r\left( r f_s(r)\right),
    \end{equation}

and the Schroedinger-like equation in \eqref{SchroedingerLike}  reads
    \begin{equation}\label{schint}
        - \frac{r}{N}\frac{d}{dr}\left(r f_{s}(r) \frac{d\Psi}{dr}\right)  + \left(V(r) + \frac{n^2}{f_{s}(r)}\left(\frac{2\pi}{L_{\varphi}}\right)^{2} \right)\Psi(r) = M^{2} \Psi(r).
    \end{equation}

%


We can move \eqref{schint} to more manageable forms. Moving to the tortoise coordinate $\rho$ by
    \begin{equation}
        r  = \cosh\left(\frac{\rho}{\sqrt{N}}\right)\sqrt{r^{2}_{+} -r^{2}_{-}\tanh^{2}\left(\frac{\rho}{ \sqrt{N}}\right)},
    \end{equation}

which maps $r\in \left[r_{+},+\infty\right[$ to $\rho\in \left[0,+\infty\right[$. Using the change of variables%
\footnote{Note that in the tortoise coordinate $\sqrt{\det(\hat{g}_{ij})} = \sqrt{f_{s}(\rho)}\Vol(S^{3})$ and therefore the Sturm-Liouville norm, see Appendix \ref{AppendixCChangeofVariables}, is     \begin{equation*}
        ||\bar{\psi}||^{2} = \int d\varphi\, d\rho \Vol(S^{3}) \sqrt{f_{s}(\rho)} 
            \left(  f_s(r)^{-1/4}\Theta(\rho) \right)^{2} 
            = \int d\varphi\, d\rho \Vol(S^{3}) |\Theta(\rho)|^{2}.
    \end{equation*} 
Therefore, when imposing boundary conditions, it is enough to require $\Theta(\rho)$ to be normalisable.}%
 $\Psi(r)= f_s(r)^{-1/4}\Theta(\rho)$, leads to the equation (see Appendix \ref{SchroedingerAppendix})
    \begin{eqnarray}\label{sch2}
         - \frac{d^2\Theta}{d\rho^2} +\tilde{V}(\rho) \Theta = M^2\Theta,\;\;\;\
        \tilde{V}(\rho)= \left( V(r) +  \frac{n^{2}}{f_{s}(r)}\left(\frac{2\pi}{L_{\varphi}}\right)^{2} -\frac{r}{N} f_s^{1/4} \frac{d}{dr} \left[ r f_s \frac{d f_s^{-1/4}}{dr}  \right]\right)\bigg|_{r=r(\rho)}.
    \end{eqnarray}  
            
Explicitly, the effective potential reads
    \begin{equation}\label{schrodinger effective potential bg1}
        \tilde{V}(\rho) = \frac{1}{N} 
        - \frac{1}{N}\frac{1}{\sinh^{2}\left(\frac{2\rho}{\sqrt{N}}\right)} 
        + \frac{n^2}{N} \coth^{2}\left(\frac{\rho}{\sqrt{N}}\right)
        \left( 1- \frac{r^{2}_{-}}{r^{2}_{+}}\tanh^{2}\left(\frac{\rho}{\sqrt{N}}\right) \right)^{2}, 
    \end{equation}
which has the following asymptotic expansions
        \begin{equation}
        \begin{aligned}
            \tilde{V}(\rho\rightarrow 0) &= 
            \left(n^{2}-\frac{1}{4}\right)\frac{1}{\rho^{2}}
            + \frac{4}{3N} + n^{2}\frac{2(r^{2}_{+}-3r^{2}_{-})}{3Nr^{2}_{+}}
            +\mathcal{O}\left( \rho ^{2}\right) \ ,\\
            \tilde{V}(\rho\rightarrow +\infty) &=
            \frac{1}{N} - n^{2}\frac{2r^{2}_{-}}{Nr^{2}_{+}}
            +\mathcal{O}\left( e^{-2N^{-1/2}\rho }\right) \, .
        \end{aligned}
        \end{equation}

Note that the equation only depends on the parameters $r_{\pm}$ through the ratio $\lambda = -\frac{r^{2}_{-}}{r^{2}_{+}}$. The potential is unbounded from below when the leading term of the effective potential close to $\rho \rightarrow 0$ is negative. Requiring a  potential that is bounded below leads to a bound in the angular momentum along the circle $n^2 \geq 1 $.  The only excitation for which the effective potential is unbounded from below is the S-wave. In the later case, the potential near the origin can be approximated by $V\sim -1/4\rho^{2}$, which, as highlighted in \cite{DeLuca:2023kjj}, needs to be studied carefully.

The previous bound is not enough to guarantee a discrete spectrum of excitations. For $0\leq \lambda \leq 1 $ the effective potential has a runaway behaviour (the minimum is at infinity), while for $\lambda >1$ the potential has a minimum. We have a transition between a spectrum with discrete states ($\lambda>1$) or without discrete states ($\lambda\leq 1$). Note that $\lambda=1$ is precisely the SUSY point.

In what follows, we study the discrete spectrum of glueballs for $\lambda >1$. Also, note that the potential \eqref{schrodinger effective potential bg1} is invariant under $n\rightarrow -n$. Without loss of generality, we consider $n\geq 1$ (we include the $n\leq-1$ cases when writing the full solution, see below). Using the change of variables 
    \begin{equation}
        z=\cosh \left( \frac{\rho}{\sqrt{N}} \right) \ ,
    \end{equation}
which maps the region $\rho \in \lbrack 0,\infty \lbrack $ to $z\in \lbrack 1,+\infty \lbrack $, we can write the profile of the metric perturbation $e^{2A}\tilde{\psi} = e^{-2A} f_{s}^{-1/4} \Theta(z) e^{i \frac{2\pi}{L_{\varphi}}n\varphi}= {\cal H}(z) e^{i \frac{2\pi}{L_{\varphi}}n\varphi} $ as
    \begin{equation}\label{general solution bg1}
    \begin{aligned}
       {\cal H}
        &=  (z^{2}-1)^{-\frac{n}{2}}\left(z^{2}+\lambda(z^{2}-1)\right)^{\frac{1}{4}}\\
        &\phantom{=} \times 
        \left[   
         C_{1} z^{n\lambda} \hypgeom{a_{+}}{a_{-}}{c_{-}}{z^{2}} 
         + C_{2} z^{-n\lambda} \hypgeom{b_{+}}{b_{-}}{c_{+}}{z^{2}}  \right],
    \end{aligned} 
    \end{equation}

where
    \begin{equation}
    \begin{aligned}
        a_{\pm} &= \frac{1}{2}\left( 1- n(\lambda+1) \pm \sqrt{1- M^{2}N+n^{2}(1+\lambda)^{2}} \right),\\
        b_{\pm} &= \frac{1}{2}\left( 1+ n(\lambda-1) \pm \sqrt{1- M^{2}N+n^{2}(1+\lambda)^{2}} \right),\\
        c_{\pm} &= 1\pm \lambda n.
    \end{aligned}
    \end{equation}

In what follows we consider $0 \leq M^{2}N < 1+n^{2}(1+\lambda)^{2}$. We show below that this condition leads to a discrete spectrum of states. On the other hand, $M^{2}N > 1+n^{2}(1+\lambda)^{2}$ leads to a continuous spectrum. We explain the origin of the continuous spectrum in the following subsection.

We now expand around $z\rightarrow 1$ and $z\rightarrow +\infty$ to find normalisable solutions.  At $z\rightarrow 1$ we have (see Appendix \ref{appendixglue})
    \begin{equation}\label{GlueballsOrigin}
    \begin{aligned}
        \lim_{z\rightarrow 1} {\cal H} 
        &= (z^{2}-1)^{\frac{n}{2}} \Gamma(-n) \left(C_{1} \frac{\Gamma(c_{-})}{\Gamma(a_{+})\Gamma(a_{-})} + C_{2}\frac{\Gamma(c_{+})}{\Gamma(b_{+})\Gamma(b_{-})} \right) 
        \\ 
        &\phantom{=} + (z^{2}-1)^{-\frac{n}{2}} \Gamma(n)\left( 
        C_{1} \frac{\Gamma(c_{-})}{\Gamma(c_{-}-a_{+})\Gamma(c_{-}-a_{-})} +
        C_{2}\frac{\Gamma(c_{+})}{\Gamma(c_{+}-b_{+})\Gamma(c_{+}-b_{-})} \right),
    \end{aligned}    
    \end{equation}

while at  $z\rightarrow +\infty$ the expansion reads\footnote{From here we see that for $M^{2}N > 1+n^{2}(1+\lambda)^{2}$ both branches are regular at infinity. There is no need to implement a boundary condition at infinity and hence, no quantisation condition for $M^{2}$.}
    \begin{equation}\label{GlueballsInfinity}
    \begin{aligned}
        \lim_{z\rightarrow +\infty} {\cal H} 
        &= z^{-\frac{1}{2}-\sqrt{1-M^{2}N+n^{2}(1+\lambda)^{2}}} \left(C_{1} \frac{\Gamma(a_{-}-a_{+})\Gamma(c_{-})}{\Gamma(a_{-})\Gamma(c_{-}-a_{+})} + C_{2}\frac{\Gamma(b_{+}-b_{-})\Gamma(c_{+})}{\Gamma(b_{+})\Gamma(c_{+}-b_{-})} \right) 
        \\ 
        &\phantom{=} +z^{-\frac{1}{2}+\sqrt{1-M^{2}N+n^{2}(1+\lambda)^{2}}} \left(C_{1} \frac{\Gamma(a_{+}-a_{-})\Gamma(c_{-})}{\Gamma(a_{+})\Gamma(c_{-}-a_{-})} + C_{2}\frac{\Gamma(b_{-}-b_{+})\Gamma(c_{+})}{\Gamma(b_{-})\Gamma(c_{+}-b_{+})} \right).
    \end{aligned}    
    \end{equation}  

We require the leading term in both cases to vanish. This is, we impose regularity condition at the origin ($z=1$) and the normalisability of the excitations (fast decay at $z\rightarrow +\infty$). Let us start with \eqref{GlueballsInfinity}. In order to make the leading term vanish we impose
    \begin{equation}\label{BoundaryCondition}
        C_{2} = -C_{1} \frac{\Gamma(a_{+}-a_{-})\Gamma(c_{-})}{\Gamma(a_{+})\Gamma(c_{-}-a_{-})} \frac{\Gamma(b_{-})\Gamma(c_{+}-b_{+})}{\Gamma(b_{-}-b_{+})\Gamma(c_{+})}.
    \end{equation}

We now move to \eqref{GlueballsOrigin}. It might seem that there is an unavoidable divergence due to the factor of $\Gamma(-n)$ but we will show that this divergence can be cancelled by a precise choice of $M^{2}$ which also sets the leading term to zero.

Since we are considering $n\geq1$, the leading term in \eqref{GlueballsOrigin} is the second one.  The coefficient of the leading term in \eqref{GlueballsOrigin} is
    \begin{equation}
      \text{coeff}_{\text{leading}}=  \frac{\Gamma(c_{-})}{\Gamma(c_{-}-a_{-})}\left(
        \frac{1}{\Gamma(c_{-}-a_{+})} 
        - \frac{\Gamma(a_{+}-a_{-})}{\Gamma(a_{+})} \frac{\Gamma(b_{-})}{\Gamma(b_{-}-b_{+})}\frac{1}{\Gamma(c_{+}-b_{-})}\right),
    \end{equation}

We can set the coefficient to zero by imposing $c_{-}-a_{-}=-p$, with $p\in \mathbb{N}$, which leads to a quantisation condition for the glueball mass provided $p$ satisfies the following bound
    \begin{equation}
        p \leq \frac{1}{2}\left( n(\lambda-1)-1 \right)  = p_{ \max}.
    \end{equation}

From this is clear that for $0\leq\lambda\leq1$ does not lead to a discrete spectrum: in that case is impossible to satisfy $p>0$ and $p<p_{\max}$ simultaneously. The same holds for $n=0$. 

The spectrum for the glueball masses reads (notice that this includes the $n\leq -1$ cases)
    \begin{equation}
        M^{2} = \frac{1}{N}\left( 1+ n^{2}(\lambda+1)^{2} -4(p_{\max}-p)^{2} \right).
    \end{equation}

\begin{figure}[h!]
    \centering
\includegraphics[scale=0.7]{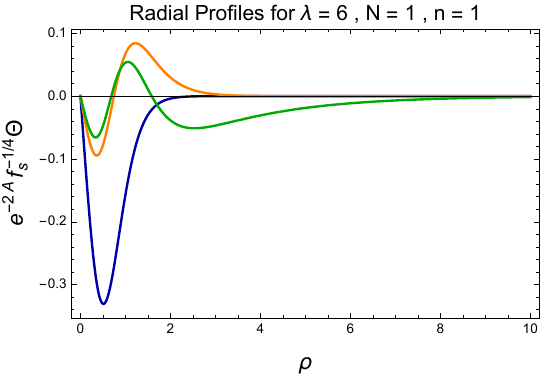} \includegraphics[scale=0.7]{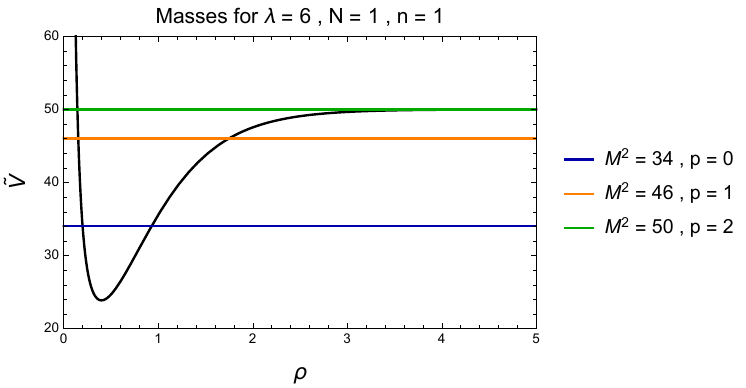}
    \caption{Glueball profile for a particular configuration $\lambda=6,N=1,n=1$. Left: Radial profile for the three allowed values of $p=0,1,2.$. Right: In black the Schroedinger effective potential \eqref{schrodinger effective potential bg1} and in colors the value of the glueball masses associated to the excitations $p=0,1,2$.}
    \label{plots}
\end{figure}

On the other hand, noting that (see Appendix \ref{appendixglue}) $c_{-}-a_{n}=a_{+}+n$ we have that $a_{+} = -n-p$, so that the factor of $1/\Gamma(a_{+})$ cancels the divergence in $\Gamma(-n)$ in \eqref{GlueballsOrigin}. Furthermore, with this choice for $M^{2}$, from \eqref{BoundaryCondition} we see that $C_{2}=0$. The solution satisfying the boundary conditions is 
    \begin{equation}
         e^{2A}\tilde{\psi} = \sum^{\IntegerPart{p_{\max}}}_{p = 0}\sum_{n\neq 0} C_{p,n} e^{i \frac{2\pi n}{L_{\varphi}}\varphi}  
             \frac{z^{|n|\lambda}\left(z^{2}+\lambda(z^{2}-1)\right)^{\frac{1}{4}}}{(z^{2}-1)^{\frac{|n|}{2}}}
            \hypgeom{-|n|-p}{1+p-|n|\lambda}{1-|n|\lambda}{z^{2}} 
    \end{equation}
In Figure \ref{plots} we plot the radial profiles for this functions. The figure also shows the potential in eq.(\ref{sch2}) and the masses of the first few glueballs.

\subsubsection{On the Continuous Spectrum}

Here we discuss the origin of the continuous spectrum that appears for the spin-2 glueballs discussed in the previous section. Let us begin by computing the spin-2 glueball spectrum on the pure D5-brane background, which in Einstein frame reads
    \begin{equation}
        ds^{2}_{E} = \sqrt{r} \left( dx^{2}_{1,5} + N\frac{dr^{2}}{r^{2}} + N ds^{2}(S^{3}) \right)
    \end{equation}

Following Appendix \ref{appendixglue}, we rewrite the metric as
    \begin{equation}
        ds^{2}_{E} = e^{2A}\left( dx^{2}_{1,5} +  ds^{2}_{\hat{y}}\right), \quad 
        ds^{2}_{\hat{y}} =  N\frac{dr^{2}}{r^{2}} + N ds^{2}(S^{3}), \quad e^{2A} = \sqrt{r}.
    \end{equation}

Before proceeding with the computation of the glueball solution, let us first go into the tortoise coordinate by $r = e^{\rho}$, which maps $r\in [0,+\infty[$ to $\rho \in \mathbb{R}$. The metric in the transverse space then reads
    \begin{equation}
        ds^{2}_{\hat{y}} =  Nd\rho^{2} + N ds^{2}(S^{3}), \quad e^{2A(\rho)} = e^{\frac{\rho}{2}}.
    \end{equation}

Using \eqref{vvv} we have $V(\rho) = 1/N$. Separating variables $\Psi(y) = \Theta(\rho)Y_{\ell\,m}$, where $Y_{\ell\,m}$ are the spherical harmonics on the 3-sphere satisfying $\nabla^{2}_{S^{3}}Y_{\ell\,m} = - \ell(\ell+2) Y_{\ell\,m}$, then, eq.\eqref{SchroedingerLike} takes the form
    \begin{equation}
        -\frac{d^{2}\Theta}{d\rho^{2}} 
        +\left[ \ell(\ell+2) + 1 \right]\Theta(\rho) = M^{2}N\Theta(\rho)
    \end{equation}

This is a Schroedinger-like equation with a constant potential $V_{eff} = \ell(\ell+2) + 1$. Defining $E = M^{2}N -  \ell(\ell+2) - 1$, the solutions of the problem depend on the sign of $E$ (recall that $M^{2}$ is always positive) and are
    \begin{equation}
        \Theta(\rho)  =      
         \begin{cases} 
      c_{1} e^{i \sqrt{E}\rho} + c_{2} e^{-i \sqrt{E}\rho} & E> 0 \\
      c_{1} e^{ \sqrt{-E}\rho} + c_{2} e^{- \sqrt{-E}\rho} & E< 0 \\
   \end{cases}
    \end{equation}

For $E<0$, imposing regularity at $\rho\rightarrow \pm \infty$ sets $c_{1}=c_{2}=0$ so that solution is not relevant. The only solution is the plane wave and we have a continuum of states. 

The continuous spectrum for the spin-two excitations of the D5 brane suggests a natural  interpretation for the spectrum we have found in Section \ref{glueballssection}. In fact, we have there a spectrum that
(for the parameter $\lambda>1$) starts being discrete and then turns continuous. On the other hand, for $1\leq \lambda$ the spectrum is solely continuous.

The way to think about this is the following. The size of the compactification circle $\varphi$ is according to \eqref{CirclePeriod},
    \begin{equation}
        \frac{L_\varphi}{\pi\sqrt{N}}= 1+ \frac{m}{\sqrt{m^2 + 8 Q^2}}
    \end{equation}

In this expression the quantity $\pi\sqrt{N}$ is the size characteristic of the LST--see for example, the comment below \eqref{LapproxWL}.

For $m>0$ we have $L_\varphi > L_{LST}$, for $m<0$ we have $L_\varphi < L_{LST}$. The case $m=0$ implies $\lambda=-\frac{r_-^2}{r_+^2}=1$, which is the SUSY situation.

In the case $L_{\varphi}<L_{LST}$ there are some field theoretical discrete states. The energy of these states is less than the energy for which little strings take over the dynamics. Hence, the discrete set of states is followed by a continuous spectrum. This continuum is identified with the one encountered in the flat D5 brane case. 

On the other hand, for $L_\varphi>L_{LST}$ the little string theory with its continuous spectrum controls the dynamics at all energies (for this particular spin-two observable).

We propose that this behaviour (a discrete of states, followed by a continuous spectrum) is characteristic of spin-two fluctuations like the ones considered here, see \eqref{pert}, \eqref{HarmonicExpansion} in holographic duals to QFTs constructed based on compactified D5 branes.

This feature should be absent in duals obtained compatifying D$_{p<5}$ branes. For example, the analog glueball in a dual to field theoretical system written in \cite{Fatemiabhari:2024aua} is purely discrete.

We close this section dealing with a dual to a $(4+1)$ dimensional confining field theory. We move to the study of a similar $(2+1)$ dimensional QFT.

\section{A dual to a (2+1) dimensional QFT}\label{section2+1qft}
In this section we present a different solution to the Type IIB equations of motion--written in Appendix \ref{TypeIIBapp}. It describes D5 branes wrapping a two sphere. 

We  use coordinates $[t, x_1,x_2, \mu, r, \vartheta, \varphi, \theta_1,\phi_1,\psi_1]$. The coordinates $[\theta_1,\phi_1,\psi_1]$ are used to parameterise a three sphere, using the left invariant forms defined in eq.(\ref{su2-left}). There is a cigar-like direction that we parameterise with $\mu$, Minkowski directions ($t,x_1,x_2$) and a radial coordinate $r$. There is also a two sphere described by $(\vartheta,\varphi)$.

To  define the background it is convenient to introduce the following one-forms,
\begin{eqnarray}
& & A^{(1)}= \sin\vartheta \cos\vartheta \cos\varphi d\varphi +\sin\varphi d\vartheta,\label{gaugemeronfield}\\
& & A^{(2)}= -\cos\varphi d\vartheta + \cos \vartheta \sin\vartheta \sin\varphi d\varphi,~~A^{(3)}=-\sin^2\vartheta d\varphi,\nonumber\\
& & \Theta^{(1)}=(\omega_1-A^{(1)}), ~ \Theta^{(2)}=(\omega_2-A^{(2)}), ~\Theta^{(3)}=(\omega_3-A^{(3)}), \nonumber
\end{eqnarray}
The one-form $A=\lambda A^{(i)}\sigma^i$ is a meron gauge field (a $\lambda$-fraction of an instanton). This type of gauge field has an important role
in mechanism of confinement in QCD. Here, we use it on the holographic dual background. Interesting comments about merons in  context similar to ours can be found in \cite{Canfora:2012ap}.
{Note that $(A^{(1)})^{2} + (A^{(2)})^{2}  + (A^{(3)})^{2} = d\vartheta^2+ \sin^2\vartheta d\varphi^2$}.
In terms of these definitions the type IIB configuration reads,
\begin{eqnarray}
ds^2_{st}  &=& r\left\{-dt^2+ dx_1^2 +dx_2^2 + \left(1-\frac{m}{r^{2}}\right) d\mu^{2} +
\frac{2dr^{2}}{r^{2}\left( 1-
\frac{m}{r^{2}}\right) }+d\vartheta ^{2}+\sin ^{2}\vartheta d\varphi
^{2}\right.  \nonumber \\
&&\left. +\sum_{i}\left( {\Theta}^{(i)}\right) ^{2}\right\} \ ,
\label{background28}\\
F_{3} &=& 2 N\, \Vol(S^3)+\frac{N}{4} d\left[\omega_1\wedge A^{(1)}  +\omega_2\wedge A^{(2)} +\omega_3\wedge A^{(3)}\right] \ ,  \nonumber \\
\Phi &=&\log \left( \frac{4}{N} r\right) \ ,\nonumber
\end{eqnarray}

we can also write the electric $F_{7}$ flux
    \begin{eqnarray}\label{F7Background2}
        C_6&=& \frac{\sqrt{2}N r^2}{8} dt\wedge dx_1\wedge dx_2\wedge d\mu\wedge\Big[\sin\vartheta d\vartheta\wedge d\varphi + f_2  \Big],\\
f_2&=&\cos\vartheta ~\Theta^{(1)}\wedge\Theta^{(2)} +  \sin\vartheta\sin\varphi ~\Theta^{(1)}\wedge\Theta^{(3)} + \sin\vartheta\cos \varphi ~\Theta^{(2)}\wedge\Theta^{(3)},\nonumber\\
F_7 &=& dC_6.\nonumber
    \end{eqnarray}

 This background does not preserve any supercharges, even for $m=0$.
We can expand the $(\mu,r)$ space around $r^2=m$ by setting $r= \sqrt{m} + \zeta^2$ and expanding  for small $\zeta$. We find
\begin{equation}
r (1-\frac{m}{r^2}) d\mu^2+ \frac{2 dr^2}{r(1-\frac{m}{r^2})}\sim 4 \left[d\zeta^2+ \frac{\zeta^2}{2} d\mu^2\right].
\end{equation}
This implies that we should choose the coordinate $\mu$ in the range $[0,\sqrt{8}\pi] $ to avoid conical singularities.

In this form we have a background that can be though as dual to a $(2+1)$-dimensional QFT. This field theory is the compactification of  $(5+1)$ Super Yang-Mills on $S^2[\theta,\varphi]\times S^1[\mu]$, with the $\mu$-circle shrinking smoothly at small values of the radial coordinate ($r\sim \sqrt{m}$).

It is interesting to discuss two features that differentiate the background in eq.(\ref{background28}) and that in eq.(\ref{BI}). The first difference, is the absence of the one form mixing the three sphere and the shrinking $S^1$. On the other hand, two coordinates of the five brane are compactified on a two sphere $(\vartheta,\varphi)$. For this compactification to be possible, the gauge field $A^{(i)}$ in eq.(\ref{gaugemeronfield}) is introduced. This one-form fibres the  two-sphere $S^2[\vartheta,\varphi]$ over the three sphere $S^3[\theta_1,\phi_1,\psi_1]$. It is also noticeable that whilst the circle $S^1[\mu]$ is shrinking, the two sphere is not. Actually, it has 'the same size' as the Minkowski $\mathbb{R}^{1,2}$.  The non-shrinking character of $S^2[\vartheta,\varphi]$ is responsible for various physical behaviours that we discuss below. The case $m=0$ is singular. In fact, the Ricci scalar for the background in eq.(\ref{background28}) is,
    \begin{equation}
        R= -\frac{6}{r} + \frac{2m}{r^{3}}
    \end{equation}

For $m=0$  the spacetime reaches $r=0$ the Ricci scalar and the dilaton diverge there, the whole configuration is not trustable (for $m=0$). Otherwise, for $m>0$ we have an everywhere smooth space. When the dilaton becomes $e^{\Phi}\sim 1$, we need to S-dualise, move to the NS-five brane frame and work with a LST. The analysis is parallel to that in Section \ref{sectionBIxx}. The only difference is that the $(5+1)$-d Super Yang-Mills is now compactified on the circle $\mu$ and on the two sphere $S^2(\vartheta,\varphi)$. As we mentioned, the two-sphere does not shrink, hence the QFT is never actually $(2+1)$ dimensional. Also, the wrapping on $S^2$ is not SUSY preserving.

Note that the charge quantisation works similarly in this background as in the one of eq.(\ref{BI}), giving
\begin{equation}
N_{D5}= \frac{1}{4\pi^2}\int_{S^3}F_3= N.
\end{equation}

\subsection{Observables }

We calculate some field theory observables using the background of eq.(\ref{background28}). 
We start with a probe D5 that extends along $[t,x_1,x_2,\mu,\vartheta,\varphi]$, keeping the rest of the coordinates fixed. We switch on a gauge field $F_{\mu\nu}$ in the $R^{1,2}$ directions. We follow now the procedure explained in Section \ref{gaugecoupling1}.
The induced metric and the Born-Infeld action expanded for small field are given by,
\begin{eqnarray}
& &ds_{ind}^2= r\left[ dx_{1,2}^2 + f(r) d\mu^2 + d\theta^2+\sin^2\theta d\varphi^2 \right], ~~~f(r)= 1 -\frac{m}{r^2}.\nonumber\\
& &  S_{BI}\approx T_{D5} N \pi  L_\mu r^2 \sqrt{f(r)} \int d^{3}x  - \frac{N \pi}{2}  T_{D5} L_\mu  \sqrt{f(r)} \int d^3 x F_{\mu\nu}^2.\label{couplingback28}
\end{eqnarray}
The effective 3d gauge coupling is
\begin{equation}
\frac{1}{g_{YM,3}^2 N}= 2\pi T_{D5}L_\mu \sqrt{1 -\frac{m}{r^2} }.\label{cani}
\end{equation}
  This coupling is very large in the IR, that for $r^2\sim m$ and asymptotes to a constant at high energies. Note that the singular solution with $m=0$ has a constant gauge coupling.
  
 We can calculate the Wess-Zumino term using the $C_6$ in eq.\eqref{F7Background2}. As mentioned, the background is not SUSY. We can nevertheless compare the charge and the tension of the probe D5 and obtain that they scale similarly for large values of the radial coordinate, but they are not exactly equal.
  
  \subsubsection{Baryon vertex}
  We can probe the background with other branes. For example, consider a D3 brane that extends in $[t, \theta_1,\phi_1,\psi_1]$ (that is, a point particle in the QFT and extends over the $S^3$ associated with the R-symmetry of the six dimensional dynamics). This probe can be associated with a baryon vertex. We can switch on a gauge field on this probe, to find that its Born-Infeld part is given at leading order by
  \begin{equation}
  S_{BI} =T_{D3}   16\pi^2 r_*
   \int d t .\label{barver}
  \end{equation}
 We have performed the integral over the angles and obtained an action for a point particle. The tension of this object (in this case, the mass) is finite, as it should be evaluated at $r_*=\sqrt{m}$ (where this tension is minimised). The Wess-Zumino part of the action (that we associate with the charge of this particle) is given by
  \begin{equation}
   S_{WZ}=- T_{D3} \int F_2\wedge C_2=T_{D_3}\int_{S^3} F_3 \int dt A_t = - T_{D3} 4\pi^2 N \int dt A_t.
  \end{equation}
  In summary, we see a particle with charge $Q=T_{D3} 4\pi^2 N$  and mass 
  $M=T_{D3 } 16\pi^2  r_* $ that propagates in time. This can be identified with a baryon vertex. Note that the action of fundamental strings emanating from the baryoon vertex should be added to eq.(\ref{barver}).

  {Notice that a  similar calculation can be done for the background in eq.(\ref{BI}). In fact, in that case the Baryon vertex is also a D3 brane extended on $[t, \theta,\phi,\psi]$.}


\subsubsection{Wilson and 't Hooft loops}
Let us now study Wilson and 't Hooft loops. We follow the usual treatment, already summarised in Section \ref{wilsonsectionx}.
Consider a fundamental string in the configuration $t=\tau, x=\sigma$ and $r(\sigma)$. The Nambu Goto action for such string (after the time integral is performed) is,
\begin{equation}
S_{NG}= T_{F1}T_\tau \int dx \sqrt{F^2+ G^2 r'^2 },~~~F= r,~~ G= \frac{\sqrt{2 }}{\sqrt{1-\frac{m}{r^2}   } }.\label{NGF1back28}
\end{equation}
We calculate the approximate separation between the quark pair $\hat{L}_{QQ}$, using eq.(\ref{laprox}), and the function $Z(r_0)$ defined in eq.(\ref{zeff}). This last function  indicates stability (if negative). The results are,
\begin{eqnarray}
& & L_{app}= \frac{\pi G}{F'}\big|_{r_0}=  \sqrt{2}\pi\frac{r_0}{\sqrt{r_0^2-m}}, ~~Z= L_{app}'\big|_{r_0}= -  \frac{\sqrt{2} \pi m}{(r_0^2-m)^{3/2}} <0. \label{Lapp background2 wilsonloop}
\end{eqnarray}
This indicates that the chosen configuration is stable. Also, the separation between the quark pair diverges in the IR (suggesting confinement)
 and stabilised towards the UV, to the value set by the LST. Compare this behaviour with the one obtained in Section \ref{wilsonsectionx} for the background of eq.(\ref{BI}).
 
 We can write the exact analytic expressions for the quantity $V_{eff}$, the separation $L_{QQ}$ and energy $E_{QQ}$ of the quark-antiquark pair, following eqs.(\ref{Veff}),(\ref{lengthwilson}),(\ref{energywilson}) respectively. These read,
 \begin{eqnarray}
 V_{eff}&=& \frac{1}{\sqrt{2} r_0}\sqrt{(r^2-m)(r^2-r_0^2)}, \label{integrals background2 wilsonloop}\\
   L_{QQ}(r_0)&=& \sqrt{8}r_0\int_{r_0}^\infty \frac{dr}{\sqrt{(r^2-m)(r^2-r_0^2)}}=  2\sqrt{2}\mathbf{F}\left( \arcsin \left( \frac{r_{0}}{%
\sqrt{m}}\right) ;\frac{m}{r_{0}^{2}}\right) +i2\sqrt{2}\mathbf{K} \left( 1-\frac{m}{r_{0}^{2}}\right) \nonumber\\
 E_{QQ}(r_0)&=& r_0 L_{QQ} + \sqrt{8}\left[ \int_{r_0}^\infty dr \sqrt{\frac{r^2-r_0^2}{r^2-m}} -\int_{m}^\infty \frac{r dr}{\sqrt{r^2-m} }\right]\nonumber\\
 &=&  r_0 L_{QQ}(r_0) + \sqrt{8}r_{0}\mathbf{E}\left( \frac{m}{r_{0}^{2}}\right) -\sqrt{8m}\mathbf{%
E}\left( \frac{r_{0}^{2}}{m}\right) +2i\sqrt{8m}\mathbf{E}\left( 1-\frac{%
r_{0}^{2}}{m}\right)  \nonumber\\
&&-\sqrt{8}r_{0}\mathbf{E}\left( \arcsin \left( \frac{r_{0}}{m}\right) ;%
\frac{m}{r_{0}^{2}}\right) +\sqrt{\frac{8}{m}}\left( m-r_{0}^{2}\right) 
\mathbf{K}\left( \frac{r_{0}^{2}}{m}\right)  \nonumber\\
&&-2ir_{0}^{2}\sqrt{\frac{8}{m}}\mathbf{K}\left( 1-\frac{r_{0}^{2}}{m}%
\right) .\nonumber
\end{eqnarray}
Here, $\mathbf{F}(\phi;x)$,$\mathbf{K}(x)$, $\mathbf{E}(\phi;x)$ and $\mathbf{E}(x)$ and are given by
    \begin{equation}
    \begin{aligned}
        &\mathbf{F}(\phi;x) = \int^{\phi}_{0} d\theta \frac{1}{\sqrt{1-x\sin ^{2}\theta }} ,\quad
        \mathbf{K}(x) = \mathbf{F}\left(\frac{\pi}{2};x\right), \\
        &\mathbf{E}(\phi;x) = \int^{\phi}_{0} \sqrt{1-m\sin ^{2}\theta} d\theta,\quad
        \mathbf{E}(x) = \mathbf{E}\left(\frac{\pi}{2};x\right)
    \end{aligned}
    \end{equation}

\begin{figure}[h]
    \centering
    \includegraphics[scale=0.7]{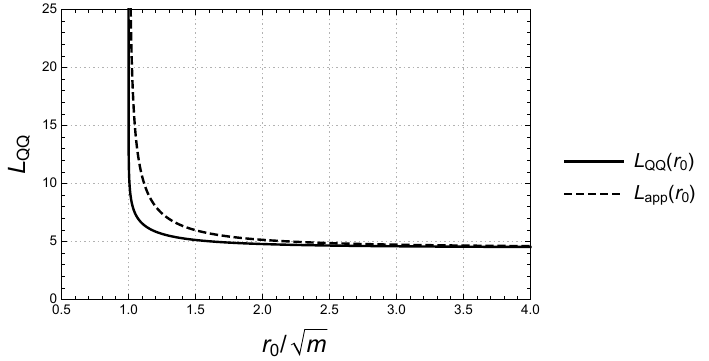}\quad
    \includegraphics[scale=0.7]{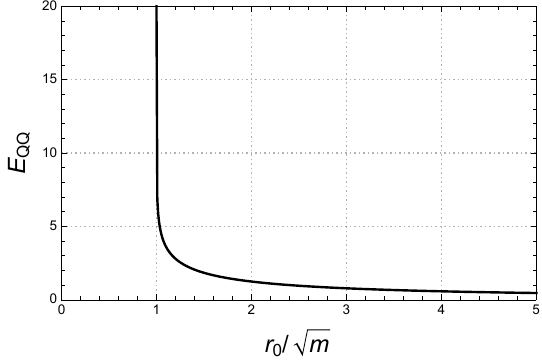}\\ \vspace{0.5cm}
    \includegraphics[scale=0.7]{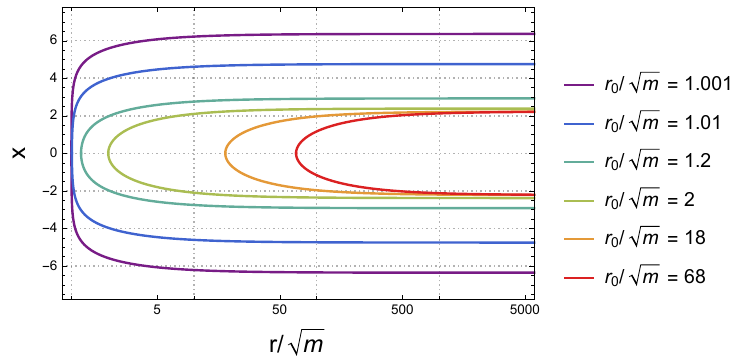}
    \includegraphics[scale=0.7]{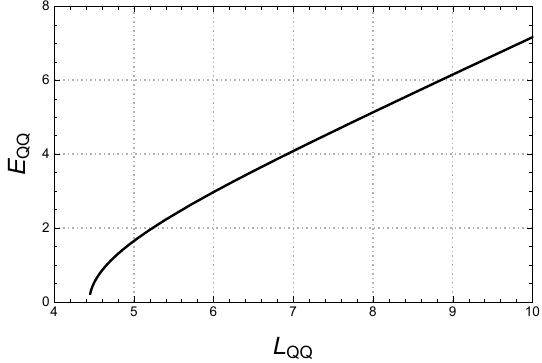}
    \caption{The integrals of the quark-antiquark Wilson loop \eqref{integrals background2 wilsonloop}. Upper left: Plot comparing the exact expression for the monopole separation \eqref{integrals background2 wilsonloop} with the approximate expression $\hat{L}_{QQ}$ in \eqref{Lapp background2 wilsonloop}. Upper right: The energy of the Wilson loop as a function of the $r_0$. Bottom left: Different  profiles of the macroscopic strings as a function of the turn-around point $r_0$. Bottom Left: Plot of the function $E_{QQ}(L_{QQ})$.}
    \label{fig:WilsonLoopMeron}
\end{figure}

\subsubsection{'t Hooft loops}
We now study 't Hooft loops. The string between two magnetic monopoles is modelled using a D5 brane extending in $[t, x_1,\mu, \theta_1,\phi_1,\psi_1]$, with $r(x_1)$ and all other coordinates set to constant.
The action for the effective magnetic string (after integrating over the internal space) is given by,
\begin{equation}
S_{eff}= T_{D5} L_\mu 16\pi^2 \int dx_1 \sqrt{F^2+ G^2 r'^2},~~~F= r\sqrt{r^2-m},~~G=\sqrt{2} r.
\end{equation}
The expressions of the approximate length and the function $Z$ are,
\begin{equation}
\hat{L}_{app}= \frac{\sqrt{2}\pi r \sqrt{r^2-m}}{2r^2-m},~~~Z= \frac{\sqrt{2} m^2\pi }{(2r^2-m)^2 \sqrt{r^2-m}}.
\end{equation}
This suggest that the embedding is unstable. In fact, in agreement with what was discussed above, the tension of the monopole-anti monopole string vanishes for $r\to \sqrt{m}$. This tension is associated with the function $F(r)$. The instability of the embedding suggest that the configuration of a monopole pair connected by a magnetic string should be replaced by the pair moving freely, at no energy expense. In other words, there should be screening of the monopoles, in agreement with the confinement of the quarks.

\subsubsection{Holographic central charge or free energy}
Now we compute the holographic central charge of the background in eq.(\ref{background28}), in the conventions that we introduced in \eqref{conventions central charge}-(\ref{chol}).
\begin{equation}
d=2\ ,\quad \alpha =r\ ,\quad \beta =\frac{2}{r^{2}- m }\ ,\quad e^{-4\Phi }=\frac{N^4}{4^{4}r^{4}}\ .
\end{equation}%
and%
\begin{equation}
g_{ij}d\theta ^{i}d\theta ^{j}=r \left[ (1-\frac{m}{r^2}) d\mu ^{2}+d\vartheta ^{2}+\sin ^{2}\vartheta d\varphi
^{2}+\sum_{i}\left( \Theta^{(i)}\right)^2 \right]
\end{equation}%
The functions $V_{\text{int}},H$ and the
holographic central charge are%
\begin{equation}
V_{\text{int}}=\mathcal{N}r\sqrt{r^2-m }\ ,\quad 
\mathcal{N}=4 \pi^3 N^2 L_{\mu }\ ,\quad c_{%
\text{hol}}=\frac{2\mathcal{N} r^{3} (r^2-m)^{3/2}}{G_{N} (2 r^2-m)^2}\ .\label{chol2+1}
\end{equation}
In line with the confining behaviour indicated by the Wilson and 't Hooft loops the number of states vanishes for $r\sim \sqrt{m}$. Hence the system is gapped. On the other hand, for large energies, the number of states grows unbounded. Note that the growth for large values of the radial coordinate is the same for this QFT and it is for the QFT dual to the background in eq.(\ref{BI}). In fact, both densities of states, the one in eq.(\ref{chol2+1}) and that in eq.(\ref{chol1}) diverge similarly.
This is not strange as both theories are very similar at high energies.

\subsubsection{Spin-two glueballs}

We follow Section \ref{glueballssection} and study glueballs of spin-two in this confining and gapped $(2+1)$ dimensional QFT.
Indeed, according to eqs.(\ref{einsteinresc})-(\ref{vvv}), we find
\begin{eqnarray}
& & ds^2_{E,int}= \frac{\sqrt{N r}}{2}\left\{ \left(1-\frac{m}{r^{2}}\right) d\mu^{2} +
\frac{2dr^{2}}{r^{2}\left( 1-\frac{m}{r^{2}}\right) }+d\vartheta ^{2}+\sin ^{2}\vartheta d\varphi
^{2}  +\sum_{i}\left( {\Theta}^{(i)}\right) ^{2}\right\} \ ,\nonumber\\
& &e^{2A}=\frac{\sqrt{Nr}}{2},~~\sqrt{\det{\hat{g}_{ab}}}=\frac{\sqrt{2}}{r} \sin\theta_1\sin\vartheta,~~V(r)= \frac{1}{2} \partial_r\left[ \frac{r^2-m}{r}\right].
\end{eqnarray}

When computing the potential above, we consider fluctuations that do not vary on the 3-sphere. Hence, $\Psi = \Psi(r,\mu)$. In the tortoise coordinate
    \begin{equation}
        r = \sqrt{\mu}\cosh\left(\frac{\rho}{2}\right),
    \end{equation}

which maps $r\in [\sqrt{m},+\infty[$ to $\rho \in [0,+\infty[$, and considering fluctuations of the form 
    \begin{equation}
        \Psi(r(\rho),\varphi) = \left(1-\frac{m}{r(\rho)^{2}}\right)^{-\frac{1}{4}} e^{i\, \frac{n\,\varphi}{\sqrt{2}}} \Theta(\rho),
    \end{equation}

from \eqref{SchroedingerLikeAppendix} we obtain a Schroedinger like equation that reads
    \begin{equation}
        -\frac{d^{2}\Theta}{d\rho^{2}} + \tilde{V}(\rho)\Theta = M^{2} \Theta, 
    \end{equation}

where the effective potential reads
    \begin{equation}
        \tilde{V}(\rho) = \frac{1}{4} - \frac{1}{4} \frac{1}{\sinh^{2}(\rho)} + \frac{n^{2}}{2} \coth^{2}\left(\frac{\rho}{2}\right).
    \end{equation}

This potential has the following asymptotic expansions
    \begin{equation}
    \begin{aligned}
        \tilde{V}(\rho\rightarrow 0) &= \left( 2n^{2}-\frac{1}{4}\right)\frac{1}{\rho^{2}} + \frac{1}{3}(1+n^{2}) + \mathcal{O}(\rho^{2}),\\
        \tilde{V}(\rho\rightarrow +\infty) &= \frac{1}{4}(1+2 n^2) + \mathcal{O}(\rho^{-1})
    \end{aligned}
    \end{equation}
The potential does not dhave a minimum. For different values of $n\geq 1$ we find a continuous spectrum.

\section{Comparison between different backgrounds}\label{sectioncomparison}
{
In this section we study three backgrounds, two of which are already present in the bibliography and the third one is new. The goal of the section is to compare the UV-behaviour and some observables computed using the large-$r$ regime of the geometries. Even when the metrics quoted below present a singular behaviour for small values of the $r$-coordinate this is not a concern for us: on the one hand we do not use the small $r$ region in our calculations. Aside from this, the backgrounds have a smooth ''IR completion'' that we do not quote in this section, just to keep the expressions simple.}
\\
{
As stated, the goal of this section is to make some general observations about the UV-behaviour of different backgrounds describing compactified five branes.
The common feature is that the five branes wrap a two sphere, with different field theory interpretation. The outcome being that in some cases, the system is not describing a QFT, but  behaviours characteristic of the Little String are found. As we discuss below this can be diagnosed by inspecting the dilaton and the fate of the compactification manifold.
}
\\
{ In the second part of this section we compare the SUSY preserving mechanism described in Section \ref{sectionBIxx} and in Appendix \ref{BIsusy} with the partial topological twist. Although conceptually different, we highlight their similarities as can be seen from the holographic perspective}


\subsection{D5-Branes on  \texorpdfstring{$S^{2}$}{S2}}

Let us start by quoting three backgrounds.
The first background represents a twisted compactification of a stack of $N$ D5 brane on a two sphere in the resolved conifold, preserving four supercharges
\cite{Maldacena:2000yy}. Using the left invariant forms defined in eq.(\ref{su2-left}), it reads,
\begin{eqnarray}
& & ds^2_{st}= e^{\Phi}\Big[dx_{1,3}^2 +dr^2+ e^{2h}(d\alpha^2+\sin^2\alpha d\beta^2) +\frac{1}{4}\left({\omega}_1^2+ \omega_2^2+ (\omega_3-\cos\alpha d\beta)^2 \right) \Big],\nonumber\\
& & F_3= \frac{N}{4}\Big[ \sin\alpha~ d\alpha\wedge d\beta\wedge \omega_3 -\omega_1\wedge \omega_2\wedge\omega_3 \Big],\nonumber\\
& & e^{4\Phi}=\frac{e^{4\Phi_0+ 4 r}}{r},\;\;\; e^{2h}= r\label{d5ons2}
\end{eqnarray}
We emphasise that whilst the background above is singular at $r=0$, the singularity has been resolved in \cite{Maldacena:2000yy}. We are interested in the comparison with other backgrounds at large values of $r$, or analogously, in the UV regime of the QFT dual.

The second background was written in Section 4.5 of \cite{Casero:2006pt} and reanalysed in \cite{Casero:2007jj}, \cite{Hoyos-Badajoz:2008znk}. This is a solution proposed to be dual to ${\cal N}=1$ SQCD with gauge group $SU(N)$ and with $N_f=2N$ flavours. The QFT is deformed by a higher dimension operator that breaks the flavour group to the diagonal $SU(N_f)$. The background reads,
\begin{eqnarray}
& & ds^2_{st}= e^{\Phi}\Big[dx_{1,3}^2+ N \left( dr^2+ \frac{1}{\xi} d\Omega_2(\theta,\varphi) + \frac{1}{4-\xi} d\Omega_2(\tilde{\theta},\tilde{\varphi}) +\frac{1}{4}(d\psi+\cos\theta d\varphi+\cos\tilde{\theta} d\tilde{\varphi})^2\right) \Big]\nonumber\\
& &  F_3=-\frac{N}{2} \left[\sin\theta d\theta\wedge d\varphi +  \sin\tilde{\theta} d\tilde{\theta} \wedge d\tilde{\varphi} \right] \wedge (d\psi+\cos\theta d\varphi+cos\tilde{\theta} d\tilde{\varphi}),\nonumber\\
& & e^{4\Phi}=e^{4\Phi_0+ 4r}.\label{nf=2nc}
\end{eqnarray}
Here $\xi \in (0,4)$. This is a SUSY  solution of the equations of motion of Type IIB in the presence of back-reacting sources. By this we mean a solution of the equations of motion derived from the Type IIB action supplemented by (smeared) sources. In fact, one can check that 
\begin{equation}
dF_3= N \sin\theta \sin\tilde{\theta}~ d\theta\wedge d\varphi\wedge d\tilde{\theta}\wedge d\tilde{\varphi}.\nonumber
\end{equation}
The third background reads, \footnote{We find this background taking the $m\to 0$ limit 
 and changing $r=e^{\frac{\rho}{\sqrt{2}}}$ in eq.(\ref{MeronBH}). Similar operations can be implemented in the background of eq.(\ref{background28}), also relabelling $\mu=x_3$.}

\begin{equation}\label{MeronBHeq2}
    \begin{aligned}
        ds^2_{st}  &= \frac{N}{4} e^{\Phi}\left[ -dt^{2} + \sum^{3}_{i=1} dx^{2}_{i} +
                        d\rho^2 
                        +d\vartheta ^{2}+\sin ^{2}\vartheta d\varphi^{2} +\sum_{i}\left( {\Theta}^{(i)}\right) ^{2}\right]\\
        F_{3} &= 2 N \Vol(S^{3}) + \frac{N}{4} d\left[\omega_1\wedge A^{(1)}  +\omega_2\wedge A^{(2)} +\omega_3\wedge A^{(3)}\right], \\
    e^{4\Phi} &=\frac{4^4}{N^4}e^{\frac{4\rho}{\sqrt{2}}}
    \end{aligned}
    \end{equation}

The three backgrounds have a linear dilaton and compactify D5 branes on $S^2$. The first two (\ref{d5ons2}),(\ref{nf=2nc}) preserve four supercharges, the third one in eq.(\ref{MeronBHeq2}) preserves no supercharges. We can calculate observables as we did in the previous sections. 

For the background in eq.(\ref{d5ons2}), we calculate a gauge coupling defined using a probe D5 extended on $R^{1,3}$ and wrapping the manifold $\theta=\alpha,\; \varphi= 2\pi-\beta,\;\psi=\psi_{*}$. 
The result found is
\begin{equation}
 \frac{1}{g_{eff,4}^2}= 4\pi T_{D5}e^{2h} = 4\pi T_{D5} r.\label{gaugecouplingd5ons2}
 \end{equation}
On the other hand, for the background in eq.(\ref{nf=2nc}) an analog computation to the one above finds the gauge coupling to be constant (see \cite{Casero:2006pt}),
\begin{equation}
\frac{1}{g_{4,eff}^2 N}\sim \frac{1}{\xi(4-\xi)}.
\end{equation}
Similarly, for the background in eq.(\ref{MeronBHeq2}), the probe D5 extended on $R^{1,3}$ and wrapping $(\theta,\varphi)$ we find a constant gauge coupling, as can be seen from eq.(\ref{cani}) in the $m\to 0$ limit. In summary, we find a similar behaviour for the backgrounds in eqs.(\ref{nf=2nc}),(\ref{MeronBHeq2}), different to the running behaviour found for the background of eq.(\ref{d5ons2}).

We can calculate Wilson loops following the formulas in Section \ref{wilsonsectionx}. For the background in eq.(\ref{d5ons2}), we have,
\begin{eqnarray}
& & F^2(r)=G^2(r)=e^{2\Phi(r)}= \frac{e^{2\Phi_0+2r}}{\sqrt{r}}, ~~V_{eff}=e^{-\Phi(r_0)} \sqrt{ e^{2\Phi(r)}    - e^{2\Phi(r_0)} },\\
& & L_{app}= \frac{\pi}{\Phi'}=\frac{4\pi r}{4 r-1},~~Z(r)= -\frac{4\pi}{(4r-1)^2}.\nonumber
\end{eqnarray}
These expressions are not trustable near to $r=0$, close to the singularity. We analyse them for large values of $r$, indicating a minimal separation that tends to grow as we approach the interior of the geometry. The negativity of $Z(r)$ indicates that the embedding is stable.
The relevant integrals for the exact expressions,
\begin{eqnarray}
& & L_{QQ}(r_0)= 2 e^{\Phi(r_0)}\int_{r_0}^\infty \frac{dr}{\sqrt{e^{2\Phi(r)}  -e^{2\Phi(r_0)}}}, \label{LQQ and EQQ background 4.1} \\
& & E_{QQ}= e^{\Phi(r_0)}L_{QQ}(r_0) + 2 \int_{r_0}^\infty dr \sqrt{e^{2\Phi(r)}  -e^{2\Phi(r_0)}} -2 \int_{0}^\infty e^{\Phi(r)} dr.\nonumber
\end{eqnarray}
\begin{figure}[h]
\centering
     \includegraphics[scale=0.5]{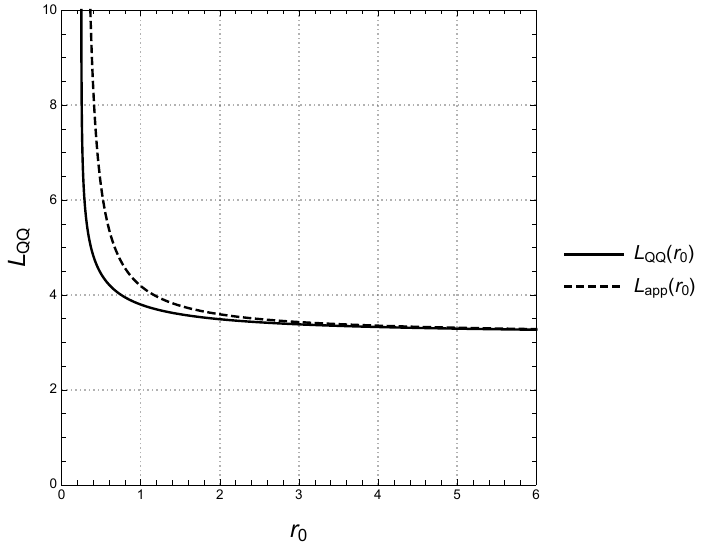}\includegraphics[scale=0.5]{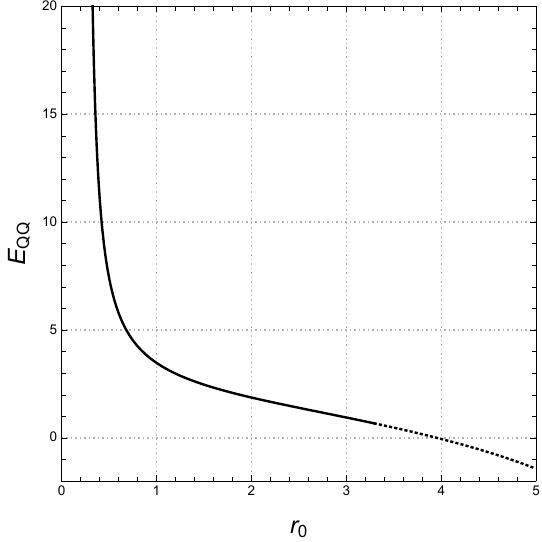} \quad \includegraphics[scale=0.5]{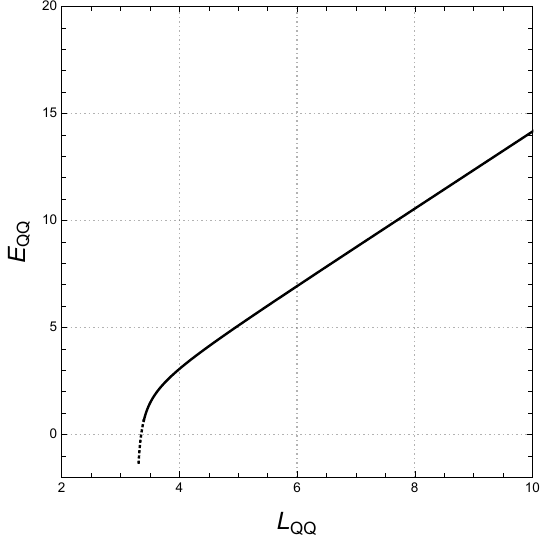}
    \caption{Plots the integrals \eqref{LQQ and EQQ background 4.1} associated to the Wilson loop of the background \eqref{d5ons2} for $\Phi_0=0$. Left: Comparison between the exact expression and approximated 
 expression for the quark-anti quark separation. Center: Plot of the energy $E_{QQ}$ as a function of $r_0$ Right: Plot of the function $E_{QQ}(L_{QQ})$. The dotted lines in the center and right panels indicate regions where the numerical integration is not faithful.}
    \label{fig:enter-label}
\end{figure}

The Wilson loop for the background in eq.(\ref{nf=2nc}) give the functions,
\begin{eqnarray}
& &F^2(r)= G^2(r)=e^{2\Phi(r)}=e^{2r}, ~~~V_{eff}= e^{-r_0}\sqrt{e^{2r}. -e^{2r_0}},\nonumber\\
& & L_{QQ}= \pi,~~~E_{QQ}= 0.
\end{eqnarray}
The integrals associated with separation $L_{QQ}$ and Energy $E_{QQ}$ can be explicitly performed, as shown in \cite{Casero:2006pt}. One can check that $L_{app}=L_{QQ}$ and $Z(r)=0$.

For the background in eq.(\ref{MeronBHeq2}) we find

\begin{equation}
    F^2(r)=G^2(r)= e^{\sqrt{2}\rho},~~ V_{eff}=\sqrt{e^{\sqrt{2}(\rho-\rho_0)} -1}.~~L_{app}=\text{constant}, ~~Z=0.
\end{equation}
Comparing these observables (something similar occurs for other observables), we find that the phenomenology of the backgrounds (\ref{nf=2nc}),(\ref{MeronBHeq2}) is similar, whilst the background in eq.(\ref{d5ons2}) has a different dynamical behaviour. We observe that in background (\ref{d5ons2}) the dilaton is not exactly linear, whilst it is linear in the other two. Also that for the backgrounds in (\ref{nf=2nc})-(\ref{MeronBHeq2}), the sphere is not shrinking. These different features explain the differences in the ensuing dynamics. Something analogous occurs between the backgrounds in Sections 2.1 and 2.2 of the paper \cite{Nunez:2023nnl}.

We believe that backgrounds like (\ref{nf=2nc})-(\ref{MeronBHeq2}) are not representing the dynamics of a four dimensional QFT. Instead, they reproduce the dynamics of  the LST with two directions on a fixed size two manifold. Notice that something different happens with the backgrounds of Sections \ref{sectionBIxx} and \ref{section2+1qft} {(even when in those backgrounds the compactification manifold is $S^1$)}. In those cases there is a direction on the D5 branes that is actually shrinking, by virtue of the function $f_s(r)$ in eq.(\ref{functions4+1}) and $f(r)=1-\frac{m}{r^2}$ in eq.(\ref{background28}).

\subsection{Comparison of  the SUSY preservation mechanisms}

Let us now make a comparative study between the background in eq.(\ref{BI}), a dual to a $(4+1)$ confining QFT preserving eight supercharges and another background written in \cite{Maldacena:2000yy}. This second background is proposed as a dual to $(3+1)$ dimensional confining QFT with four preserved supercharges. The description above suggest that there are few commonalities between these solutions. To see some common features, let us quote the two metrics (a similar analysis can be made for the dilaton and Ramond fields). For the background in eq.(\ref{BI}) we have
\begin{eqnarray}
      & &  ds^{2}_{st} = r \left[ dx_{1,4}^2 + f_{s}(r)d\varphi^{2} 
            + \frac{Ndr^{2}}{ r^{2}f_{s}(r)} 
            + \frac{N}{4}\left(\omega^{2}_{1} + \omega^{2}_{2} + 
        \left(\omega_{3} - \sqrt{\frac{8}{N}} Q \zeta(r) d\varphi\right)^{2}\right) \right],\nonumber\\
        & & f_s(r)=\frac{(r^2-r_+^2)(r^2-r_-^2)}{r^4},~~~\zeta(r)= \frac{1}{r^2}-\frac{1}{r_+^2}.\label{BII}
        \end{eqnarray}
        For the background in \cite{Maldacena:2000yy}. we have
\begin{eqnarray}
& &        ds^{2}_{st} = N e^{\Phi(r)} \Bigg[ \frac{1}{N} dx_{1,3}^2 + e^{2h(r)}( d\theta^2 + \sin^2\theta d\varphi^{2}) 
            + dr^{2} +\nonumber\\
& &            + \frac{1}{4} (\omega_{1} +a(r) d\theta)^2+ \frac{1}{4}(\omega_{2}-a(r)\sin\theta d\varphi )^2 + \frac{1}{4} 
        (\omega_{3} - \cos\theta d\varphi)^{2} \Bigg],\nonumber\\
        & & a(r)=\frac{2r}{\sinh(2r)},~~  e^{2h(r)}= r \coth(2r) -\frac{1}{4}(a(r)^2-1),~~e^{-4\Phi(r)}= \frac{4 e^{-4\Phi_0 +2h(r)}}{\sinh^2(2r)}.\label{MN} 
        \end{eqnarray}
The points we wish to emphasise are the following:
\begin{itemize}
\item{The solution in eq.(\ref{MN}) is SUSY thanks to a twisting procedure, namely a mix between the  Lorentz group and the $SO(4)$ R-symmetry group of the D5 brane theory. See  \cite{Andrews:2005cv} for a detailed account of the twisting in this QFT. The twisting is reflected by the fibration of the $S^3(\omega_i)$ on the $S^2(\theta,\varphi)$. This fibre does not vanish at large values of the radial coordinate (the far UV of the QFT). The function $a(r)$ takes the form in eq.(\ref{MN}) to allow a smooth ending of the space at $r\sim 0$. Close to the end of the space, the dilaton presents a minimum and the $(r, S^2)$ shrinks as flat space. The $SU(2)$-valued one form performing the fibre $A= -a(r) d\theta~ T^{1} + a(r)\sin\theta d\varphi~ T^{2} + \cos\theta d\varphi~ T^{3}$ has zero curvature at the origin of the space.}
\item{The background in eq.(\ref{BII}) preserves SUSY due to a mechanism nicely explained recently  in  \cite{Anabalon:2021tua}, \cite{Anabalon:2022aig}, see also Appendix \ref{BIsusy}. Whilst this is not a twisting procedure, it shares some  features with it. In fact, the fibre of the $S^3$ on the $S^1$ does not vanish for large values of the radial coordinate--by virtue of the $\frac{1}{r_+^2}$ term in the function $\zeta(r)$. This allows for spinors with anti-periodic boundary conditions to exists in this configuration. The rest of the one form performing the fibre (the part dependent on $r$) is what allows the configuration to have a smooth closure near $r\sim r_+$. In fact, the dilaton presents a minimum and the sub-space $(r,\varphi)$ shrinks as flat space. The $U(1)$-valued one form performing the fibre $A=\sqrt{\frac{8}{N}} Q\zeta(r) d\varphi$, vanish at the end of the space $r=r_+$.}
\end{itemize}
In other words, whilst both dual QFTs are different, some common features are shared. This is clear using the holographic (gravitational) description. The common characteristics of metrics and background fields are more apparent in that language.

\section{Black Membranes}\label{bhsectionsx}

 In this section we present  two new type IIB solutions containing black membranes. These are obtained by Wick rotating the solutions in Sections \ref{sectionBIxx} and \ref{section2+1qft}. In order to study thermodynamics aspects of the solutions, we use the expressions obtained in \cite{Nunez:2023nnl}, where the Noether-Wald formalism \cite{Wald:1993nt} was used to compute the gravitational charges in backgrounds with dilaton and RR 3-form switched-on. We summarise these results below

 \subsection{Gravitational Charges in \texorpdfstring{$F_{3}$}{F3}-Dilaton Backgrounds}

The Noether prepotential for this type of configurations is given by
    \begin{equation}
        q^{\mu \nu }\left( \xi \right) 
        =-\frac{1}{\kappa ^{2}}\left( \nabla^{\lbrack \mu }\xi ^{\nu ]} +  
        e^{\Phi }C_{\lambda \rho }\xi^{\lambda }F^{\mu \nu \rho }\right) \ .
    \end{equation}

By integrating the Hodge dual of the prepotential
    \begin{equation}
        \boldsymbol{Q}\left[ \xi \right] =\frac{1}{2}\frac{1}{8!}\sqrt{-g}\epsilon
        _{\mu \nu \rho _{1}\dots \rho _{8}}q^{\mu \nu }dx^{\rho _{1}}\wedge \dots
        \wedge dx^{\rho _{8}}\ ,
    \end{equation}
we can obtain the charges of the configurations by choosing different Killing vector fields $\xi$. In order to obtain finite mass and have a well posed action principle we include the boundary term
    \begin{equation}
        S_{\text{full}}=S_{\text{IIB,bulk}}+\int_{\partial M}d^{9}x\sqrt{-h}\frac{1}{\kappa ^{2}}\left( 
        \mathcal{K} +\mathcal{L}_{\rm ct} \right) \ , \label{Sfull IIB}
    \end{equation}
As usual, the extrinsic curvature is defined in terms of the normal unit outwards  vector $n^{\mu }$
to the boundary of the spacetime by
    \begin{equation}
        \mathcal{K}_{\mu \nu }=h_{\ \mu }^{\rho }h_{\ \nu }^{\sigma }\nabla _{\rho
        }n_{\sigma }\ .
    \end{equation}
In \eqref{Sfull IIB} $\mathcal{L}_{\rm ct}$ stands for the conterterms that depend on intrinsic quantities defined in the boundary of the spacetime. For asymptotically AdS spacetimes there is a well established procedure to systematically find the counter terms that accommodate the boundary conditions for the fields, leading to finite on-shell action, finite charges and an extermum of the action principle \cite{Emparan:1999pm},\cite{Balasubramanian:1999re},\cite{Bianchi:2001kw}. This renormalization procedure relies on the existence of a Fefferman-Graham expansion of the fields close to the boundary. For asymptotically flat configurations, as appear in D5 systems \cite{Boonstra:1998mp}, the situation is more subtle because instead of having a boundary manifold, as in the AdS case, one has a \textit{family of boundary manifolds} \cite{Mann:2005yr}. Different proposals have been made to construct a general family of counterterms for asymptotically flat spacetimes with intrinsically geometric quantities \cite{Kraus:1999di}. However, in presence of matter fields, such as scalars, one is able to consider a function of the scalar field at the boundary as a counterterm. We will consider this type of counterterms in what follows. A careful analysis using the standard holographic renormalisation scheme will be left  for future research.

If the background has a a timelike Killing $\boldsymbol{{t}}$ and a rotation generator $\boldsymbol{\psi}$ then, following \cite{Wald:1993nt} the energy, angular momentum and entropy are defined by%
\begin{eqnarray}
\mathcal{E}\left[ \boldsymbol{t}\right]  &=&\int_{\infty }\left( \boldsymbol{Q}\left[ 
\boldsymbol{t}\right] -\boldsymbol{t} \lnot \boldsymbol{B}\right) \ ,
\label{def energy} \\
\mathcal{J}\left[ \boldsymbol{\psi }\right]  &=&-\int_{\infty }\boldsymbol{Q}\left[ 
\boldsymbol{\psi }\right] \ ,  \label{def J} \\
S\left[ \boldsymbol{\xi }\right]  &=&\frac{1}{T}\int_{\mathcal{H}}%
\boldsymbol{Q}\left[ \boldsymbol{\xi }\right] \ ,  \label{def S}
\end{eqnarray}
where $\lnot$ is the contraction operator, $\boldsymbol{B}$ is a 9-form given by  
    \begin{equation}
        \boldsymbol{B}=-\frac{1}{\kappa ^{2}}\left( \mathcal{K} +\mathcal{L}_{\rm ct}\right) 
                \star n\ ,
    \end{equation}
and  $\boldsymbol{\xi}$ is the horizon generator
    \begin{equation}
        \boldsymbol{\xi }=\boldsymbol{t}+\Omega \boldsymbol{\psi }.
    \end{equation}

Here, the angular velocity $\Omega$ is such that $\boldsymbol{\xi }$ is null at the horizon and also that satisfies the geodesic equation at the horizon
    \begin{equation}\label{geodesic}
        \xi ^{\mu }\nabla _{\mu }\xi ^{\nu }=\kappa _{s}\xi ^{\nu }\ ,
    \end{equation}
defining the surface gravity $\kappa _{s}$ that is related to the
temperature as $T=\frac{\kappa _{s}}{2\pi }$.

For now on we consider the boundary counterterm as
\begin{equation}
    \mathcal{L}_{\rm ct}=-e^{-\Phi/4} \ \, , \label{L_ct black holes}
\end{equation}
which will allows us to renormalize the physical quantities for a fixed value of $\Phi_0$ that we introduce latter.

\subsection{Rotating Black Brane}
In this section we study the black membrane solutions constructed from the
double analytic continuation of the background \eqref{BI}%
\begin{equation}
\varphi \rightarrow it\ ,\quad t\rightarrow ix_{5}\ ,\qquad Q\rightarrow
-iQ\ .
\end{equation}%
For simplicity of the computations we move to Einstein frame where the black membrane
configuration has the following form%
\begin{eqnarray}
ds_{E}^{2} &=&\sqrt{r}\left[ -dv\left( f_{s}\left( r\right) dv-2\sqrt{N}%
r^{-1}dr\right) +dx_{5}^{2}\right]   \label{black hole from cigar} \\
&&+\frac{N\sqrt{r}}{4}\left[ d\theta ^{2}+\sin ^{2}\theta d\phi +\left(
d\psi ^{\prime }+\cos \theta d\phi -\sqrt{8N^{-1}}Q\zeta \left( r\right)
dv\right) ^{2}\right] \ ,  \notag \\
F_{3} &=&e^{-\frac{\Phi _{0}}{2}}d\left( \frac{N}{4}\cos \theta d\psi
^{\prime }\wedge d\phi +\sqrt{\frac{N}{2}}Q\zeta \left( r\right) \left(
d\psi ^{\prime }+\cos \theta d\phi \right) \wedge dv\right) \ , \\
\Phi  &=&\log \left( r\right) +\Phi _{0}\ , \\
f\left( r\right)  &=&1-\frac{m}{r^{2}}+\frac{2Q^{2}}{r^{4}}\ ,\qquad r_{\pm
}^{2}=\frac{m}{2}\pm \frac{1}{2}\sqrt{m^{2}-8Q^{2}}\ ,
\end{eqnarray}%
written in in-going Eddingtong-Finkelstein coordinates defined by%
\begin{equation}
dv=dt+\frac{\sqrt{N}}{rf_{s}\left( r\right) }dr\ ,\qquad d\psi ^{\prime
}=d\psi +\frac{\sqrt{8}Q\zeta \left( r\right) }{rf_{s}\left( r\right) }dr\ .
\end{equation}%

We fixed $\Phi _{0}=\log \frac{N^{2}}{16}$ in order to obtain a finite mass. Considering the horizon generator%
\begin{equation}
\boldsymbol{\xi }=\boldsymbol{t}+\Omega \boldsymbol{\psi }
\end{equation}%
which is null Killing vector at the horizon located at $r=r_{+}$ for the
following quantities%
\begin{equation}
\boldsymbol{t}=\partial _{v}\ ,\qquad \boldsymbol{\psi }=\frac{%
\partial }{\partial \psi }\ ,\qquad \Omega =\sqrt{\frac{8}{N}}\frac{Q}{%
r_{+}^{2}}\ .
\end{equation}%
The mass, angular momentum, entropy and Hawking temperature for the configuration  are given by%
\begin{eqnarray}
M &=&\frac{N\left( r_{+}^{4}+2Q^{2}\right) }{\kappa ^{2}r_{+}^{2}}%
L_{x_{1}}L_{x_{2}}L_{x_{3}}L_{x_{4}}L_{x_{5}}\pi ^{3}\ , \\
J &=&\frac{\sqrt{2N^{3}}Q}{\kappa ^{2}}%
L_{x_{1}}L_{x_{2}}L_{x_{3}}L_{x_{4}}L_{x_{5}}\pi ^{3}\ , \\
S &=&\frac{4N^{3/2}r_{+}^{2}\left( r_{+}^{4}+2Q^{2}\right) }{%
\left( r_{+}^{4}-2Q^{2}\right) }%
L_{x_{1}}L_{x_{2}}L_{x_{3}}L_{x_{4}}L_{x_{5}}\pi ^{4}\ , \label{entropy rot black brane} \\
T &=&\frac{r_{+}^{4}-2Q^{2}}{2\sqrt{N}\pi r_{+}^{4}}\ .
\end{eqnarray}
The non-trivial angular momentum indicates that the spacetime is rotating. Note that the entropy receives a contribution from the $F_3$ term in \eqref{def S} leading an entropy which is not one-quarter of the horizon area. This is also an indication of the non-triviality of the matter fields in this configuration. One could try to perform a gauge transformation $C_2\mapsto C_2+d\lambda_1$ in order to obtain vanishing $C_2$ at the horizon. A 1-form gauge parameter which does the job is $\lambda_1=(d\psi'+\cos(\theta) d\phi)v/r_0^2$, but it changes the mass of the spacetime because the gauge parameter does not vanish at infinity.

Another subtlety regarding the charges of this spacetime is that if we insist in considering the entropy as one-quarter of the horizon area, we find that the following differential relation between the thermodynamic quantities holds $dM=TdS-2\Omega dJ$. 

For this black hole configuration, the Euclidean on-shell action \eqref{Sfull IIB} with the counterterm \eqref{L_ct black holes} is finite, but it is zero. This suggests that a detailed analysis on the renormalisation scheme is needed. {We should propose a set of boundary conditions for the fields in the asymptotic region such that our black hole configuration belongs to this family. Then we must ensure that the variation of the fields in this family leads to an extremum of the action principle. However, even imposing the above condition and having a finite on-shell action, there is still an ambiguity in the holographic renormalisation procedure reflected in the fact that one is able to include local and finite counterterms to the renormalised action, see \cite{Anabalon:2023kcp},\cite{Caceres:2023gfa} for a recent discussion. These counter-terms modify the renormalised on-shell action and hence they must have an effect on the dual-field theory. A careful analysis along these lines could point out the correct counterterms which lead to charges linked to standard thermodynamics interpretation and a well posed on-shell action.}

\subsection{Meron Black Membrane}\label{meronBH}

As before, here we study the black hole solution obtained by performing a double Wick rotation of the configuration \eqref{background28}. For this we take
    \begin{equation}
        t \rightarrow i x^{3}, \quad \mu \rightarrow i\tau,
    \end{equation}

which, after moving to Einstein frame, leads to
    \begin{equation}\label{MeronBH}
    \begin{aligned}
        ds^2_{E}  &= \sqrt{\frac{Nr}{4}}\left[ -\left(1-\frac{m}{r^{2}}\right) d\tau^{2} + \sum^{3}_{i=1} dx^{2}_{i} +
                        \frac{2dr^{2}}{r^{2}\left( 1-\frac{m}{r^{2}}\right)}
                        +d\vartheta ^{2}+\sin ^{2}\vartheta d\varphi^{2} +\sum_{i}\left( {\Theta}^{(i)}\right) ^{2}\right]\\
        F_{3} &= 2 N \Vol(S^{3}) + \frac{N}{4} d\left[\omega_1\wedge A^{(1)}  +\omega_2\wedge A^{(2)} +\omega_3\wedge A^{(3)}\right], \\
    \Phi &=\log \left( \frac{4}{N} r\right).
    \end{aligned}
    \end{equation}
As in the cigar case, SUSY is not preserved even when $m=0$, hence, this solution is always non-SUSY.

In this case, since the configuration is static, the only non-trivial black-membrane charges are the ones associated with time translations, namely, the energy and the entropy. Defining the Killing vector $\mathbf{\xi} = \partial_{\tau}$, and noting that $C_{\mu\nu}\xi^{\mu}=0$, we see that  only the Einstein-Hilbert action has a non-trivial contribution to the Noether prepotential. However, there is a non-trivial contribution of the Dilaton in boundary term to obtain a finite mass.  
    \begin{equation}
    \begin{aligned}
        M &= \frac{N^{2}\pi^{2}}{2\sqrt{2} G_{N}} L_{x^{1}}L_{x^{2}}L_{x^{3}}m ,\\
        S &= \frac{N^{2}\pi^{3}}{G_{N}} L_{x^{1}}L_{x^{2}}L_{x^{3}}m ,\\
        T &= \frac{1}{2\sqrt{2}\pi}
    \end{aligned}
    \end{equation}

These quantities satisfy the first law of thermodynamics.

\section{Conclusions}\label{sectionconclusions}

To close the paper, this section starts with a summary of the material in this work. Then we propose some lines of research for future study.

We write four new backgrounds of type IIB, all describing D5 branes. The background in Section \ref{sectionBIxx} is SUSY for some choice of parameters. It describes a QFT in $(4+1)$ dimensions, that is confining, gapped and has a spectrum of spin-two glueballs that is either continuous or first discrete and then continuos (like the hydrogen atom) as the energy increases.

The background in Section \ref{section2+1qft} is dual to a $(2+1)$ dimensional QFT, obtained by compactifying the theory on D5 branes on $S^2\times S^1$, being the size of $S^2$ fixed, whilst the size of $S^1$ vanishes close to the end of the space. Whilst Wilson loops suggest a confining behaviour at low energies, other observables display a non-field theoretic behaviour.

Section \ref{sectioncomparison} is a comparative study between solutions already present in the bibliography and the solution of Sections \ref{sectionBIxx} and  \ref{section2+1qft}. The size of the two-sphere, either shrinking or stabilised towards the end of the space is responsible for the behaviour of some observables. Also we compared SUSY preservation via topological twist with the insertion of the Wilson line presented in Section \ref{sectionBIxx}: the key similarities being the non-vanishing of the fibre at infinity, and the radially dependent part  making the IR regular. Finally, Section \ref{bhsectionsx} presents two asymptotically locally flat black hole backgrounds supported by a logarithmic dilaton and a non-trivial RR $F_3$ field. We considered a counterterm depending on the value of the dilaton field at the boundary leading finite charges computed with the Noether-Wald method and finite but vanishing on-shell action. Thermodynamic quantities are calculated.

In the future, it would be interesting to study the following topics:
\begin{itemize}
\item{The $SU(2)$ G-structure of the background of Section  \ref{sectionBIxx}. This may lead to a clearer picture on the mechanism of SUSY preservation. }

\item{More in depth study of spin two glueball like excitations. What happens when allowing excitations on $S^3$. Does the continuum part of the spectrum exist  for non-s-waves.}

\item{Some of the non-field theoretic behaviour of the models here presented is due to the high energy (large $r$) being controlled by a LST. It would be interesting to generalise these types of solutions to other Dp branes. In particular D4 of D3 branes would lead to interesting new geometries dual to lower dimensional field theories.}

\item{On this line, it would be of interest to find solutions representing compactification on circles, that have a large $r$ behaviour asymptotic to AdS-space.}

\item{Some of the solutions are constructed using a meron gauge field. It would be interesting to probe these backgrounds with two-D5 branes. This leads to an $SU(2)$ charged scalar in the presence of a meron. This might lead to fermionic excitations constructed in terms of bosons. The QFT dual to this effect (Jackiw, Rebbi, Hasenfratz, 't Hooft \cite{Jackiw:1976xx}) would be interesting to understand.}
\item{Further analysis on the renormalisation scheme for asymptotically locally flat configuration with non-trivial matter fields is required in order to construct a consistent and non-trivial on-shell action.}
\item{For the SUSY solution in Section \ref{sectionBIxx}, various supersymmetric observables can be computed. It would be interesting to study the (in)dependence of these observables on the size of the $S^1_\varphi$ that shrinks. In the present case, techniques like those developed in \cite{Bobev:2019bvq} may become relevant. }
\end{itemize}
We hope to report on these and other issues in the near future.
\newpage
\section*{Acknowledgments}
We would like to thank: Andr\'es Anabal\'on, Riccardo Argurio, Adi Armoni, Marcelo Rezende Barbosa, Nikolay Bobev, Fabrizio Canfora, Yolanda Lozano, Niall Macpherson, Horatiu Nastase, Julio Oliva, Alfonso V. Ramallo, Diego Rodriguez-Gomez, Simon Ross and Alessandro Tomasiello for various discussions and for sharing with us their ideas and knowledge. This work is supported by STFC grants ST/X000648/1 and ST/T000813/1. The work of M.O. is partially funded by Beca ANID de Doctorado 21222264. The work of R.S. is supported by STFC grant ST/W507878/1.
\\
\\
C.N.  would like to add few words: Fidel A. Schaposnik passed away on July 29th 2023. Fidel taught me very many interesting things (bosonisation, topological defects, WZW, TQFTs, Path Integral, etc). Not very successfully, he tried to teach me how to be a good professional physicist. He would have liked to like some of my works. This paper is dedicated to the memory of Fidel A. Schaposnik. Hoping that when Laura and Fidelito read this, remember their father and smile.

\appendix
\section{Type IIB: a summary of equations and conventions}\label{TypeIIBapp}

In this work we consider background in the Metric-Dilaton$-H_{3}-F_{3}$
sector of type IIB. The action is
    \begin{equation}
        S_{IIB}=\frac{1}{2\kappa ^{2}}\int d^{10}x\sqrt{-g}\left[ 
        e^{-2\Phi }\left( R + 4\partial_{\mu}\Phi\partial^{\mu}\Phi 
        -\frac{1}{2}|H_{3}|^{2}\right) 
        -\frac{1}{2}|F_{3}|^{2} \right]
    \end{equation}
with equations
\begin{eqnarray}
\nabla ^{2}\Phi -\partial _{\mu }\Phi \partial ^{\mu }\Phi +\frac{1}{4}R-%
\frac{1}{8}\left\vert H_{3}\right\vert ^{2} &=&0\ , \\
d\left( e^{-2\Phi }H_{3}\right) &=&0\ , \\
d\star F_{3} &=&0\ , \\
R_{\mu \nu }+2\nabla _{\mu }\nabla _{\nu }\Phi -\frac{1}{2}\left\vert
H_{3}\right\vert _{\mu \nu }^{2}-\frac{e^{2\Phi }}{2}\left( \left\vert
F_{3}\right\vert _{\mu \nu }^{2}-\frac{1}{2}g_{\mu \nu }\left\vert
F_{3}\right\vert ^{2}\right) &=&0\ ,
\end{eqnarray}
We have denoted 
\begin{equation}
\left\vert F_{p}\right\vert ^{2}=\frac{1}{p!}F_{\mu _{1}\dots \mu
_{p}}F^{\mu _{1}\dots \mu _{p}}\ ,\qquad \left\vert F_{p}\right\vert _{\mu
\nu }^{2}=\frac{1}{\left( p-1\right) !}F_{\mu \sigma _{1}\dots \sigma
_{p-1}}F_{\nu }^{\ \ \sigma _{1}\dots \sigma _{p-1}}\ .
\end{equation}

The SUSY transformations for fermions are%
\begin{eqnarray}
\delta \lambda &=&\frac{1}{2}\left( \not{\partial}\Phi +\frac{1}{2}\not%
{H}_{3}\sigma _{3}-\frac{1}{2}e^{\Phi }\not{F}_{3}\sigma _{1}\right)
\epsilon \ ,\nonumber \\
\delta \Psi _{\mu }dx^{\mu } &=&d\epsilon +\frac{1}{4}\left( \omega
_{ab}\Gamma ^{ab}+\frac{1}{2}H_{\mu ab}dx^{\mu }\Gamma ^{ab}\sigma _{3}+%
\frac{e^{\Phi }}{2}\not{F}_{3}\sigma _{1}\Gamma _{\mu }dx^{\mu }\right)
\epsilon\label{susyIIB}
\end{eqnarray}%
where%
\[
\not{F}_{p}=\frac{1}{p!}F_{a_{1}\dots a_{p}}\Gamma ^{a_{1}\dots a_{p}}\ , 
\]%
for any p-form. Also we use the fact that the killing spinor in type IIB can be
written as
    \begin{equation}
    \epsilon =\left( 
        \begin{array}{c}
        \epsilon _{1} \\
        \epsilon _{2}
        \end{array}
        \right) \ ,
    \end{equation}
    
where $\epsilon _{1}$ and $\epsilon _{2}$ are 32-components spinors.

\section{SUSY of the background in eq.(\ref{BI})}\label{BIsusy}

In what follows we use the following representation of the Dirac matrices
    \begin{equation}
    \begin{aligned}
        \Gamma^{0} &= i \sigma^{2} \otimes \sigma^{3} \otimes \sigma^{3} \otimes \sigma^{3} \otimes \sigma^{3}, \\
        \Gamma^{1} &= \sigma^{1} \otimes \sigma^{3} \otimes \sigma^{3} \otimes \sigma^{3} \otimes \sigma^{3}, \\
        \Gamma^{2} &= -\Id_{2} \otimes \sigma^{1} \otimes \sigma^{3} \otimes \sigma^{3} \otimes \sigma^{3}, \\
        \Gamma^{3} &= -\Id_{2} \otimes \sigma^{2} \otimes \sigma^{3} \otimes \sigma^{3} \otimes \sigma^{3}, \\
        \Gamma^{4} &= \Id_{2} \otimes \Id_{2} \otimes \sigma^{1} \otimes \sigma^{3} \otimes \sigma^{3}, \\
        \Gamma^{5} &= \Id_{2} \otimes \Id_{2} \otimes \sigma^{2} \otimes \sigma^{3} \otimes \sigma^{3}, \\
        \Gamma^{6} &= -\Id_{2} \otimes \Id_{2} \otimes \Id_{2} \otimes \sigma^{1} \otimes \sigma^{3}, \\
        \Gamma^{7} &= -\Id_{2} \otimes \Id_{2} \otimes \Id_{2} \otimes \sigma^{2} \otimes \sigma^{3}, \\
        \Gamma^{8} &= \Id_{2} \otimes \Id_{2} \otimes \Id_{2} \otimes \Id_{2} \otimes \sigma^{1}, \\
        \Gamma^{9} &= \Id_{2} \otimes \Id_{2} \otimes \Id_{2} \otimes \Id_{2} \otimes \sigma^{2},
    \end{aligned}
    \end{equation}

where $(\sigma^{1},\sigma^{2},\sigma^{3})$ are Pauli matrices and $\Id_{2}$ is the $2\times2$ identity matrix.

\subsection{SUSY on the NS5-Brane and anti-periodic Boundary Conditions}\label{AppendixSpinor}

Let us first quickly review how SUSY in the pure NS5-brane background, that is 
    \begin{equation}
    \begin{aligned}
        ds^{2}_{st} &=  dx_{1,5}^2  + \frac{Ndr^{2}}{ r^{2}} 
            + \frac{N}{4}\left(\omega^{2}_{1} + \omega^{2}_{2} + \omega_{3}^{2}\right),\\
        H_{3} &= 2N \Vol(S^{3}),\\
        \Phi &= -\log(r),\\
    \end{aligned}
    \end{equation}    

recall that
    \begin{equation}
        \omega^{2}_{1} + \omega^{2}_{2} = d\theta^{2} + \sin^{2}(\theta)d\phi^{2}, \quad
        \omega_{3} = d\psi + \cos(\theta)d\phi.
    \end{equation}

using the following choice of vielbeins 
    \begin{equation}
    \begin{aligned}
        & e^{0} = dt, \quad 
        e^{m} = dx^{m}, \quad 
        e^{6} = \frac{\sqrt{N}}{r\sqrt{f_{s}(r)}}dr, \\
        &e^{7} = \frac{\sqrt{N}}{2} d\theta, \quad
        e^{8} = \frac{\sqrt{N}}{2} \sin(\theta)d\phi, \quad 
        e^{9} = \frac{\sqrt{N}}{2} \omega_{3}
    \end{aligned}
    \end{equation}

where $m=1,...,5$, we find the SUSY spinor to be
    \begin{equation}
        \epsilon_{1} = \begin{pmatrix}
                        c_{1}\, e^{i \frac{\psi}{2}} + c_{2}\, e^{-i \frac{\psi}{2}}\\
                        0 \\
                        0\\
                        c_{1}\, e^{i \frac{\psi}{2}} - c_{2}\, e^{-i \frac{\psi}{2}}\\
                        0\\
                        0\\
                        0\\
                        0\\
                        0\\
                        0\\
                        0\\
                        0\\
                        c_{3}\, e^{i \frac{\psi}{2}} + c_{4}\, e^{-i \frac{\psi}{2}}\\
                        0\\
                        0\\
                        c_{3}\, e^{i \frac{\psi}{2}} - c_{4}\, e^{-i \frac{\psi}{2}}\\
                        0\\
                        0\\
                        0\\
                        0\\
                        c_{5}\, e^{i \frac{\psi}{2}} + c_{6}\, e^{-i \frac{\psi}{2}}\\
                        0\\
                        0\\
                        c_{5}\, e^{i \frac{\psi}{2}} - c_{6}\, e^{-i \frac{\psi}{2}}\\
                        c_{7}\, e^{i \frac{\psi}{2}} + c_{8}\, e^{-i \frac{\psi}{2}}\\
                        0\\
                        0\\
                        c_{7}\, e^{i \frac{\psi}{2}} - c_{8}\, e^{-i \frac{\psi}{2}}\\
                        0\\
                        0\\
                        0\\
                        0
                        \end{pmatrix}, \quad
        \epsilon_{2} = \begin{pmatrix}
                        0\\
                        0\\
                        0\\
                        0\\
                        0\\
                        c_{9}\, e^{\frac{i}{2}(\theta+\phi)} + c_{10}\, e^{\frac{i}{2}(\theta-\phi)}\\
                        c_{9}\, e^{\frac{i}{2}(-\theta+\phi)} - c_{10}\, e^{\frac{i}{2}(-\theta-\phi)}\\
                        0\\
                        0\\
                        c_{11}\, e^{\frac{i}{2}(\theta+\phi)} + c_{12}\, e^{\frac{i}{2}(\theta-\phi)}\\
                        c_{11}\, e^{\frac{i}{2}(-\theta+\phi)} - c_{12}\, e^{\frac{i}{2}(-\theta-\phi)}\\
                        0\\
                        0\\
                        0\\
                        0\\
                        0\\
                        0\\
                        c_{13}\, e^{\frac{i}{2}(\theta+\phi)} + c_{14}\, e^{\frac{i}{2}(\theta-\phi)}\\
                        c_{13}\, e^{\frac{i}{2}(-\theta+\phi)} - c_{14}\, e^{\frac{i}{2}(-\theta-\phi)}\\
                        0\\
                        0\\
                        0\\
                        0\\
                        0\\
                        0\\
                        0\\
                        0\\
                        0\\
                        0\\
                        c_{15}\, e^{\frac{i}{2}(\theta+\phi)} + c_{16}\, e^{\frac{i}{2}(\theta-\phi)}\\
                        c_{15}\, e^{\frac{i}{2}(-\theta+\phi)} - c_{16}\, e^{\frac{i}{2}(-\theta-\phi)}\\
                        0
                        \end{pmatrix},
    \end{equation}

which has 16 independent solutions, as expected. 

If we take one of the field theory directions to be compact $x^{5}\rightarrow \bvarphi$ (with $\bvarphi\sim \bvarphi+2\pi$) and take anti-periodic boundary conditions for the spinor in $\bvarphi$, we see that none of the spinor components satisfy this boundary conditions, so that SUSY is completely broken.

There is a way of preserving some amount of SUSY: in the above parametrisation of the 3-sphere there is an explicit $U(1)$ isometry along $\psi$. Replacing
    \begin{equation}
        \psi \rightarrow \psi + \alpha \, \bvarphi,
    \end{equation}

then the spinor components that include $\psi$ will be shifted to $e^{\pm i \frac{\psi}{2}} \rightarrow e^{\pm \frac{i}{2}(\psi + \alpha\, \bvarphi)}$. Note that this is not a coordinate transformation: we are allowing $\alpha$ to take any possible values, not necessarily the one that respects the periodicity of $\psi$. This "shift" by $\bvarphi$ corresponds to the insertion of a Wilson Loop in the boundary theory.

Imposing anti-periodic boundary conditions in $\bvarphi$ fixes $\alpha = \pm 1$. After boundary conditions are imposed the spinor reads
    \begin{equation}\label{SpinorWilson1}
        \epsilon_{1} = \begin{pmatrix}
                        c_{1}\, e^{ \frac{i}{2}(\psi \pm \bvarphi)} + c_{2}\, e^{ -\frac{i}{2}(\psi \pm \bvarphi)}\\
                        0 \\
                        0\\
                        c_{1}\, e^{ \frac{i}{2}(\psi \pm \bvarphi)} - c_{2}\,e^{ -\frac{i}{2}(\psi \pm \bvarphi)}\\
                        0\\
                        0\\
                        0\\
                        0\\
                        0\\
                        0\\
                        0\\
                        0\\
                        c_{3}\, e^{ \frac{i}{2}(\psi \pm \bvarphi)} + c_{4}\,e^{ -\frac{i}{2}(\psi \pm \bvarphi)}\\
                        0\\
                        0\\
                        c_{3}\, e^{ \frac{i}{2}(\psi \pm \bvarphi)} - c_{4}\, e^{ -\frac{i}{2}(\psi \pm \bvarphi)}\\
                        0\\
                        0\\
                        0\\
                        0\\
                        c_{5}\, e^{ \frac{i}{2}(\psi \pm \bvarphi)} + c_{6}\, e^{ -\frac{i}{2}(\psi \pm \bvarphi)}\\
                        0\\
                        0\\
                        c_{5}\, e^{ \frac{i}{2}(\psi \pm \bvarphi)} - c_{6}\, e^{ -\frac{i}{2}(\psi \pm \bvarphi)}\\
                        c_{7}\, e^{ \frac{i}{2}(\psi \pm \bvarphi)} + c_{8}\, e^{ -\frac{i}{2}(\psi \pm \bvarphi)}\\
                        0\\
                        0\\
                        c_{7}\, e^{ \frac{i}{2}(\psi \pm \bvarphi)} - c_{8}\, e^{ -\frac{i}{2}(\psi \pm \bvarphi)}\\
                        0\\
                        0\\
                        0\\
                        0
                        \end{pmatrix}, \quad
        \epsilon_{2} = 0.
    \end{equation}

In this way, by charging the spinor under the compact field theory directions, we manage to preserve half of the supersymmetries. SUSY in the case where the coordinate $\bvarphi$ shrinks to zero works in the exact same way. We present this now

\subsection{SUSY on 4+1}

We now move to the study of the SUSY variations on \eqref{BI}. As before we work on the S-dual frame
    \begin{equation}
    \begin{aligned}
        ds^{2}_{st} &=  dx_{1,4}^2 + f_{s}(r)d\varphi^{2} 
            + \frac{Ndr^{2}}{ r^{2}f_{s}(r)} 
            + \frac{N}{4}\left(\omega^{2}_{1} + \omega^{2}_{2} + 
        \left(\omega_{3} - \sqrt{\frac{8}{N}} Q \zeta(r) d\varphi\right)^{2}\right),\\
        H_{3} &= 2N \Vol(S^{3}) + \sqrt{\frac{N}{2}}Q d\left( \zeta(r) \omega_{3}\wedge d\varphi \right),\\
        \Phi &= -\log(r),\\
    \end{aligned}
    \end{equation}

We choose the vielbein basis
    \begin{equation}
    \begin{aligned}
        & e^{0} = dt, \quad 
        e^{l} = dx^{l}, \quad 
        e^{5} = \sqrt{f_{s}(r)} d\varphi,\quad 
        e^{6} = \frac{\sqrt{N}}{r\sqrt{f_{s}(r)}}dr, \\
        &e^{7} = \frac{\sqrt{N}}{2} d\theta, \quad
        e^{8} = \frac{\sqrt{N}}{2} \sin(\theta)d\phi, \quad 
        e^{9} = \frac{\sqrt{N}}{2} \left(\omega_{3} - \sqrt{\frac{8}{N}} Q \zeta(r) d\varphi\right),
    \end{aligned}
    \end{equation}

where now $l=1,...,4$. The background in eq.(\ref{BI}) is BPS when $m=0$. The most general solution of the Killing spinor equation is
    \begin{equation}\label{spinor1}
           \epsilon_{1} = \frac{\Omega(r)}{r}  
                    \begin{pmatrix}
                        \epsilon^{(1)}_{1}\\
                        \epsilon^{(2)}_{1}\\
                        \epsilon^{(3)}_{1}\\
                        \epsilon^{(4)}_{1}
                    \end{pmatrix}, \quad 
            \epsilon_{2}=\vec{0}\ ,
    \end{equation}

where $\epsilon^{(i)}_{1}$ are 8 component spinors given by
    \begin{equation}
        \epsilon^{(1)}_{1} = 
                \begin{pmatrix}
                      c_{1}\, e^{ \frac{i}{2}\left(\psi + \frac{2\sqrt{2}Q}{\sqrt{N}r^{2}_{+}}\varphi \right)} + c_{2}\, e^{ -\frac{i}{2}\left(\psi + \frac{2\sqrt{2}Q}{\sqrt{N}r^{2}_{+}}\varphi \right)}\\
                        0 \\
                        0\\
                        c_{1}\, e^{ \frac{i}{2}\left(\psi + \frac{2\sqrt{2}Q}{\sqrt{N}r^{2}_{+}}\varphi \right)} - c_{2}\, e^{ -\frac{i}{2}\left(\psi + \frac{2\sqrt{2}Q}{\sqrt{N}r^{2}_{+}}\varphi \right)}\\
                        0\\ 
                        -\frac{r^{2}}{\sqrt{2}Q}\left(1+ \sqrt{1-\frac{2Q^{2}}{r^{4}}}\right) \left(c_{1}\, e^{ \frac{i}{2}\left(\psi + \frac{2\sqrt{2}Q}{\sqrt{N}r^{2}_{+}}\varphi \right)} + c_{2}\, e^{ -\frac{i}{2}\left(\psi + \frac{2\sqrt{2}Q}{\sqrt{N}r^{2}_{+}}\varphi \right)}\right)\\
                        \frac{r^{2}}{\sqrt{2}Q}\left(1+ \sqrt{1-\frac{2Q^{2}}{r^{4}}}\right) \left(c_{1}\, e^{ \frac{i}{2}\left(\psi + \frac{2\sqrt{2}Q}{\sqrt{N}r^{2}_{+}}\varphi \right)} - c_{2}\, e^{ -\frac{i}{2}\left(\psi + \frac{2\sqrt{2}Q}{\sqrt{N}r^{2}_{+}}\varphi \right)}\right)\\
                        0
                \end{pmatrix}
    \end{equation}

    \begin{equation}
        \epsilon^{(2)}_{1} = 
                \begin{pmatrix}
                        0\\
                        \frac{1}{\Omega^{2}(r)}\left( c_{3}\, e^{ \frac{i}{2}\left(\psi + \frac{2\sqrt{2}Q}{\sqrt{N}r^{2}_{+}}\varphi \right)} + c_{4}\, e^{ -\frac{i}{2}\left(\psi + \frac{2\sqrt{2}Q}{\sqrt{N}r^{2}_{+}}\varphi \right)} \right)\\
                        \frac{1}{\Omega^{2}(r)}\left( c_{3}\, e^{ \frac{i}{2}\left(\psi + \frac{2\sqrt{2}Q}{\sqrt{N}r^{2}_{+}}\varphi \right)} - c_{4}\, e^{ -\frac{i}{2}\left(\psi + \frac{2\sqrt{2}Q}{\sqrt{N}r^{2}_{+}}\varphi \right)} \right)\\
                        0\\
                        -\frac{r^{2}}{\sqrt{2} Q\Omega^{2}(r)}\left(-1+ \sqrt{1-\frac{2Q^{2}}{r^{4}}}\right)\left( c_{3}\, e^{ \frac{i}{2}\left(\psi + \frac{2\sqrt{2}Q}{\sqrt{N}r^{2}_{+}}\varphi \right)} + c_{4}\, e^{ -\frac{i}{2}\left(\psi + \frac{2\sqrt{2}Q}{\sqrt{N}r^{2}_{+}}\varphi \right)} \right)\\
                        0\\
                        0\\
                        \frac{r^{2}}{\sqrt{2} Q\Omega^{2}(r)}\left(-1+ \sqrt{1-\frac{2Q^{2}}{r^{4}}}\right)\left( c_{3}\, e^{ \frac{i}{2}\left(\psi + \frac{2\sqrt{2}Q}{\sqrt{N}r^{2}_{+}}\varphi \right)} - c_{4}\, e^{ -\frac{i}{2}\left(\psi + \frac{2\sqrt{2}Q}{\sqrt{N}r^{2}_{+}}\varphi \right)} \right)
                \end{pmatrix}
    \end{equation}

    \begin{equation}
        \epsilon^{(3)}_{1} = 
                \begin{pmatrix}
                        0\\
                        \frac{1}{\Omega^{2}(r)}\left( c_{5}\, e^{ \frac{i}{2}\left(\psi + \frac{2\sqrt{2}Q}{\sqrt{N}r^{2}_{+}}\varphi \right)} + c_{6}\, e^{ -\frac{i}{2}\left(\psi + \frac{2\sqrt{2}Q}{\sqrt{N}r^{2}_{+}}\varphi \right)} \right)\\
                        \frac{1}{\Omega^{2}(r)}\left( c_{5}\, e^{ \frac{i}{2}\left(\psi + \frac{2\sqrt{2}Q}{\sqrt{N}r^{2}_{+}}\varphi \right)} - c_{6}\, e^{ -\frac{i}{2}\left(\psi + \frac{2\sqrt{2}Q}{\sqrt{N}r^{2}_{+}}\varphi \right)} \right)\\
                        0\\
                        -\frac{r^{2}}{\sqrt{2} Q\Omega^{2}(r)}\left(-1+ \sqrt{1-\frac{2Q^{2}}{r^{4}}}\right)\left( c_{5}\, e^{ \frac{i}{2}\left(\psi + \frac{2\sqrt{2}Q}{\sqrt{N}r^{2}_{+}}\varphi \right)} + c_{6}\, e^{ -\frac{i}{2}\left(\psi + \frac{2\sqrt{2}Q}{\sqrt{N}r^{2}_{+}}\varphi \right)} \right)\\
                        0\\
                        0\\
                        \frac{r^{2}}{\sqrt{2} Q\Omega^{2}(r)}\left(-1+ \sqrt{1-\frac{2Q^{2}}{r^{4}}}\right)\left( c_{5}\, e^{ \frac{i}{2}\left(\psi + \frac{2\sqrt{2}Q}{\sqrt{N}r^{2}_{+}}\varphi \right)} - c_{6}\, e^{ -\frac{i}{2}\left(\psi + \frac{2\sqrt{2}Q}{\sqrt{N}r^{2}_{+}}\varphi \right)} \right)
                \end{pmatrix}
    \end{equation}

    \begin{equation}
        \epsilon^{(4)}_{1} = 
                \begin{pmatrix}
                        c_{7}\, e^{ \frac{i}{2}\left(\psi + \frac{2\sqrt{2}Q}{\sqrt{N}r^{2}_{+}}\varphi \right)} + c_{8}\, e^{ -\frac{i}{2}\left(\psi + \frac{2\sqrt{2}Q}{\sqrt{N}r^{2}_{+}}\varphi \right)}\\
                        0 \\
                        0\\
                        c_{7}\, e^{ \frac{i}{2}\left(\psi + \frac{2\sqrt{2}Q}{\sqrt{N}r^{2}_{+}}\varphi \right)} - c_{8}\, e^{ -\frac{i}{2}\left(\psi + \frac{2\sqrt{2}Q}{\sqrt{N}r^{2}_{+}}\varphi \right)}\\
                        0\\ 
                        -\frac{r^{2}}{\sqrt{2}Q}\left(1+ \sqrt{1-\frac{2Q^{2}}{r^{4}}}\right) \left(c_{7}\, e^{ \frac{i}{2}\left(\psi + \frac{2\sqrt{2}Q}{\sqrt{N}r^{2}_{+}}\varphi \right)} + c_{8}\, e^{ -\frac{i}{2}\left(\psi + \frac{2\sqrt{2}Q}{\sqrt{N}r^{2}_{+}}\varphi \right)}\right)\\
                        \frac{r^{2}}{\sqrt{2}Q}\left(1+ \sqrt{1-\frac{2Q^{2}}{r^{4}}}\right) \left(c_{7}\, e^{ \frac{i}{2}\left(\psi + \frac{2\sqrt{2}Q}{\sqrt{N}r^{2}_{+}}\varphi \right)} - c_{8}\, e^{ -\frac{i}{2}\left(\psi + \frac{2\sqrt{2}Q}{\sqrt{N}r^{2}_{+}}\varphi \right)}\right)\\
                        0
                \end{pmatrix}
    \end{equation}

where $c_{i}$ are integration constants and
    \begin{equation}
        \Omega(r) = \left(\frac{ 1-\sqrt{1-\frac{2Q^{2} }{ r^{4} } } }{1+\sqrt{1-\frac{ 2Q^{2} }{r^{4}}}}\right)^{\frac{1}{4}}.
    \end{equation}

In order to see the connection with the spinor in the previous section, first recall that when $m=0$ we have $r^{2}_{+} = \sqrt{2}|Q|$, so that
    \begin{equation}
        \frac{i}{2}\frac{2\sqrt{2}Q}{\sqrt{N}r^{2}_{+}}\varphi 
            = i\text{sign}(Q)\frac{\varphi}{\sqrt{N}},
    \end{equation}

also recall that $\varphi \sim \varphi \sqrt{N}\pi$. Defining $\bvarphi = \frac{2}{\sqrt{N}}\varphi$, we recover what we had in the next section, that is, the dependence of the spinor on $\bvarphi$ is of the form 
    \begin{equation}
        \epsilon \sim e^{\frac{i}{2}(\psi \pm \bvarphi)},
    \end{equation}

so that the spinor is anti-periodic in $\bvarphi$.

\section{Spin-two fluctuations. Detailed derivations.}\label{appendixglue}

In this appendix we review some general formulas for spin-2 fluctuations, previously obtained in  \cite{Bachas:2011xa}. We adapt those result to have the macroscopic space to be $(p+1)$-dimensional.

The equations of motion of Type IIB Supergravity in Einstein frame, only with $F_{3}$ flux are given by
    \begin{align}
        &R_{\mu\nu} = \frac{1}{2}\partial_{\mu}\Phi\partial_{\nu}\Phi 
            +\frac{g_{s}e^{\Phi}}{4}\left( F_{\mu\lambda\rho}F_{\nu}^{\phantom{\nu}\lambda\rho} 
            -\frac{1}{12} g_{\mu\nu} F_{\lambda\rho\sigma}F^{\lambda\rho\sigma}\right), \\
        &\nabla^{2}\Phi = \frac{g_{s}e^{\Phi}}{12}F_{\mu\nu\lambda}F^{\mu\nu\lambda},\\
        &\partial_{\mu}\left(\sqrt{g} e^{\Phi}F^{\mu\nu\lambda}\right)=0, \\
        &\partial_{[\mu}F_{\nu\lambda\rho]}=0 
    \end{align}

In what follows we will consider a spacetime of the following form
    \begin{equation}\label{AnsatzMetric}
    \begin{aligned}
        ds^{2}_{10D} &= e^{2A(r)}ds^{2}(\mathcal{M}_{p+1})  + g_{ij}(y^{k})dy^{i}dy^{k},\\
        F_{3} &= \frac{1}{3!}f_{ijk}(y)dy^{i} \wedge  dy^{j} \wedge  dy^{k},\\ 
        \Phi &= \Phi(y)
    \end{aligned}
    \end{equation}

where $\mathcal{M}_{p+1}$ is a space of constant curvature. We will split the 10D indexes in the following way
    \begin{equation}
    \begin{aligned}
        a,b &= 0,1,..,p \\
        i,j &= y^{1},...,y^{9-p}.
    \end{aligned}
    \end{equation} 

We consider metric fluctuations of the form
    \begin{equation}\label{pertAppendix}
        \delta g_{\mu\nu} = e^{2A} \bar{h}_{\mu\nu}, \quad 
        \bar{h}_{\mu\nu} = \begin{pmatrix}
                        h_{ab}(x)\psi(y)& 0 \\
                         0 & 0
                        \end{pmatrix},
    \end{equation}

i.e. metric fluctuations parallel to the macroscopic space $\mathcal{M}$. We work in the transverse-traceless gauge
    \begin{equation}\label{GaugeCond}
        \Trh{a} = 0, \quad \nabla_{a}h^{a}_{\phantom{a}b}=0.
    \end{equation}
Note that this conditions are expressed in the macroscopic space, but due to the form of the ansatz, they also can be written as a 10D condition
    \begin{equation}
        h^{\mu}_{\phantom{\mu}\mu} = 0, \quad \nabla_{\mu}h^{\mu}_{\phantom{\mu}\nu}=0.
    \end{equation}

As showed in \cite{Bachas:2011xa}, with this choice of fluctuation, it is consistent to take
    \begin{equation}
        \delta F_{\mu\nu\lambda}=0,\quad \delta \Phi  =0 
    \end{equation}

since any fluctuation of those field will not source the fluctuation \eqref{pertAppendix}. 

First, we study the variation of the energy-stress tensor. For the configurations of interest the later is given by
    \begin{equation}
        T_{\mu\nu} = \frac{1}{2}\left( \partial_{\mu}\Phi\partial_{\nu}\Phi - \frac{1}{2}g_{\mu\nu}(\partial\Phi)^{2} \right)
        + \frac{g_{s}e^{\Phi}}{4}\left( F_{\mu\lambda\rho}F_{\nu}^{\phantom{\nu}\lambda\rho} 
            -\frac{1}{3!} g_{\mu\nu} F_{\lambda\rho\sigma}F^{\lambda\rho\sigma} \right).
    \end{equation}

Within the configurations of the form \eqref{AnsatzMetric}
    \begin{equation}
        T^{c}_{\phantom{c}c} = (p+1)\left( \frac{1}{4} (\partial\Phi)^{2} 
                        + \frac{g_{s}e^{\Phi}}{24} F_{\lambda\rho\sigma}F^{\lambda\rho\sigma} \right),
    \end{equation}

so that with perturbations of the form \eqref{pertAppendix} we have
    \begin{equation}\label{Tmunupert}
        \delta T_{\mu\nu} = \frac{1}{p+1}  T^{c}_{\phantom{c}c} \delta g_{\mu\nu}.
    \end{equation}

For $(\mu,\nu)=(a,b)$ Einstein's equations give
    \begin{equation}
        R_{ab} - \frac{1}{2}g_{ab}R = T_{ab}
    \end{equation}

and tracing over $a,b$ we get
    \begin{equation}
        R^{a}_{\phantom{a}a} - \frac{(p+1)}{2} R = T^{a}_{\phantom{a}a}
    \end{equation}    

From where we can write the $(\mu,\nu)=(a,b)$ components of \eqref{Tmunupert} as
    \begin{equation}
        \delta T_{ab} = \frac{1}{p+1} R^{c}_{\phantom{c}c}\delta g_{ab} - \frac{1}{2} R\delta g_{ab}
    \end{equation}

We now turn to the variation of Einstein's equations. From the Ricci tensor variation we have
    \begin{equation}
        \delta R_{\mu\nu} = \frac{1}{2} \left( \nabla_{\lambda}\nabla_{\mu}\dg^{\lambda}_{\phantom{\lambda}\nu}
        + \nabla_{\lambda}\nabla_{\nu}\dg^{\lambda}_{\phantom{\lambda}\mu}
        - \nabla^{2}\dg_{\mu\nu}\right)
    \end{equation}

Note that since $\dg_{\mu\nu}$ is transverse 
    \begin{equation}
         \left[ \nabla_{\lambda},\nabla_{\mu} \right] \dg^{\lambda}_{\phantom{\lambda}\nu} 
        = \RUddd{\lambda}{\sigma}{\lambda}{\mu}\dg^{\sigma}_{\phantom{\sigma}\nu} 
          -\RUddd{\sigma}{\nu}{\lambda}{\mu} \dg^{\lambda}_{\phantom{\lambda}\sigma}
    \end{equation}

is equivalent to
    \begin{equation}
           \nabla_{\lambda}\nabla_{\mu}\dg^{\lambda}_{\phantom{\lambda}\nu} 
        = R_{\sigma\mu}\dg^{\sigma}_{\phantom{\sigma}\nu} 
          -\RUddd{\sigma}{\nu}{\lambda}{\mu} \dg^{\lambda}_{\phantom{\lambda}\sigma}.
    \end{equation}

Then we have
    \begin{equation}\label{deltaR}
        \delta R_{\mu\nu} = \frac{1}{2} \left(R_{\sigma\mu}\dg^{\sigma}_{\phantom{\sigma}\nu} 
          -\RUddd{\sigma}{\nu}{\lambda}{\mu} \dg^{\lambda}_{\phantom{\lambda}\sigma}
          +R_{\sigma\nu}\dg^{\sigma}_{\phantom{\sigma}\mu} 
          -\RUddd{\sigma}{\mu}{\lambda}{\nu} \dg^{\lambda}_{\phantom{\lambda}\sigma}
        - \nabla^{2}\dg_{\mu\nu}\right)
    \end{equation}

In what follows, we rewrite the background metric as 
    \begin{equation}
        ds^{2} = e^{2A}\left( ds^{2}(\mathcal{M}_{p+1})) + e^{-2A}ds^{2}_{\text{Int}} \right) = e^{2A} \hat{ds}^{2},
    \end{equation}

so that it takes the form of a Weyl rescaling of the metric $\hat{ds}^{2}$ of a product space. Since we are dealing with a Weyl rescaling we can use
    \begin{equation}
    \begin{aligned}
        \GUdd{\lambda}{\mu}{\nu} &= \hGUdd{\lambda}{\mu}{\nu}  
            +\left( \delta^{\lambda}_{\nu} \partial_{\mu}A  
            +\delta^{\lambda}_{\mu} \partial_{\nu}A
            -\hat{g}_{\mu\nu} \partial^{\lambda}A\right),\\
        \RUddd{\mu}{\nu}{\lambda}{\rho} &= \hRUddd{\mu}{\nu}{\lambda}{\rho} 
            - \delta^{\mu}_{\lambda} \hat{g}_{\nu\rho} (\partial A)^{2}
            + \delta^{\mu}_{\rho} \hat{g}_{\nu\lambda} (\partial A)^{2} + ...,\\
        R_{\mu\nu}  &= \hat{R}_{\mu\nu} - \hat{g}_{\mu\nu}\left( \hat{\square} A  -8 (\partial A)^{2} \right) + ...
    \end{aligned}
    \end{equation}

from where
    \begin{equation}
        \nabla_{\lambda}\dg_{\mu\nu} =
        e^{2A} \left(\hat{\nabla}_{\lambda}\bar{h}_{\mu\nu} 
        - \bar{h}_{\lambda\nu}\partial_{\mu}A
        - \bar{h}_{\lambda\mu}\partial_{\nu}A\right), 
    \end{equation}

and after a bit of algebra
    \begin{equation}
        \nabla^{\lambda}\nabla_{\lambda}\dg_{\mu\nu} 
        = \hat{\square}\bar{h}_{\mu\nu} +8 \partial^{\sigma}A \hat{\nabla}_{\sigma}\bar{h}_{\mu\nu}
        - 2\bar{h}_{\mu\nu}(\partial A)^{2}
    \end{equation}

we also use
    \begin{equation}
    \begin{aligned}
        R_{\sigma\mu}\dg^{\sigma}_{\phantom{\sigma}\nu}
        &= \hat{R}_{\sigma\mu} \bar{h}^{\sigma}_{\phantom{\sigma}\nu} - \bar{h}_{\mu\nu}\left( \hat{\square} A  -8 (\partial A)^{2} \right),\\
        \RUddd{\sigma}{\nu}{\lambda}{\mu} \dg^{\lambda}_{\phantom{\lambda}\sigma}
        &=\hRUddd{\sigma}{\nu}{\lambda}{\mu} \bar{h}^{\lambda}_{\phantom{\lambda}\sigma}
           + \bar{h}_{\mu\nu} (\partial A)^{2}
    \end{aligned}
    \end{equation}

Finally, \eqref{deltaR} reads
    \begin{equation}
        \delta R_{\mu\nu} =\frac{1}{2}\left(
        \hat{R}_{\sigma\mu} \bar{h}^{\sigma}_{\phantom{\sigma}\nu} 
        +\hat{R}_{\sigma\nu} \bar{h}^{\sigma}_{\phantom{\sigma}\mu}
        - 2\bar{h}_{\mu\nu}\left( \hat{\square} A  -8 (\partial A)^{2} \right)
        -2\hRdddd{\sigma}{\nu}{\lambda}{\mu} \bar{h}^{\lambda\sigma}
         - \hat{\square}\bar{h}_{\mu\nu} 
        -8 \partial^{\sigma}A \hat{\nabla}_{\sigma}\bar{h}_{\mu\nu}
        \right),
    \end{equation}

Now we use the fact that $\hat{ds}^{2}$ is a product space and that the macroscopic space is of constant curvature
    \begin{equation}
    \begin{aligned}
        \hat{R}_{abcd} &= k \left( \hat{g}_{ac}\hat{g}_{bd}
        -\hat{g}_{ad}\hat{g}_{bc} \right),\\
        R_{ab} &= kp \hat{g}_{ab}
    \end{aligned}
    \end{equation}

then
    \begin{equation}
        \delta R_{ab} =\frac{1}{2}\left(
        2k(p+1) \bar{h}_{ab} 
        - 2\bar{h}_{ab}\left( \hat{\square} A  -8 (\partial A)^{2} \right)
         - \hat{\square}\bar{h}_{ab} 
        -8 \partial^{\sigma}A \hat{\nabla}_{\sigma}\bar{h}_{ab}
        \right)
    \end{equation}

Also, note that
    \begin{equation}
        \delta (g_{\mu\nu}R) = \dg_{\mu\nu} R + g_{\mu\nu}\dg^{\lambda\rho}R_{\lambda\rho} + g_{\mu\nu}g^{\lambda\rho}\delta R_{\lambda\rho}
    \end{equation}

The second and third term do not contribute since $\dg_{\mu\nu}$ is traceless. Then, specialising to $(\mu,\nu)=(a,b)$
    \begin{equation}
        \delta (g_{ab}R) = \dg_{ab} R 
    \end{equation}

Putting everything together, from the variation of Einstein's equations
    \begin{equation}
        \delta R_{ab} - \frac{1}{2}R \dg_{ab}  = \delta T_{ab},
    \end{equation}
we have
    \begin{equation}\label{eqPert}
        \frac{1}{2}\left( 2k(p+1) \bar{h}_{ab} 
        - 2\bar{h}_{ab}\left( \hat{\square} A  -8 (\partial A)^{2} \right)
         - \hat{\square}\bar{h}_{ab} 
        -8 \partial^{\sigma}A \hat{\nabla}_{\sigma}\bar{h}_{ab}
        \right)
        = \frac{1}{p+1} R^{c}_{\phantom{c}c}\delta g_{ab}.
    \end{equation}

Finally, by noting that in our ansatz
    \begin{equation}
        R^{c}_{\phantom{c}c}
        = e^{-2A}(p+1)\left(kp - \hat{\square}A - 8(\partial A)^{2}\right),
    \end{equation}

we can write \eqref{eqPert} as
    \begin{equation}
        2k \bar{h}_{ab} 
         - \hat{\square}\bar{h}_{ab} 
        -8 \partial^{\sigma}A \hat{\nabla}_{\sigma}\bar{h}_{ab}
        = 0.
    \end{equation}

Writing $\bar{h}_{ab}$ in terms of $h_{ab}$ and $\psi$ leads to
    \begin{equation}\label{eqPert1}
        2k h_{ab}\psi - \psi \hat{\square}_{x}h_{ab} - h_{ab} \hat{\square}_{y} \psi - 8 h_{ab}\partial^{i}A \hat{\nabla}_{i}\psi = 0
    \end{equation}

Suppose that $h_{ab}$ satisfies
    \begin{equation}
        \hat{\square}_{x}h_{ab} - \lambda h_{ab}=0,
    \end{equation}

Then \eqref{eqPert1}
    \begin{equation}
        (2k-\lambda) h_{ab}\psi - h_{ab} \hat{\square}_{y} \psi - 8 h_{ab}\partial^{i}A \hat{\nabla}_{i}\psi = 0
    \end{equation}

from where (here we define $M^{2} = \lambda-2k$)
    \begin{equation}\label{eqFluctuation1}
        \hat{\square}_{y} \psi  + 8 \hat{g}^{ij}\partial_{i}A \partial_{j}\psi +M^{2} \psi = 0
    \end{equation}

\subsection{Change of Variables}\label{AppendixCChangeofVariables}

We will check the two changes of variables used in \cite{Bachas:2011xa}. First we rewrite \eqref{eqFluctuation1} in terms of $g_{ij}=e^{2A}\hat{g}_{ij}$
    \begin{equation}
        e^{(1-p)A} \frac{1}{\sqrt{g}}\partial_{i}\left(\sqrt{g}e^{(p+1)A} g^{ij}\partial_{j}\psi\right)
         + M^{2}\psi = 0.
    \end{equation}

From here, a well defined Sturm-Liouville norm
    \begin{equation}
        ||\psi||^{2} = \int d^{9-p}y \sqrt{\det(g_{ij})}\, e^{(p-1)A} |\psi|^{2},
    \end{equation}
leads to the a positiveness condition for $M^{2}$
    \begin{equation}\label{Norm1}
        M^{2}||\psi||^{2} 
        =  -\int d^{9-p}y \psi\partial_{i}\left(\sqrt{\det(g_{ij})} 
            e^{(p+1)A} g^{ij}\partial_{j}\psi\right) 
        = \int d^{9-p}y \sqrt{\det(g_{ij})}e^{(p+1)A} (\partial\psi)^{2} \geq 0,
    \end{equation}

provided
    \begin{equation}
        \int d^{9-p}y\, \partial_{i}\left(\psi\sqrt{\det(g_{ij})}e^{(p+1)A} g^{ij}\partial_{j}\psi\right)=0,
    \end{equation}
which is satisfied by imposing Dirichlet or Neumann boundary conditions.

Another useful change of variables is $\psi = e^{-4A}\Psi$, which leads to
    \begin{equation}\label{BeforeSchroedinger}
        \hat{\square}_{y} \Psi
             + \Psi e^{4A}\hat{\square}_{y}  e^{-4A}
             -32 \Psi\hat{g}^{ij}\partial_{i}A\partial_{j}A + M^{2}\Psi = 0.
    \end{equation}

Noting that
    \begin{equation}
        - e^{4A}\hat{\square}_{y}  e^{-4A} + 32\hat{g}^{ij}\partial_{i}A\partial_{j}A = e^{-4A}\hat{\square}_{y}  e^{4A} 
    \end{equation}

we can write \eqref{BeforeSchroedinger} in a Schroedinger like form
    \begin{equation}\label{SchroedingerLikeAppendix}
        - \hat{\square}_{y} \Psi + V(y)\Psi =  M^{2}\Psi,
    \end{equation}

where the effective potential is given by
    \begin{equation}
        V(y) = e^{-4A}\hat{\square}_{y}  e^{4A}.
    \end{equation}

Note that in terms of $\Psi$ and $\hat{g}_{ij}$, \eqref{Norm1} takes the form (here we use $\sqrt{\det(g_{ij})} = e^{(9-p)A}\sqrt{\det(\hat{g}_{ij})}$ and $|\psi|^{2} = e^{-8A}|\Psi|^{2}$)
    \begin{equation}
        ||\psi||^{2} = \int d^{9-p}y \sqrt{\det(\hat{g}_{ij})}\, |\Psi|^{2},
    \end{equation}

\subsection{Schroedinger Like Equation}\label{SchroedingerAppendix}

Note that although \eqref{SchroedingerLikeAppendix} has the form of a Schroedinger equation, the Laplace operator can still contain first derivatives of $\Psi$. Here we show how to obtain an equation that only contains second derivatives of the function. This is achieved by using a change of variables and a tortoise coordinate for the radial direction.

In what follows we consider internal spaces of the form
    \begin{equation}
        ds^{2}_{\text{Int}} = f(r)d\Omega^{2}_{d} + \frac{dr^{2}}{g(r)}  + ds^{2}(\tilde{M}_{8-p-d}),
    \end{equation}

where $\Omega^{2}_{d}$ is a $d$-dimensional compact space and  and $\tilde{M}_{8-p-d}$ is a $(8-p-d)$-dimensional space, possibly including fibrations over $\Omega_{d}$. We will denote the unfibered version as $M_{8-p-d}$, and we assume that the fibration is such that $\text{Vol}(\tilde{M}_{8-p-d})=\text{Vol}(M_{8-p-d})$. Also, we assume $A=A(r)$.

In what follows we will consider metric fluctuations that only depend on $\Omega_{d}$ and $r$, that is, we take the S-wave $M_{8-p-d}$. Then
    \begin{equation}
        \hat{\square}_{y} \Psi  
        = \frac{1}{f(r)^{\frac{d}{2}}g(r)^{-\frac{1}{2}}} \frac{d}{dr} \left( 
        f(r)^{\frac{d}{2}}g(r)^{\frac{1}{2}}  \frac{d\Psi}{dr} \right) + \frac{1}{f(r)}\nabla^{2}_{\Omega}\Psi,
    \end{equation}

with $\nabla^{2}_{\Omega}$ the Laplace operator in $\Omega_{d}$. By using the tortoise coordinate $d\rho = dr/\sqrt{g(r)}$ and using the change of variables
    \begin{equation}\label{VariableChange}
        \Psi = f(r)^{-\frac{d}{4}} \Theta(\rho) Y_{l}(\Omega),
    \end{equation}

with $Y_{l}(\Omega)$ the eigenfunction of $\nabla^{2}_{\Omega}$, i.e. $\nabla^{2}_{\Omega}Y_{l} = -l(l+d-1)$, we can write \eqref{SchroedingerLikeAppendix} as
    \begin{equation}
        - \frac{d^{2}\Theta}{d\rho^{2}}
         +\tilde{V}(\rho)\, \Theta(\rho) =  M^{2} \Theta(\rho),
    \end{equation}

where the effective potential $\tilde{V}$ is given by
    \begin{equation}\label{EffectivePotentialAppendix}
        \tilde{V}(\rho) = \left( V(r) + \frac{l(l+d-1)}{f(r)}   - \frac{g(r)^{\frac{1}{2}}}{f(r)^{\frac{d}{4}}}\frac{d}{dr}\left( 
         f(r)^{\frac{d}{2}}g(r)^{\frac{1}{2}}  \frac{d}{dr}\left(f(r)^{-\frac{d}{4}}\right) \right) \right) \bigg|_{r = r(\rho)}.
    \end{equation}

\subsection{Glueball Spectrum in the (4+1)d Confining Theory}

We start from \eqref{einstein}. For this case we have
    \begin{eqnarray}
  & &  e^{4A(\vec{y})}= r,\;\;\;\;\; ds^{2}(\mathcal{M}_{p+1})= dx_{1,4}^2,\\
        & & \hat{g}_{ab}dy^a dy^b=  f_{s}(r)d\varphi^{2} 
            + \frac{Ndr^{2}}{ r^{2}f_{s}(r)} 
            + \frac{N}{4}\left(\omega^{2}_{1} + \omega^{2}_{2} + 
        \left(\omega_{3} - \sqrt{\frac{8}{N}} Q \zeta(r) d\varphi\right)^{2}\right) . \nonumber
        \end{eqnarray}

We identify $d\Omega^{2}_{d} = d\varphi^{2}$, hence $d=1$. From \eqref{SchroedingerLikeAppendix} and \eqref{EffectivePotentialAppendix} we have
 \begin{eqnarray}
         - \frac{d^2\Theta}{d\rho^2} +\tilde{V}(\rho) \Theta = m^2\Theta,\;\;\;\
        \tilde{V}(\rho)= \left( V(r) +  \frac{j^{2}}{f_{s}(r)}\left(\frac{2\pi}{L_{\varphi}}\right)^{2} -\frac{r}{N} f_s^{1/4} \frac{d}{dr} \left[ r f_s \frac{d f_s^{-1/4}}{dr}  \right]\right)\bigg|_{r=r(\rho)}.
    \end{eqnarray}  
            
Explicitly, the effective potential reads
    \begin{equation}
        \tilde{V}(\rho) = \frac{1}{N} 
        - \frac{1}{N}\frac{1}{\sinh^{2}\left(\frac{2\rho}{\sqrt{N}}\right)} 
        + \frac{n^2}{N} \coth^{2}\left(\frac{\rho}{\sqrt{N}}\right)
        \left( 1- \frac{r^{2}_{-}}{r^{2}_{+}}\tanh^{2}\left(\frac{\rho}{\sqrt{N}}\right) \right)^{2}.
    \end{equation}

In order to solve the Schroedinger equation, it is convenient to perform the change of variables
    \begin{equation}
        z=\cosh \left( \frac{\rho}{\sqrt{N}}\right),
    \end{equation}

which maps the region $\rho \in \lbrack 0,\infty \lbrack $ to $z\in \lbrack
1,\infty \lbrack $, which leads to the following equation for the radial profile
    \begin{equation}
        -\left( z^{2}-1\right) \frac{d^{2}\Theta }{dz^{2}}-z\frac{d\Theta }{dz}
        +\left( 1 - \frac{1}{4}\frac{1}{z^{2}(z^{2}-1)} + n^{2} \frac{\left( z^{2} + \lambda (z^{2}-1) \right)^{2}}{z^{2}(z^{2}-1)}\right) \Theta \left( z\right) =NM^{2}\Theta
        \left( z\right).
    \end{equation}

This equation is solved in terms of hypergeometric functions as
    \begin{equation}\label{SolutionRadialProfile}
    \begin{aligned}
        \Theta(z)
        &=  (z^{2}-1)^{\frac{1}{4}-\frac{n}{2}}\left(z^{2}+\lambda(z^{2}-1)\right)^{\frac{1}{4}}\\
        &\phantom{=} \times 
        \left[   
         C_{1} z^{n\lambda} \hypgeom{a_{+}}{a_{-}}{c_{-}}{z^{2}} 
         + C_{2} z^{-n\lambda} \hypgeom{b_{+}}{b_{-}}{c_{+}}{z^{2}}  \right].
    \end{aligned} 
    \end{equation}

where
    \begin{equation}
    \begin{aligned}
        a_{\pm} &= \frac{1}{2}\left( 1- n(\lambda+1) \pm \sqrt{1- M^{2}N+n^{2}(1+\lambda)^{2}} \right),\\
        b_{\pm} &= \frac{1}{2}\left( 1+ n(\lambda-1) \pm \sqrt{1- M^{2}N+n^{2}(1+\lambda)^{2}} \right),\\
        c_{\pm} &= 1\pm \lambda n.
    \end{aligned}
    \end{equation}

In order to impose boundary conditions, we need the full radial profile of the metric fluctuation. Recall that $\delta g_{\mu\nu} = e^{2A}\bar{h}_{\mu\nu} = e^{2A}h_{ab}(x)\psi(r,\varphi)$, and also that $\psi = e^{-4A}\Psi= e^{-4A}f_{s}^{-\frac{1}{4}} \Theta e^{i \frac{2\pi}{L_{\varphi}}n\varphi}$, so that the complete radial profile is
    \begin{equation}\label{RadialProfileAppendix}
    \begin{aligned}
        e^{2A}\psi &=(z^{2}-1)^{-\frac{n}{2}}\left(z^{2}+\lambda(z^{2}-1)\right)^{\frac{1}{4}}\\
        &\phantom{=} \times 
        \left[   
         C_{1} z^{n\lambda} \hypgeom{a_{+}}{a_{-}}{c_{-}}{z^{2}} 
         + C_{2} z^{-n\lambda} \hypgeom{b_{+}}{b_{-}}{c_{+}}{z^{2}}  \right]
    \end{aligned}
    \end{equation}

To impose boundary conditions at $z=1$ and $z\rightarrow +\infty$ we use the following Kumar identities for the hypergeometric functions 
    \begin{align}
        &\begin{aligned}
            \hypgeom{a}{b}{c}{z}&= \frac{\Gamma(c)\Gamma(a+b-c)}{\Gamma(a)\Gamma(b)}(1-z)^{c-a-b} \hypgeom{c-a}{c-b}{c-a-b+1}{1-z} \\
            &\phantom{=} + \frac{\Gamma(c)\Gamma(c-a-b)}{\Gamma(c-a)\Gamma(c-b)}(1-z)^{c-a-b} \hypgeom{a}{b}{a+b-c+1}{1-z}
        \end{aligned} \label{Kumar1} ,\\
        &\begin{aligned}
            \hypgeom{a}{b}{c}{z}&= \frac{\Gamma(b-z)\Gamma(c)}{\Gamma(b)\Gamma(c-a)}(-z)^{-a} \hypgeom{a}{a-c+1}{a-b+1}{\frac{1}{z}} \\
            &\phantom{=} + \frac{\Gamma(a-b)\Gamma(c)}{\Gamma(a)\Gamma(c-b)}(-z)^{-b} \hypgeom{b}{b-c+1}{-a+b+1}{\frac{1}{z}}
        \end{aligned}.\label{Kumar2}
    \end{align}

These are convenient since they allow us to center the Hypergeometric functions at zero (recall that $\hypgeom{a}{b}{c}{0}=1$) when expanding a around the $z=1$ and $z\rightarrow +\infty$. Using \eqref{Kumar1} and expanding \eqref{RadialProfileAppendix} around $z=1$ we obtain 
    \begin{equation}\label{GlueballsOriginAppendix}
    \begin{aligned}
        \lim_{z\rightarrow 1}e^{2A}\tilde{\psi} 
        &= (z^{2}-1)^{\frac{n}{2}} \Gamma(-n) \left(C_{1} \frac{\Gamma(c_{-})}{\Gamma(a_{+})\Gamma(a_{-})} + C_{2}\frac{\Gamma(c_{+})}{\Gamma(b_{+})\Gamma(b_{-})} \right) 
        \\ 
        &\phantom{=} + (z^{2}-1)^{-\frac{n}{2}} \Gamma(n)\left( 
        C_{1} \frac{\Gamma(c_{-})}{\Gamma(c_{-}-a_{+})\Gamma(c_{-}-a_{-})} +
        C_{2}\frac{\Gamma(c_{+})}{\Gamma(c_{+}-b_{+})\Gamma(c_{+}-b_{-})} \right),
    \end{aligned}    
    \end{equation}

while for  $z\rightarrow +\infty$ we use \eqref{Kumar2} the expansion reads
    \begin{equation}\label{GlueballsInfinityAppendix}
    \begin{aligned}
        \lim_{z\rightarrow +\infty}e^{2A}\tilde{\psi} 
        &= z^{-\frac{1}{2}-\sqrt{1-M^{2}N+n^{2}(1+\lambda)^{2}}} \left(C_{1} \frac{\Gamma(a_{-}-a_{+})\Gamma(c_{-})}{\Gamma(a_{-})\Gamma(c_{-}-a_{+})} + C_{2}\frac{\Gamma(b_{+}-b_{-})\Gamma(c_{+})}{\Gamma(b_{+})\Gamma(c_{+}-b_{-})} \right) 
        \\ 
        &\phantom{=} +z^{-\frac{1}{2}+\sqrt{1-M^{2}N+n^{2}(1+\lambda)^{2}}} \left(C_{1} \frac{\Gamma(a_{+}-a_{-})\Gamma(c_{-})}{\Gamma(a_{+})\Gamma(c_{-}-a_{-})} + C_{2}\frac{\Gamma(b_{-}-b_{+})\Gamma(c_{+})}{\Gamma(b_{-})\Gamma(c_{+}-b_{+})} \right).
    \end{aligned}    
    \end{equation}


\begin{thebibliography}{99}

\bibitem{Maldacena:1997re}
J.~M.~Maldacena,
Adv. Theor. Math. Phys. \textbf{2}, 231-252 (1998)
[arXiv:hep-th/9711200 [hep-th]].
\bibitem{Gubser:1998bc}
S.~S.~Gubser, I.~R.~Klebanov and A.~M.~Polyakov,
Phys. Lett. B \textbf{428}, 105-114 (1998)
[arXiv:hep-th/9802109 [hep-th]].
\bibitem{Witten:1998qj}
E.~Witten,
Adv. Theor. Math. Phys. \textbf{2}, 253-291 (1998)
[arXiv:hep-th/9802150 [hep-th]].

\bibitem{Itzhaki:1998dd}
N.~Itzhaki, J.~M.~Maldacena, J.~Sonnenschein and S.~Yankielowicz,
Phys. Rev. D \textbf{58}, 046004 (1998)
[arXiv:hep-th/9802042 [hep-th]].

\bibitem{Witten:1998zw}
E.~Witten,
Adv. Theor. Math. Phys. \textbf{2}, 505-532 (1998)
[arXiv:hep-th/9803131 [hep-th]].

\bibitem{Boonstra:1998mp}
H.~J.~Boonstra, K.~Skenderis and P.~K.~Townsend,
JHEP \textbf{01}, 003 (1999)
[arXiv:hep-th/9807137 [hep-th]].


\bibitem{Klebanov:1998hh}
I.~R.~Klebanov and E.~Witten,
Nucl. Phys. B \textbf{536}, 199-218 (1998)
[arXiv:hep-th/9807080 [hep-th]].
\bibitem{Klebanov:2000nc}
I.~R.~Klebanov and A.~A.~Tseytlin,
Nucl. Phys. B \textbf{578}, 123-138 (2000)
[arXiv:hep-th/0002159 [hep-th]].
\bibitem{Klebanov:2000hb}
I.~R.~Klebanov and M.~J.~Strassler,
JHEP \textbf{08}, 052 (2000)
doi:10.1088/1126-6708/2000/08/052
[arXiv:hep-th/0007191 [hep-th]].

\bibitem{Dymarsky:2005xt}
A.~Dymarsky, I.~R.~Klebanov and N.~Seiberg,
JHEP \textbf{01}, 155 (2006)
[arXiv:hep-th/0511254 [hep-th]].
\bibitem{Gubser:2004qj}
S.~S.~Gubser, C.~P.~Herzog and I.~R.~Klebanov,
JHEP \textbf{09}, 036 (2004)
[arXiv:hep-th/0405282 [hep-th]].

\bibitem{Butti:2004pk}
A.~Butti, M.~Grana, R.~Minasian, M.~Petrini and A.~Zaffaroni,
JHEP \textbf{03}, 069 (2005)
[arXiv:hep-th/0412187 [hep-th]].









\bibitem{Maldacena:2000yy}
J.~M.~Maldacena and C.~Nunez,
Phys. Rev. Lett. \textbf{86}, 588-591 (2001)
[arXiv:hep-th/0008001 [hep-th]].
See also 
A.~H.~Chamseddine and M.~S.~Volkov,
Phys. Rev. Lett. \textbf{79}, 3343-3346 (1997)
[arXiv:hep-th/9707176 [hep-th]].

\bibitem{Maldacena:2001pb}
J.~M.~Maldacena and H.~S.~Nastase,
JHEP \textbf{09}, 024 (2001)
[arXiv:hep-th/0105049 [hep-th]].

\bibitem{Gauntlett:2001ps}
J.~P.~Gauntlett, N.~Kim, D.~Martelli and D.~Waldram,
Phys. Rev. D \textbf{64}, 106008 (2001)
[arXiv:hep-th/0106117 [hep-th]].

\bibitem{Bigazzi:2001aj}
F.~Bigazzi, A.~L.~Cotrone and A.~Zaffaroni,
Phys. Lett. B \textbf{519}, 269-276 (2001)
[arXiv:hep-th/0106160 [hep-th]].

\bibitem{Apreda:2001qb}
R.~Apreda, F.~Bigazzi, A.~L.~Cotrone, M.~Petrini and A.~Zaffaroni,
Phys. Lett. B \textbf{536}, 161-168 (2002)
[arXiv:hep-th/0112236 [hep-th]].

\bibitem{Bigazzi:2002gyi}
F.~Bigazzi, A.~L.~Cotrone, M.~Petrini and A.~Zaffaroni,
Riv. Nuovo Cim. \textbf{25N12}, 1-70 (2002)
[arXiv:hep-th/0303191 [hep-th]].

\bibitem{Gomis:2001aa}
J.~Gomis and J.~G.~Russo,
JHEP \textbf{10}, 028 (2001)
[arXiv:hep-th/0109177 [hep-th]].

\bibitem{Aharony:2002up}
O.~Aharony,
[arXiv:hep-th/0212193 [hep-th]].


\bibitem{Maldacena:2009mw}
J.~Maldacena and D.~Martelli,
JHEP \textbf{01}, 104 (2010)
[arXiv:0906.0591 [hep-th]].

\bibitem{Gaillard:2010qg}
J.~Gaillard, D.~Martelli, C.~Nunez and I.~Papadimitriou,
Nucl. Phys. B \textbf{843}, 1-45 (2011)
[arXiv:1004.4638 [hep-th]].

\bibitem{Gaillard:2010gy}
J.~Gaillard and D.~Martelli,
JHEP \textbf{05}, 109 (2011)
[arXiv:1008.0640 [hep-th]].





\bibitem{Anabalon:2021tua}
A.~Anabalon and S.~F.~Ross,
JHEP \textbf{07}, 015 (2021)
[arXiv:2104.14572 [hep-th]].

\bibitem{Canfora:2021nca}
F.~Canfora, J.~Oliva and M.~Oyarzo,
JHEP \textbf{02} (2022), 057
[arXiv:2111.11915 [hep-th]].


\bibitem{Anabalon:2022aig}
A.~Anabal\'on, A.~Gallerati, S.~Ross and M.~Trigiante,
JHEP \textbf{02}, 055 (2023)
[arXiv:2210.06319 [hep-th]].

\bibitem{Bobev:2020pjk}
N.~Bobev, A.~M.~Charles and V.~S.~Min,
JHEP \textbf{10}, 073 (2020)
[arXiv:2006.01148 [hep-th]].

\bibitem{Nunez:2023nnl}
C.~Nunez, M.~Oyarzo and R.~Stuardo,
JHEP \textbf{09}, 201 (2023)
[arXiv:2307.04783 [hep-th]].



\bibitem{Sonnenschein:1999if}
J.~Sonnenschein,
[arXiv:hep-th/0003032 [hep-th]].

\bibitem{Nunez:2009da}
C.~Nunez, M.~Piai and A.~Rago,
Phys. Rev. D \textbf{81}, 086001 (2010)
[arXiv:0909.0748 [hep-th]].


\bibitem{Kol:2014nqa}
U.~Kol, C.~Nunez, D.~Schofield, J.~Sonnenschein and M.~Warschawski,
JHEP \textbf{06}, 005 (2014)
[arXiv:1403.2721 [hep-th]].



\bibitem{Faedo:2014naa}
A.~F.~Faedo, M.~Piai and D.~Schofield,
Phys. Rev. D \textbf{89}, no.10, 106001 (2014)
[arXiv:1402.4141 [hep-th]].


\bibitem{Rey:1998ik}
S.~J.~Rey and J.~T.~Yee,
Eur. Phys. J. C \textbf{22}, 379-394 (2001)
[arXiv:hep-th/9803001 [hep-th]].




\bibitem{Maldacena:1998im}
J.~M.~Maldacena,
Phys. Rev. Lett. \textbf{80}, 4859-4862 (1998)
[arXiv:hep-th/9803002 [hep-th]].



\bibitem{Macpherson:2014eza}
N.~T.~Macpherson, C.~N\'u\~nez, L.~A.~Pando Zayas, V.~G.~J.~Rodgers and C.~A.~Whiting,
JHEP \textbf{02}, 040 (2015)
[arXiv:1410.2650 [hep-th]].

\bibitem{Bea:2015fja}
Y.~Bea, J.~D.~Edelstein, G.~Itsios, K.~S.~Kooner, C.~Nunez, D.~Schofield and J.~A.~Sierra-Garcia,
JHEP \textbf{05}, 062 (2015)
[arXiv:1503.07527 [hep-th]].



\bibitem{Ryu:2006bv}
S.~Ryu and T.~Takayanagi,
Phys. Rev. Lett. \textbf{96}, 181602 (2006)
[arXiv:hep-th/0603001 [hep-th]].


\bibitem{Klebanov:2007ws}
I.~R.~Klebanov, D.~Kutasov and A.~Murugan,
Nucl. Phys. B \textbf{796}, 274-293 (2008)
[arXiv:0709.2140 [hep-th]].





\bibitem{Jokela:2020wgs}
N.~Jokela and J.~G.~Subils,
JHEP \textbf{02}, 147 (2021)
[arXiv:2010.09392 [hep-th]].
N.~Jokela, H.~Ruotsalainen and J.~G.~Subils,
[arXiv:2310.11205 [hep-th]].

\bibitem{Anabalon:2023lnk}
A.~Anabal\'on and H.~Nastase,
[arXiv:2310.07823 [hep-th]].



\bibitem{Caceres:2005yx}
E.~Caceres and C.~Nunez,
JHEP \textbf{09}, 027 (2005)
doi:10.1088/1126-6708/2005/09/027
[arXiv:hep-th/0506051 [hep-th]].





\bibitem{Bachas:2011xa}
C.~Bachas and J.~Estes,
JHEP \textbf{06}, 005 (2011)
doi:10.1007/JHEP06(2011)005
[arXiv:1103.2800 [hep-th]].




\bibitem{DeLuca:2023kjj}
G.~B.~De Luca, N.~De Ponti, A.~Mondino and A.~Tomasiello,
JHEP \textbf{09}, 127 (2023)
doi:10.1007/JHEP09(2023)127
[arXiv:2306.05456 [hep-th]].

\bibitem{Fatemiabhari:2024aua}
A.~Fatemiabhari and C.~Nunez,
[arXiv:2401.04158 [hep-th]].


\bibitem{Canfora:2012ap}
F.~Canfora, F.~Correa, A.~Giacomini and J.~Oliva,
Phys. Lett. B \textbf{722}, 364-371 (2013)
doi:10.1016/j.physletb.2013.04.029
[arXiv:1208.6042 [hep-th]].




\bibitem{Casero:2006pt}
R.~Casero, C.~Nunez and A.~Paredes,
Phys. Rev. D \textbf{73}, 086005 (2006)
[arXiv:hep-th/0602027 [hep-th]]. 

\bibitem{Casero:2007jj}
R.~Casero, C.~Nunez and A.~Paredes,
Phys. Rev. D \textbf{77}, 046003 (2008)
[arXiv:0709.3421 [hep-th]].

\bibitem{Hoyos-Badajoz:2008znk}
C.~Hoyos-Badajoz, C.~Nunez and I.~Papadimitriou,
Phys. Rev. D \textbf{78}, 086005 (2008)
[arXiv:0807.3039 [hep-th]].


\bibitem{Andrews:2005cv}
R.~P.~Andrews and N.~Dorey,
Phys. Lett. B \textbf{631}, 74-82 (2005)
[arXiv:hep-th/0505107 [hep-th]].
Nucl. Phys. B \textbf{751}, 304-341 (2006)
[arXiv:hep-th/0601098 [hep-th]].




\bibitem{Wald:1993nt}
R.~M.~Wald,
``Black hole entropy is the Noether charge,''
Phys. Rev. D \textbf{48} (1993) no.8, R3427-R3431
doi:10.1103/PhysRevD.48.R3427
[arXiv:gr-qc/9307038 [gr-qc]].

\bibitem{Emparan:1999pm}
R.~Emparan, C.~V.~Johnson and R.~C.~Myers,
Phys. Rev. D \textbf{60} (1999), 104001
doi:10.1103/PhysRevD.60.104001
[arXiv:hep-th/9903238 [hep-th]].
\bibitem{Balasubramanian:1999re}
V.~Balasubramanian and P.~Kraus,
Commun. Math. Phys. \textbf{208} (1999), 413-428
doi:10.1007/s002200050764
[arXiv:hep-th/9902121 [hep-th]].
\bibitem{Bianchi:2001kw}
M.~Bianchi, D.~Z.~Freedman and K.~Skenderis,
Nucl. Phys. B \textbf{631} (2002), 159-194
doi:10.1016/S0550-3213(02)00179-7
[arXiv:hep-th/0112119 [hep-th]].

\bibitem{Mann:2005yr}
R.~B.~Mann and D.~Marolf,
Class. Quant. Grav. \textbf{23} (2006), 2927-2950
doi:10.1088/0264-9381/23/9/010
[arXiv:hep-th/0511096 [hep-th]].




\bibitem{Kraus:1999di}
P.~Kraus, F.~Larsen and R.~Siebelink,
Nucl. Phys. B \textbf{563} (1999), 259-278
doi:10.1016/S0550-3213(99)00549-0
[arXiv:hep-th/9906127 [hep-th]].
\bibitem{Anabalon:2023kcp}
A.~Anabal\'on, \'A.~Arboleya and A.~Guarino,
[arXiv:2312.13955 [hep-th]].

\bibitem{Caceres:2023gfa}
N.~C\'aceres, C.~Corral, F.~Diaz and R.~Olea,
[arXiv:2311.04054 [hep-th]].

\bibitem{Jackiw:1976xx}
R.~Jackiw and C.~Rebbi,
Phys. Rev. Lett. \textbf{36}, 1116 (1976)
P.~Hasenfratz and G.~'t Hooft,
Phys. Rev. Lett. \textbf{36}, 1119 (1976)

\bibitem{Bobev:2019bvq}
N.~Bobev, P.~Bomans, F.~F.~Gautason, J.~A.~Minahan and A.~Nedelin,
JHEP \textbf{03}, 047 (2020)
[arXiv:1910.08555 [hep-th]].

\end{thebibliography}
\end{document}